\documentclass[twocolumn, superscriptaddress,showpacs,amsmath,amssymb,aps,pra,floatfix,longbibliography, 10pt]{revtex4-1}

\usepackage[latin1]{inputenc}
\usepackage[american]{babel}

\usepackage[T1]{fontenc}

\usepackage{graphicx}

\usepackage{amsfonts}
\usepackage{amsmath}
\usepackage{bbold}
\usepackage{mathrsfs}
\usepackage{hyperref}
\usepackage{epstopdf}
\usepackage{bm, epsfig}

\usepackage[tbtags]{mathtools}
\usepackage{afterpage}

\DeclarePairedDelimiterX\braket[2]{\langle}{\rangle}{#1 \delimsize\vert #2}

\hypersetup{colorlinks=true, urlcolor=blue, citecolor=blue, filecolor=blue, linkcolor=blue}
\newcommand*{\diff}{\mathop{}\!\mathrm{d}}

\newcommand{\uimm}{\mathrm{i}}
\newcommand{\eu}{\mathrm{e}}

\newcommand{\daga}{^{\dagger}}

\begin{document}

\title{Simulating strong-field electron--hole dynamics in solids probed by attosecond transient absorption spectroscopy}

\author{Stefano~M.~Cavaletto}
\email[Email: ]{smcavaletto@gmail.com}
\altaffiliation[Current address: ]{Departamento de Qu\'imica, Universidad Aut\'onoma de Madrid, 28049 Madrid, Spain.}
\affiliation{Department of Physics and Astronomy, Ny Munkegade 120, Aarhus University, 8000 Aarhus C, Denmark}
\author{Lars Bojer Madsen}
\email[Email: ]{bojer@phys.au.dk}
\affiliation{Department of Physics and Astronomy, Ny Munkegade 120, Aarhus University, 8000 Aarhus C, Denmark}
\date{\today}

\begin{abstract}

We investigate the ultrafast electron dynamics of a model of a wide-bandgap material with inner, valence, and conduction bands excited by an intense few-femtosecond pump and monitored by a delayed attosecond extreme-ultraviolet probe pulse. Complementary  computational methods are utilized and compared, based on the semiconductor Bloch equations (SBEs) and time-dependent density functional theory (TDDFT). TDDFT is employed to study a finite-size system, while the SBEs are utilized to investigate the corresponding solid with periodic boundary conditions imposed, with the crystal-momentum-dependent energy bands and interband couplings calculated in the parallel-transport structure gauge. The resulting strong-field electron dynamics are employed to predict experimentally accessible attosecond transient absorption spectroscopy (ATAS) signals as a function of the probe-pulse frequency and pump--probe interpulse delay. Both simulation protocols similarly capture the time-delay-dependent spectral features in the ATAS signals. The very good agreement between our TDDFT and SBE-based results allows us to interpret the \textit{ab-initio} TDDFT simulations in terms of SBEs' interband couplings, validating our SBE-based model and corroborating its conclusions.

\end{abstract}

\maketitle
%\onecolumngrid

\section{Introduction}

Over the past decade, the investigation of strong-field attosecond dynamics in solids has received increasing attention \cite{li_attosecond_2020, geneaux_transient_2019}. This was triggered, on the one hand, by the first demonstration of nonperturbative high-order harmonic generation (HHG) in solids in the infrared regime \cite{Ghimire2011}, opening prospects for novel, more compact sources of extreme-ultraviolet (XUV) radiation \cite{garg_ultimate_2018}, and for the time-resolved HHG spectroscopy of condensed-matter systems \cite{luu_extreme_2015}. In this context, solid-state HHG has been investigated for band-structure reconstruction \cite{vampa_all-optical_2015} and for accessing ultrafast electronic currents, which has led to an intense debate on the relationship between HHG and the underlying strong-field-induced inter- and intraband currents \cite{schubert_sub-cycle_2014, higuchi2014strong, wu2015high, McDonald2015, hohenleutner2015real, vampa2015all, ndabashimiye2016solid, Garg2016}. On the other hand, very intense few-femtosecond laser pulses have been shown to transiently and reversibly modify the optical and electronic properties of solids \cite{schiffrin_optical-field-induced_2013, Schultze2013, mashiko_petahertz_2016, sommer_attosecond_2016, Garg2016, higuchi_light-field-driven_2017, PhysRevResearch.3.023250}, with possible applications towards ultrafast, petahertz optoelectronic devices overcoming the speed limits of contemporary digital electronics \cite{ossiander_speed_2022}. An increasing number of studies have employed ultrafast time-resolved techniques, ranging from time- and angle-resolved photoelectron spectroscopy (ARPES) \cite{wang_observation_2013} via attosecond transient absorption and reflection spectroscopy (ATAS and ATRS) \cite{geneaux_transient_2019, DiPalo2024} through multidimentional nonlinear spectroscopy \cite{PhysRevB.103.245140}, to access the ultrafast electronic response and monitor the strong-field-induced charge injection from the valence into the conduction band in (wide-bandgap) semiconductors and insulators.

When monitoring strong-field electron dynamics in solids, ATAS and ATRS provide several advantages \cite{geneaux_transient_2019, DiPalo2024}. These transient techniques employ a pump--probe setup, where the electron dynamics induced by a few-femtosecond intense pump, typically at near-infrared frequencies, are monitored by a suitably delayed attosecond XUV probe pulse \cite{Schultze2013, Schultze2014, Lucchini2016,mashiko_petahertz_2016,  Jager2017, zurch_direct_2017, moulet_soft_2017, schlaepfer_attosecond_2018, volkov_attosecond_2019, buades_attosecond_2021, PhysRevB.107.184304, Inzani2023a}. In ATAS, the attosecond probe pulse is spectrally dispersed and its absorption spectrum is recorded as a function of the pump--probe delay, whereas ATRS utilizes a reflection geometry. Thanks to the ultrashort duration of the attosecond XUV probe, ATAS and ATRS provide high temporal resolution for monitoring the strong-field electron dynamics taking place \textit{within} an intense pulse. Transient changes in the absorption or reflection signal as a function of the interpulse delay encode information on the strong-field-driven electron dynamics. In addition to ultrashort duration and high termporal resolution, attosecond XUV pulses also provide an extremely broad spectral bandwidth, allowing one to simultaneously interrogate a large set of valence bands (VBs) and conduction bands (CBs) \cite{zurch_direct_2017}. Finally, a third significant advantage is provided by the large carrier frequency of the attosecond XUV probe. This enables the excitation of the spatially more localized inner-valence electrons, thereby providing sensitivity to their local chemical environment \cite{buades_attosecond_2021}. With advances in the generation of attosecond pulses at even higher soft- and hard-x-ray frequencies at free-electron lasers \cite{pellegrini_physics_2016, duris_tunable_2020, malyzhenkov_single-_2020, guo_experimental_2024}, x-ray ATAS promises the direct excitation of core electrons with atom-specific transition energies, providing a spatially selective probe of ultrafast electron dynamics.

Understanding and interpreting the information provided by existing and upcoming ATAS experiments in terms of the underlying strong-field electron dynamics requires suitable models able to capture the correlated electron dynamics of a many-body fermionic system excited by external laser fields. For many-electron systems, this can represent a formidable task, and several computational techniques have been developed for the simulation of strong-field dynamics in solids, with different levels of approximation. Simulations based on time-dependent density functional theory (TDDFT) allow one to map the evolution of the many-interacting-electron system into that of an auxiliary system of noninteracting ones \cite{PhysRevLett.52.997, ullrich_time-dependent_2011, marques_fundamentals_2012}. This mapping is always possible, as ensured by Kohn--Sham (KS) theory, although it requires the determination of a suitable exchange-correlation (XC) potential for the auxiliary noninteracting system. TDDFT allows one to fix the ionic potential directly in real space, by selecting the position and type of each ion in the system. This freedom has been exploited in recent studies of HHG in solids, where TDDFT has been employed to investigate the properties of the HHG signal \cite{PhysRevLett.118.087403, PhysRevA.97.011401}, focusing on its dependence on, e.g., the size of the solid \cite{hansen_finite-system_2018}, its short-range order \cite{yu_enhanced_2019, yu_high-order_2019}, the type of atoms in the solid \cite{hansen_high-order_2017}, and the presence of topological effects \cite{bauer_high-harmonic_2018}. Further studies have also employed TDDFT to investigate the dependence of the HHG signal on strong electron correlations \cite{PhysRevLett.121.097402} and spin-orbit effects \cite{tancogne2022effect, PhysRevX.13.031011}. However, due to its \textit{ab-initio} nature, TDDFT is less suitable for the development of interpretation models. An alternative, often employed method is based on the semiconductor Bloch equations (SBEs). These sets of differential equations allow one to calculate the evolution of the reduced one-particle density matrix in the presence of an external laser excitation via suitable energy and coupling terms \cite{meier_coherent_2007}. The SBEs provide a flexible framework for the modeling of strong-field-induced electron dynamics, and for the inclusion of additional effects such as phonon couplings and excitons, also for x-ray excitations \cite{Picon2019, cistaro_theoretical_2023}. However, they require knowledge of several state energies and coupling terms, which can best be calculated, e.g., in reciprocal space for models of solids with periodic boundary conditions. This limits their applications to the investigation of, e.g., finite-size effects and associated electronic edge states. At the same time, being based on several coupling terms, the SBEs allow one to selectively control which couplings are included in the simulation, and thereby assign the role of different states or energy bands to the resulting strong-field dynamics.

Here, we simulate ATAS signals for a model of a material---a wide-bandgap semiconductor or insulator---with an inner-core VB, a more highly excited VB, and several CBs, as depicted in Figs.~\ref{fig:TDDFT_states} and \ref{fig:Bloch_states}. The external pump pulse is set to be off-resonant from the bandgap between the higher VB and the lowest CB, whereas the attosecond XUV pulse is assumed to be in resonance with the transition between the inner VB and the CBs. The spectral bandwidth of the attosecond probe is chosen such that a broad range of CB states can be observed and their evolution monitored. We employ complementary computational approaches based on both the SBEs and TDDFT. In order to obtain quantitatively comparable results, we extract the transition energies and interband couplings for the SBEs directly from the ground-state (GS) potential obtained in TDDFT. At low pump intensities, our simulations predict fishbone-like spectral structures in the ATAS signal localized at the upper and lower edges of the CBs, in agreement with previous theoretical and experimental findings \cite{Lucchini2016, schlaepfer_attosecond_2018, Picon2019, cistaro_theoretical_2023, PhysRevA.100.043840, PhysRevA.106.063107, PhysRevB.107.184304}. For larger pump intensities, when the driven electron's crystal momentum spans larger regions of the Brillouin zone (BZ) and covers a correspondingly broader energy range of the CBs, we show that these intraband electronic motions are imprinted in the spectral amplitude of the ATAS features. For such large pump intensities, we predict additional spectral peaks, which we ascribe to the Floquet dressing of the energy bands by the intense pump pulse \cite{PhysRevA.100.013412, PhysRevB.107.184304}. The comparison between the SBEs and TDDFT calculations allows us to validate our results and interpret the \textit{ab-initio} TDDFT simulations in terms of SBEs' interband couplings.

The paper is structured as follows. In Sec.~\ref{Sec:EHmodel}, we derive the theory model based on the evolution of the electron--hole (EH) pairs generated by the pump and probe pulses. The equations of motion (EOMs) are derived from the time-dependent Schr\"odinger equation (TDSE) for the wavefunction of the many-body electronic system, with the restriction that for every crystal momentum $k$ at most one EH pair is generated. Our model is equivalent to the SBEs for the reduced one-particle density matrix, which can be retrieved from it by suitably tracing the many-body density matrix. Our derivation highlights the advantage of using creation and annihilation operators of the adiabatic Houston states \cite{houston_acceleration_1940, krieger_time_1986}, which allows one to circumvent issues related to the expansion of the position operator in the basis of Bloch states. In Sec.~\ref{Sec:TDDFT}, we present the TDDFT formalism which is used to obtain the parameters of the solid and validate the EH model. In Sec.~\ref{Sec:Solid}, we use imaginary-time propagation of the TDDFT equations for investigating a finite-size solid consisting of a linear chain of 40 atoms, modeling a wide-bandgap semiconductor with one inner-core VB, a more highly excited VB, and several CBs. In the same section, we employ a parallel-transport procedure to calculate the band structure and the crystal-momentum-dependent interband couplings of the associated periodic solid, obtained by imposing periodic boundary conditions. The EH model and TDDFT are employed in Sec.~\ref{Sec:ATAS} to investigate ATAS for the solid of Sec.~\ref{Sec:Solid} when it is strongly excited by an intense optical pump and monitored by a delayed attosecond XUV probe. Section~\ref{Sec:Conclusions} concludes the paper. Atomic units (a.u.) are used throughout unless otherwise stated.

\section{Electron--hole theory model}
\label{Sec:EHmodel}

In this section, we present the EOMs which will later be used to simulate strong-field dynamics in solids probed by delayed XUV attosecond pulses and for computing the associated ATAS signals. %In contrast to the TDDFT formalism employed in Sec.~\ref{Sec:TDDFT}, here, 
We consider a periodic solid by imposing Born--von K\'arm\'an periodic boundary conditions, and focus on the dynamics of the light-induced EH pairs by a suitable ansatz on the form of the many-body wavefunction---what we refer to as the EH model. We provide general EOMs for a wide-bandgap semiconductor or insulator with $N_v$ valence and $N_c$ conduction bands, without additional assumptions on the shape of the mean-field potential of the solid. %In Sec.~\ref{Sec:Solid}, we will describe the procedure to extract the required energy bands and interband couplings for a particular model system.

EOMs describing the electron dynamics of solids illuminated by external laser pulses have been presented before \cite{meier_coherent_2007, wu2015high, PhysRevA.100.043404, PhysRevB.102.155201, PhysRevLett.120.253201, Picon2019, cistaro_theoretical_2023}, and different choices of light and structure gauges have been shown to provide different advantages \cite{Silva2019, PhysRevA.101.053411, Yue_tutorial_22}. In the velocity gauge, the light--matter interaction Hamiltonian is diagonal over the eigenstates of the momentum operator, which allows one to derive EOMs that do not couple states of different $k$, but that require a large number of valence and conduction bands for convergence. In contrast, in the length gauge, the light--matter interaction Hamiltonian involves the position operator $\hat{x}$ which, if expanded over a basis of static Bloch state, leads to derivatives over the crystal momentum $k$ coupling all Bloch states with each other \cite{blount_formalisms_1962}. This complication can be circumvented by  moving to a basis of adiabatic Houston states \cite{houston_acceleration_1940, krieger_time_1986}, where the interband dynamics induced by the electromagnetic field are directly accounted for via a time-dependent crystal momentum $k(t)$, and where only states with the same $k$ are coupled by the electromagnetic field. 

Here, we derive the set of differential EOMs which will be used for the simulation of ATAS signals. Although our EOMs are in agreement with previous results, and more in general with the SBEs which can be retrieved from them as we state below, in the following we highlight the intermediate steps employed to derive our EH model, since they differ from previous approaches. In our derivation, we expand the many-body Hamiltonian in terms of creation and annihilation operators of the adiabatic Houston states, which allows us to more straightforwardly circumvent the complications associated with expanding the position operator in the basis of Bloch states. 

\subsection{Single-particle light--matter interaction Hamiltonian}
\label{Subsec:singlepartHam}

For convenience, we consider a periodic linear chain of identical, equally spaced atoms with lattice constant $a$ and atomic number $Z$. The results of this section can be readily generalized to three-dimensional space. In the absence of external electromagnetic fields, at the mean-field level and neglecting beyond-mean-field electron--electron interactions, the single-particle Hamiltonian describing the electrons in the solid is given by
\begin{equation}
\hat{H} = \frac{\hat{p}^2}{2} + U(\hat{x}),
\label{eq:singlepartHam}
\end{equation}
with momentum $\hat{p}$, position $\hat{x}$, and the periodic potential $U(\hat{x}) = U(\hat{x} + a)$. For the time being, we do not fix the spatial dependence of the periodic potential $U(\hat{x})$, which we will later extract from density functional theory (DFT) simulations as presented in the following subsections. In order to simplify the notation, we assume a large chain of $N$ atoms and length $L = N a$, and ensure periodicity by assuming Born--von K\'arm\'an periodic boundary conditions \cite{ashcroft_solid_1976}. With these assumptions, the eigenstates of the single-particle Hamiltonian
\begin{equation}
\left[-\frac{1}{2} \frac{\partial^2}{\partial x^2} + U(x)\right]\,\psi_{n,k_m}(x) = \mathcal{E}_{n,k_m}\, \psi_{n,k_m}(x)
\label{eq:SchroedingerBloch}
\end{equation}
are given by Bloch states
\begin{equation}
\psi_{n,k_m}(x) = \frac{1}{\sqrt{L}}\,\eu^{\uimm k_m x}\,u_{n,k_m}(x),
\label{eq:Blochstates}
\end{equation}
labeled by the band index $n$ and by the crystal momentum $k_m$, which we let vary inside the first BZ $-\pi/a \leq k_m < \pi/a $. The functions $u_{n,k_m}(x) = u_{n,k_m}(x + a)$ are the periodic part of the Bloch states, with period $a$. Due to the periodic boundary conditions employed here, the crystal momentum only assumes discrete values $k_m = m \Delta k$, with $\Delta k = 2\pi/L$ and with the index $m$ running over $m = -N/2,\,-N/2 + 1,\,\ldots,\,N/2-1$. The Bloch states are normalized over the length $L$ of the solid,
\begin{equation}
\int_0^L \diff x\,\psi^*_{n,k_m}(x)\,\psi_{n',k_{m'}}(x) = \delta_{nn'}\delta_{k_m k_{m'}},
\label{eq:normalizationBloch}
\end{equation}
which implies the following normalization for the periodic part of the Bloch states,
\begin{equation}
\frac{1}{a}\int_0^a\diff x\,u^*_{n,k_m}(x)\,u_{n',k_{m}}(x)  = \delta_{nn'}.
\label{eq:normofu}
\end{equation}
This can be verified by recasting the integral in Eq.~(\ref{eq:normalizationBloch}) as
\begin{equation}
\begin{aligned}
&\int_0^L \diff x\,\psi^*_{n,k_m}(x)\,\psi_{n',k_{m'}}(x) \\
= \,&\frac{1}{Na}\sum_{j=0}^{N-1}\int_0^a\diff x\,\eu^{-\uimm(k_m - k_{m'})(x+ ja)}\,u^*_{n,k_m}(x)\,u_{n',k_{m'}}(x)
\end{aligned}
\end{equation}
where we have exploited the periodicity of $u_{n,k_m}(x) $ over the unit cell, and then employing the following identity:
\begin{equation}
\frac{1}{N}\sum_{j=0}^{N-1} \eu^{-\uimm \frac{2 \pi}{N}(m - {m'})j} = \delta_{mm'}.
\end{equation}

In the presence of an external electromagnetic field $E(t)$ of vector potential $A(t)$, 
\begin{equation}
E(t) = -\frac{\diff A(t)}{\diff t},
\label{eq:dAdt}
\end{equation}
the single-particle light--matter interaction Hamiltonian in the velocity gauge reads
\begin{equation}
\hat{H}_{\mathrm{VG}}(t) = \frac{(\hat{p} + A(t))^2}{2} + U(\hat{x}).
\label{eq:lightmatterVG}
\end{equation}
In such case, one can introduce accelerated Bloch states, also called Houston states \cite{houston_acceleration_1940, krieger_time_1986}, representing the instantaneous eigenstates of the Hamiltonian in Eq.~(\ref{eq:lightmatterVG}):
\begin{equation}
\left[\frac{(\hat{p} + A(t))^2}{2} + U(\hat{x})\right]\,\varphi_{n,k_m}(x, t) = \tilde{\mathcal{E}}_{n,k_m}(t)\,\varphi_{n,k_m}(x, t).
\end{equation}
Requesting that also the Houston states satisfy Born--von K\'arm\'an periodic boundary conditions allows one to write them and their energies in terms of the Bloch eigenstates and eigenenergies,
\begin{equation}
\begin{aligned}
\varphi_{n,k_m}(x, t) &= \eu^{-\uimm A(t) x}\,\psi_{n,k_m(t)}(x),\\
\tilde{\mathcal{E}}_{n,k_m}(t) &= \mathcal{E}_{n,k_m(t)},
\end{aligned}
\end{equation}
where
\begin{equation}
k_m(t) = k_m + A(t)
\label{eq:kmt}
\end{equation}
describes the semiclassical evolution of the crystal momentum in the presence of an external vector potential, and where $k_m$ represents the crystal momentum for this particular state in the absence of laser fields. We stress that the time dependence of the Houston states is fully encoded in the periodic part of the Bloch states,
\begin{equation}
\varphi_{n,k_m}(x, t)  = \frac{1}{\sqrt{L}}\,\eu^{\uimm k_m x}\,u_{n,k_m + A(t)}(x),
\label{eq:varphi}
\end{equation}
so that their time derivative is equal to
\begin{equation}
\begin{aligned}
\frac{\diff }{\diff t}\varphi_{n,k_m}(x, t) =\,& \frac{1}{\sqrt{L}}\,\eu^{\uimm k_m x}\,\frac{\diff A(t)}{\diff t}\left.\frac{\diff u_{n,k}(x)}{\diff k}\right|_{k_m + A(t)} \\
=\,& -E(t)\,\frac{1}{\sqrt{L}}\,\eu^{\uimm k_m x}\left.\frac{\diff u_{n,k}(x)}{\diff k}\right|_{k_m + A(t)}  .
\end{aligned}
\label{eq:dvarphi}
\end{equation}

We note that Eq.~(\ref{eq:dvarphi}) relies on the assumption that $u_{n,k}(x)$ is a differentiable function of the crystal momentum $k$. This requirement might not be satisfied, especially if one bears in mind that the Bloch states $\psi_{n,k}(x)$ and their spatially periodic components $u_{n,k}(x)$ are defined, for each $k$, up to a free phase term $\eu^{\uimm\alpha(k)}$ \cite{vanderbilt_berry_2018}. In the remaining of this section, we will assume that $\diff u_{n,k}(x)/\diff k$ is well defined for every $k$. In Sec.~\ref{Sec:Solid}, we will present the procedure of parallel transport which allows one to determine a gauge in which this condition is fulfilled. 

\subsection{Many-body fermionic system and light--matter interaction Hamiltonian}

The band structure depends on the explicit properties of the potential $U(x)$. In the following, we assume a wide-bandgap semiconductor or insulator, with $N_{v}$ occupied VBs $\{v_1,\, v_2,\,\ldots,\,v_{N_{v}}\}$, and $N_{c}$ initially unoccupied higher-energy CBs $\{c_1,\, c_2,\,\ldots,\,c_{N_{c}}\}$. Every band $n$ contains $N$ Bloch states $|\psi_{n,k_m}\rangle$, each of which can be occupied by two electrons of opposite spin. For a charge- and spin-neutral solid, this results in a total number of electrons $N_{\mathrm{el}} = ZN = 2N N_{v} $. We will neglect the spin degrees of freedom in the following discussion, and therefore only focus on the $N N_{v} $ electrons with spin up, assuming that the up and down electrons undergo the same dynamics in the absence of spin effects.

In the absence of an external electromagnetic field, one can describe the many-body system in the basis of Bloch states, by introducing the associated creation $\hat{d}\daga_{n,k_m}$ and annihilation $\hat{d}_{n,k_m}$ operators of an electron in the Bloch state $|\psi_{n,k_m}\rangle$,
\begin{equation}
|\psi_{n,k_m}\rangle = \hat{d}\daga_{n,k_m}|0\rangle,
\end{equation}
where $|0\rangle$ denotes the vacuum state in the absence of any particle. In this basis, the GS $|\mathrm{GS}\rangle$ and the many-body Hamiltonian in the absence of two-body electron--electron interactions are given by
\begin{equation}
|\mathrm{GS}\rangle = \bigotimes_{k_m} |\mathrm{GS}_{k_m}\rangle = \bigotimes_{k_m} \hat{d}\daga_{v_1,k_m}\,\hat{d}\daga_{v_2,k_m}\,\cdots\,\hat{d}\daga_{v_{N_{\mathrm{v}}},k_m}\,|0\rangle
\label{eq:GSstate}
\end{equation}
and
\begin{equation}
\hat{\mathcal{H}} = \sum_{n,k_m}\mathcal{E}_{n,k_m}\,\hat{d}\daga_{n,k_m}\,\hat{d}_{n,k_m},
\label{eq:GShamiltonian}
\end{equation}
respectively.

In order to describe the light--matter interaction of the many-body system with the external vector potential $A(t)$, it is convenient to move to the adiabatic basis of Houston states, and introduce creation $\hat{c}\daga_{n,k_m}(t)$ and annihilation $\hat{c}_{n,k_m}(t)$ operators of an electron in the adiabatic state $|\varphi_{n,k_m}(t)\rangle$,
\begin{equation}
|\varphi_{n,k_m}(t)\rangle = \hat{c}\daga_{n,k_m}(t)|0\rangle.
\end{equation}
In this basis, the state of fully occupied valence states is given by
\begin{equation}
|\mathrm{GS}\rangle = \bigotimes_{k_m} |\mathrm{GS}_{k_m}(t)\rangle,
\end{equation}
with 
\begin{equation}
|\mathrm{GS}_{k_m}(t)\rangle = \hat{c}\daga_{v_1,k_m}(t)\,\hat{c}\daga_{v_2,k_m}(t)\,\cdots\,\hat{c}\daga_{v_{N_{\mathrm{v}}},k_m}(t)\,|0\rangle,
\label{eq:GSk}
\end{equation}
and the many-body Hamiltonian can be recast as
\begin{equation}
\hat{\mathcal{H}}(t) = \sum_{n,k_m}\tilde{\mathcal{E}}_{n,k_m}(t)\,\hat{c}\daga_{n,k_m}(t)\,\hat{c}_{n,k_m}(t).
\label{eq:timedepHam}
\end{equation}
Note that, in the absence of an external electromagnetic field, $A(t) = 0$, one retrieves the expressions for the time-independent GS and many-body Hamiltonian in the basis of Bloch states given by Eqs.~(\ref{eq:GSstate}) and (\ref{eq:GShamiltonian}), respectively. 

\subsection{Single electron--hole state ansatz}

We compute the evolution of the many-body system $|\varPsi(t)\rangle$ in the presence of the light--matter interaction Hamiltonian of Eq.~(\ref{eq:timedepHam}) by solving the TDSE
\begin{equation}
\uimm\frac{\diff |\varPsi(t)\rangle}{\diff t} = \hat{\mathcal{H}}(t)|\varPsi(t)\rangle.
\label{eq:TDSE}
\end{equation}
For this purpose, we expand the state $|\varPsi(t)\rangle$ in the basis of Houston states and write it as the following product state
\begin{widetext}
\begin{equation}
\begin{aligned}
|\varPsi(t)\rangle = \bigotimes_{k_m}\Biggl[ & b_{0,k_m}(t)\,|\mathrm{GS}_{k_m}(t)\rangle + \sum_{i= 1}^{N_{c}} \sum_{j= 1}^{N_{v}} b_{c_i,v_j,k_m}(t)\,\hat{c}\daga_{c_i,k_m}(t)\,\hat{c}_{v_j,k_m}(t)\,|\mathrm{GS}_{k_m}(t)\rangle\  + \\
&\sum_{i= 1}^{N_{c}} \sum_{\substack{i'= 1 \\ i'\neq i}}^{N_{c}} \sum_{j= 1}^{N_{v}} \sum_{\substack{j'= 1 \\ j'\neq j}}^{N_{v}}b_{c_i,c_{i'},v_j,v_{j'},k_m}(t)\,\hat{c}\daga_{c_i,k_m}(t)\,\hat{c}\daga_{c_{i'},k_m}(t)\,\hat{c}_{v_j,k_m}(t)\,\hat{c}_{v_{j'},k_m}(t)\,|\mathrm{GS}_{k_m}(t)\rangle +\,\ldots\,\Biggr].
\end{aligned}
\label{eq:totalPsi}
\end{equation}
\end{widetext}
The coefficient $b_{0,k_m}(t)$ in the first term of the above sum describes the amplitude with which the system is in the GS $|\mathrm{GS}_{k_m}(t)\rangle $ [Eq.~(\ref{eq:GSk})], where the VBs are fully occupied and the CBs are fully unoccupied. In the second term, the coefficients $b_{c_i,v_j,k_m}(t)$ represent the evolution of the EH pair states $\hat{c}\daga_{c_i,k_m}(t)\,\hat{c}_{v_j,k_m}(t)\,|\mathrm{GS}_{k_m}(t)\rangle$, in which an electron from the VB $v_j$ has been moved to the CB $c_i$. Two-EH states are accounted for by the third term in the above sum and by the coefficients $b_{c_i,c_{i'},v_j,v_{j'},k_m}(t)$, and the sum potentially runs over states consisting of larger numbers of electrons and holes. 

In the following, we focus on the contribution from single EH pairs, neglecting the small contribution from states consisting of two or more pairs \cite{Picon2019}. This single-EH state ansatz corresponds to truncating the sum of Eq.~(\ref{eq:totalPsi}) after the second term:
\begin{equation}
\begin{aligned}
&|\varPsi(t)\rangle =  \bigotimes_{k_m}\Biggl[  b_{0,k_m}(t)\,|\mathrm{GS}_{k_m}(t)\rangle \\
&\ \ \ \ +  \sum_{i= 1}^{N_{c}} \sum_{j= 1}^{N_{v}} b_{c_i,v_j,k_m}(t)\,\hat{c}\daga_{c_i,k_m}(t)\,\hat{c}_{v_j,k_m}(t)\,|\mathrm{GS}_{k_m}(t)\rangle \Biggr].
\end{aligned}
\label{eq:EHansatz}
\end{equation}
Solving the TDSE for the ansatz state of Eq.~(\ref{eq:EHansatz}) allows one to compute the evolution of the EH pairs generated by the external electromagnetic fields. We refer to this as the EH model.

The EH model provides an alternative formulation of the SBEs \cite{meier_coherent_2007}, which can be retrieved from it \cite{cistaro_theoretical_2023}. This can be achieved by introducing the many-body density matrix
\begin{equation}
\hat{\varrho}(t) = |\varPsi(t)\rangle\langle\varPsi(t)|
\end{equation}
and the single-particle reduced density-matrix elements
\begin{equation}
\rho_{i,j}(k_m(t)) = \mathrm{Tr}\left\{\hat{\varrho}(t)\,\hat{c}\daga_{j,k_m}(t)\,\hat{c}_{i,k_m}(t)\right\}.
\end{equation}
The SBEs are then the EOMs determining the dynamics of the elements $\rho_{i,j}(k_m(t))$. In the following, we do not derive these equations explicitly, but rather directly compute the ATAS signal in terms of the coefficients $b_{0,k_m}(t)$ and $b_{c_i,v_j,k_m}(t)$ in Eq.~(\ref{eq:EHansatz}). We stress that the two formulations are entirely equivalent.

\subsection{Inter- and intraband couplings}

We observe that the derivative of $|\varPsi(t)\rangle$ in Eq.~(\ref{eq:EHansatz}) involves the time derivative of the creation and annihilation operators, $\hat{c}\daga_{n,k_m}(t)$ and $\hat{c}_{n,k_m}(t)$, respectively. To compute these time derivatives, we first exploit the completeness of the basis formed by the eigenstates of the position operator,
\begin{equation}
\begin{aligned}
\hat{c}\daga_{n,k_m}(t)\,|0\rangle =\,& \int_0^L \diff x\,|x\rangle\langle x|\,\hat{c}\daga_{n,k_m}(t)\,|0\rangle\\
=\,& \int_0^L \diff x\,|x\rangle\,\varphi_{n,k_m}(x,t),
\end{aligned}
\end{equation}
and then the completeness of the basis of Houston states,
\begin{equation}
\begin{aligned}
|x\rangle  =\,& \sum_{n',k_{m'}} |\varphi_{n',k_{m'}}(t)\rangle\langle \varphi_{n',k_{m'}}(t)|x\rangle \\
=\,& \sum_{n',k_{m'}} \varphi^*_{n',k_{m'}}(x,t)\,\hat{c}\daga_{n',k_{m'}}(t)\,|0\rangle,
\end{aligned}
\end{equation}
which allows one to recast the time derivative of $\hat{c}\daga_{n,k_m}(t)$ as follows:
\begin{equation}
\frac{\diff\hat{c}\daga_{n,k_m}(t)}{\diff t} =\sum_{n',k_{m'}} \int_0^L \diff x\,\varphi^*_{n',k_{m'}}(x,t)\, \frac{\diff \varphi_{n,k_m}(x,t)}{\diff t} \,\hat{c}\daga_{n',k_{m'}}(t).
\label{eq:diffcintermediate}
\end{equation}
By taking advantage of Eqs.~(\ref{eq:normofu}), (\ref{eq:varphi}), and (\ref{eq:dvarphi}), Eq.~(\ref{eq:diffcintermediate}) then reduces to
\begin{equation}
\frac{\diff\hat{c}\daga_{n,k_m}(t)}{\diff t} = \uimm\sum_{n'} E(t)\,\xi_{n',n}(k_m + A(t))\,\hat{c}\daga_{n',k_{m}}(t)
\end{equation}
where we have introduced the couplings
\begin{equation}
\xi_{n',n}(k) = \frac{\uimm}{a}\int_0^a \diff x\,u^*_{n',k}(x)\frac{\diff u_{n,k}(x)}{\diff k}.
\end{equation}

For $n\neq n'$, the interband couplings $\xi_{n',n}(k)$ represent the transition elements of the position operator $\hat{x}$, 
\begin{equation}
\xi_{n',n}(k) = \langle \psi_{n',k}|\hat{x}|\psi_{n,k}\rangle = \int_0^L \diff x\,\psi^*_{n',k}(x)\,x\,\psi_{n,k}(x).
\label{eq:xin'n}
\end{equation}
This can be verified by recasting the following integral as
\begin{equation}
\begin{aligned}
&\uimm\int_0^L \diff x\,\psi^*_{n',k}(x)\frac{\diff \psi_{n,k}(x)}{\diff k} \\
=\,& \frac{\uimm}{a}\int_0^a \diff x\,u^*_{n',k}(x)\,\frac{\diff u_{n,k}(x)}{\diff k}  -  \int_0^L \diff x\,\psi^*_{n',k}(x)\,x\,\psi_{n,k}(x),
\end{aligned}
\label{eq:intprova1}
\end{equation}
where we have taken advantage of Eq.~(\ref{eq:Blochstates}) and the periodicity of $u_{n,k}(x)$, and then noting that the left-hand side of Eq.~(\ref{eq:intprova1}) has to vanish when $n\neq n'$ due to
\begin{equation}
\begin{aligned}
&\uimm\int_0^L \diff x\,\psi^*_{n',k'}(x)\frac{\diff \psi_{n,k}(x)}{\diff k} \\
=\,& \uimm\,\frac{\diff}{\diff k} \int_0^L \diff x\,\psi^*_{n',k'}(x)\psi_{n,k}(x) = 0 \ \text{for $n\neq n'$}.
\end{aligned}
\label{eq:intprova2}
\end{equation}
This emplies that the right-hand side of Eq.~(\ref{eq:intprova1}) also vanishes for $n\neq n'$, from which Eq.~(\ref{eq:xin'n}) follows. These equalities however do not hold for $n = n'$. In such case, the associated intraband couplings $\xi_{n,n}(k)$ represent the Berry connections of each band of the system \cite{blount_formalisms_1962}. While the Berry connections depend on the choice of structure gauge, the Berry phases 
\begin{equation}
\phi_n = \oint \diff k \,\xi_{n,n}(k),
\label{eq:Berryphase}
\end{equation}
i.e., the cycle integral of the Berry connections over one BZ, are gauge invariant \cite{vanderbilt_berry_2018}. In Sec.~\ref{Sec:Solid}, we will employ a parallel-transport procedure to determine a structure gauge in which the functions $u_{n,k}(x)$ are differentiable over $k$. For this particular choice of structure gauge, the Berry connections vanish identically. 

\subsection{Equations of motion of the electron--hole model}

We are now endowed with all the necessary elements to write explicitly the EOMs of the EH model. We can namely solve the TDSE given in Eq.~(\ref{eq:TDSE}) for the ansatz state of Eq.~(\ref{eq:EHansatz}) in the presence of the light--matter interaction Hamiltonian of Eq.~(\ref{eq:timedepHam}) and the couplings $\xi_{n',n}(k)$ defined in Eq.~(\ref{eq:xin'n}). This provides the following set of differential equations:
\begin{widetext}
\begin{equation}
\begin{aligned}
&\uimm\frac{\diff b_{0,k_m}}{\diff t} - \sum_{j}\mathcal{E}_{v_j,k_m(t)}\,b_{0,k_m}(t)\\
=\,& E(t) \sum_{j} \xi_{v_j,v_j}(k_m(t))\,b_{0,k_m}(t) + E(t) \sum_{i}\sum_{j} \xi_{v_j,c_i}(k_m(t))\,b_{c_i,v_j,k_m}(t), \\
&\uimm\frac{\diff b_{c_i,v_j,k_m}}{\diff t}  - \Bigl[\mathcal{E}_{c_i,k_m(t)} + \sum_{j' \neq j}\mathcal{E}_{v_{j'},k_m(t)}\Bigr]\,b_{c_i,v_j,k_m}(t) \\
=\,&  E(t)\,\Bigl[\sum_{j'\neq j}\xi_{v_{j'},v_{j'}}(k_m(t)) +  \xi_{c_i,c_i}(k_m(t))\Bigr]\,b_{c_i,v_j,k_m}(t)   + E(t)\, \xi_{c_i,v_j}(k_m(t))\,b_{0,k_m}(t) \\
&+ E(t) \sum_{j'\neq j} \xi_{v_{j'},v_{j}}(k_m(t))\,b_{c_i,v_{j'},k_m}(t) + E(t) \sum_{i'\neq i} \xi_{c_i,c_{i'}}(k_m(t))\,b_{c_{i'},v_j,k_m}(t) -\uimm\,\gamma_{c_i,v_j}\,b_{c_i,v_j,k_m}(t)  .
\end{aligned}
\label{eq:EH-EOMs}
\end{equation}
The last term in the second equation of Eq.~(\ref{eq:EH-EOMs}) accounts for the finite lifetime $1/\gamma_{c_i,v_j}$ of the EH pair between the CB $c_i$ and the VB $v_j$ by an effective decay with decay rate $\gamma_{c_i,v_j}$. By introducing the slowly varying variables
\begin{equation}
\begin{aligned}
\tilde{b}_{0,k_m}(t) &= b_{0,k_m}(t)\,\eu^{\uimm\int_{t_0}^t\diff t' \sum_{j}\mathcal{E}_{v_j,k_m(t')}},\\
\tilde{b}_{c_i,v_j,k_m}(t) &= b_{c_i,v_j,k_m}(t)\,\eu^{\uimm\int_{t_0}^t\diff t' \left[\mathcal{E}_{c_i,k_m(t')} + \sum_{j' \neq j}\mathcal{E}_{v_{j'},k_m(t')}\right]},
\end{aligned}
\end{equation}
the above EOMs can be recast in the form
\begin{equation}
\begin{aligned}
\frac{\diff \tilde{b}_{0,k_m}}{\diff t} =\,&  -\uimm\,E(t) \sum_{j} \xi_{v_j,v_j}(k_m(t))\,\tilde{b}_{0,k_m}(t) \\
&-\uimm\,E(t) \sum_{i} \sum_{j} \xi_{v_j,c_i}(k_m(t))\,\eu^{-\uimm\int_{t_0}^t\diff t'\,\left[\mathcal{E}_{c_i,k_m(t')} - \mathcal{E}_{v_j,k_m(t')}\right]}\,\tilde{b}_{c_i,v_j,k_m}(t),\\
\frac{\diff \tilde{b}_{c_i,v_j,k_m}}{\diff t} =\,&  -\uimm\,E(t)\,\Bigl[\sum_{j'\neq j}\xi_{v_{j'},v_{j'}}(k_m(t)) +  \xi_{c_i,c_i}(k_m(t))\Bigr]\,\tilde{b}_{c_i,v_j,k_m}(t)  \\
&-\uimm\,E(t)\, \xi_{c_i,v_j}(k_m(t))\,\eu^{\uimm\int_{t_0}^t\diff t'\,\left[\mathcal{E}_{c_i,k_m(t')} - \mathcal{E}_{v_j,k_m(t')}\right]}\,\tilde{b}_{0,k_m}(t) \\
&-\uimm\,E(t) \sum_{i'\neq i} \xi_{c_i,c_{i'}}(k_m(t))\,\eu^{-\uimm\int_{t_0}^t\diff t'\,\left[\mathcal{E}_{c_{i'},k_m(t')} - \mathcal{E}_{c_i,k_m(t')}\right]}\,\tilde{b}_{c_{i'},v_j,k_m}(t) \\
&-\uimm\, E(t) \sum_{j'\neq j} \xi_{v_{j'},v_{j}}(k_m(t))\,\eu^{\uimm\int_{t_0}^t\diff t'\,\left[\mathcal{E}_{v_{j'},k_m(t')} - \mathcal{E}_{v_j,k_m(t')}\right]}\,\tilde{b}_{c_i,v_{j'},k_m}(t) -\uimm\,\gamma_{c_i,v_j}\,b_{c_i,v_j,k_m}(t) .
\end{aligned}
\label{eq:EH-EOMs-tilde}
\end{equation}
\end{widetext}
Equation~(\ref{eq:EH-EOMs-tilde}) provides the EOMs for slowly varying variables, which renders them convenient for numerical calculations. These are the sets of EOMs that will be implemented in the following.

We stress that the EOMs of the EH model do not couple states and amplitudes associated with different crystal momenta $k_m$. This is a consequence of having used a basis of Houston states, which incorporate the light-induced interband dynamics in terms of time-dependent crystal momenta $k_m(t)$ [Eq.~(\ref{eq:kmt})]. In this basis, the time-dependent light--matter interaction Hamiltonian $\hat{\mathcal{H}}(t)$ of Eq.~(\ref{eq:timedepHam}) is diagonal in $k_m$. Similarly, the time derivative of the creation operator $\hat{c}\daga_{n,k_m}(t)$, which creates the Houston state $|\varphi_{n,k_m}(t)\rangle$, only involves creation operators associated with the same $k_m$. This is a clear advantage with respect to expanding the EOMs in a basis of Bloch states, which is known to involve $k$-derivatives coupling states of different $k_m$ \cite{Yue_tutorial_22}.

In Sec.~\ref{Sec:ATAS}, we will employ the EH model to study the evolution of a model of a solid, with 2 VBs and 2 CBs, suitably excited by an intense pump pulse and probed by a time-delayed attosecond XUV probe pulse. The evolution of the system thereby obtained will be employed to calculate the dipole response of the system and thus simulate the associated ATAS signal.

\section{Modeling based on time-dependent density functional theory}
\label{Sec:TDDFT}

We additionally perform simulations based on TDDFT, in order to validate and benchmark the results obtained by our EH model. In this section, we briefly present the details of our TDDFT simulations, as they are implemented in an extension \cite{hansen_high-order_2017, PhysRevA.103.053121} of the program QPROP \cite{bauer_qprop_2006}.

TDDFT allows us to describe the evolution of a many-interacting-electron system in terms of an auxiliary system of noninteracting KS orbitals $\phi_{i}(x,t)$ \cite{PhysRevLett.52.997, ullrich_time-dependent_2011, marques_fundamentals_2012}
\begin{equation}
|\varPsi(t)\rangle = \bigotimes_{i}|\phi_{i}(t)\rangle.
\label{eq:Psi-TDDFT}
\end{equation}
Also here, similarly to the approach used in the EH model, we will neglect interaction terms sensitive to the spin degrees of freedom, thus assuming that the up and down electrons undergo the same dynamics. The index $i$ therefore runs over half the total number of the electrons in the solid. We assume a charge- and spin-neutral system with $N_{\mathrm{at}}$ atoms with atomic number $Z$, implying that $i \in \{1,\,\ldots,\, ZN_{\mathrm{at}}/2\}$. Note that our TDDFT model does not assume periodic boundary conditions, in contrast to the EH model presented in Sec.~\ref{Sec:EHmodel}. Simulations based on TDDFT are therefore sensitive to the number of atoms $N_{\mathrm{at}}$, and have therefore been used before to investigate finite-size effects in strong-field solid--laser interaction, as well as to analyze the effect of topologically protected edge states in finite-size systems \cite{hansen_finite-system_2018, yu_enhanced_2019, yu_high-order_2019, hansen_high-order_2017, bauer_high-harmonic_2018}. 

For our TDDFT simulations, we assume a single-particle ionic potential given by
\begin{equation}
V_{\mathrm{ion}}(x) = -\sum_{j = 0}^{N_{\mathrm{at}}-1}\frac{Z}{\sqrt{(x - x_j)^2 + \epsilon}}
\label{eq:Vion}
\end{equation}
with $x_j = [j - (N_{\mathrm{at}}-1)/2]a$, with lattice constant $a$ and softening parameter $\epsilon$. The parameters $a$ and $\epsilon$ will be chosen in order to provide a wide-bandgap system with two VBs and a series of CBs, separated by transition energies which are suitable for excitation by optical and XUV pulses. This will be discussed in Sec.~\ref{Sec:Solid}.

In the absence of external electromagnetic fields, one can obtain the static KS orbitals by solving the time-independent KS equation
\begin{equation}
\left\{-\frac{1}{2}\frac{\partial^2}{\partial x^2} + V_{\mathrm{KS}}[\{n(x)\}](x)\right\}\phi_{i}(x) = \mathcal{E}_{i}\phi_{i}(x)
\end{equation}
with the density
\begin{equation}
n(x) = 2 \sum_{i = 1}^{ZN_{\mathrm{at}}/2} |\phi_{i}(x)|^2,
\end{equation}
where the factor 2 accounts for contributions from electrons of opposite spin. The KS potential
\begin{equation}
V_{\mathrm{KS}}[\{n(x)\}](x) = V_{\mathrm{ion}}(x) + V_{\mathrm{H}}[n(x)](x) + V_{\mathrm{XC}}[\{n(x)\}](x)
\label{eq:KSpot}
\end{equation}
is the sum of the ionic potential of Eq.~(\ref{eq:Vion}), the Hartree potential
\begin{equation}
V_{\mathrm{H}}[n(x)](x) = \int \diff x' \frac{ n(x') }{ \sqrt{(x - x')^2 + \epsilon } }
\end{equation}
treating the electron--electron interaction at the mean-field level, and the XC potential $V_{\mathrm{XC}}[\{n(x)\}](x)$, which we choose as
\begin{equation}
V_{\mathrm{XC}}[\{n(x)\}](x) = -\sqrt[3]{\frac{6}{\pi}n(x)}.
\label{eq:XCpot}
\end{equation}
A proper choice of XC potential is critical in TDDFT, as it is the key element allowing one to map the evolution of the many-body interacting system into the evolution of an auxiliary system of noninteracting KS orbitals. Here, we assume the XC potential of Eq.~(\ref{eq:XCpot}) in the local-density approximation. Other choices of XC potentials were tested and were shown not to influence the result of the simulations \cite{PhysRevA.103.053121}.

In order to calculate the evolution of the electrons in the solid in the presence of an external field of vector potential $A(t)$, we solve the time-dependent KS equation (TDKSE) in real time
\begin{equation}
\begin{aligned}
&\uimm\frac{\partial}{\partial t} \phi_{i}(x,t)\\
 =\,& \left\{-\frac{1}{2}\frac{\partial^2}{\partial x^2}- \uimm A(t)\frac{\partial}{\partial x}+ V_{\mathrm{KS}}[\{n(x, t)\}](x)\right\}\phi_{i}(x,t)
\end{aligned}
\label{eq:TDKSE}
\end{equation}
for the time-dependent density
\begin{equation}
n(x, t) = 2 \sum_{i = 1}^{ZN_{\mathrm{at}}/2} |\phi_{i}(x, t)|^2.
\end{equation}
We also solve the TDKSE in imaginary time in the absence of external fields in order to determine the GS of the many-body system of noninteracting KS orbitals. This, in turn, provides the GS density $n_{\mathrm{GS}}(x)$ and the associated GS potential $V_{\mathrm{KS}}[\{n_{\mathrm{GS}}(x)\}](x)$. 

\section{Model of solid}
\label{Sec:Solid}

In this section, we introduce the model of a solid which we will employ for our simulations, both based on the EH model and TDDFT. Our approach is as follows: We first use TDDFT, and identify suitable parameters to describe a material---a wide-bandgap semiconductor or insulator---with an inner-core VB, a second, more highly excited VB, and a series of CBs. The parameters are set to provide a model of a solid with transition energies suitable for optical and XUV laser excitation. Solving the TDKSE in imaginary time in the absence of external fields provides the single-particle GS potential of the system. Based on this potential, we construct Bloch states for a corresponding solid, for which periodic boundary conditions are imposed. This provides the energy bands $\mathcal{E}_{n,k}$ and the couplings $\xi_{n',n}(k)$ required as input parameters for the EH model.

\subsection{TDDFT parameters and ground-state potential}

\begin{figure}[t]
\centering
\includegraphics[width=\linewidth]{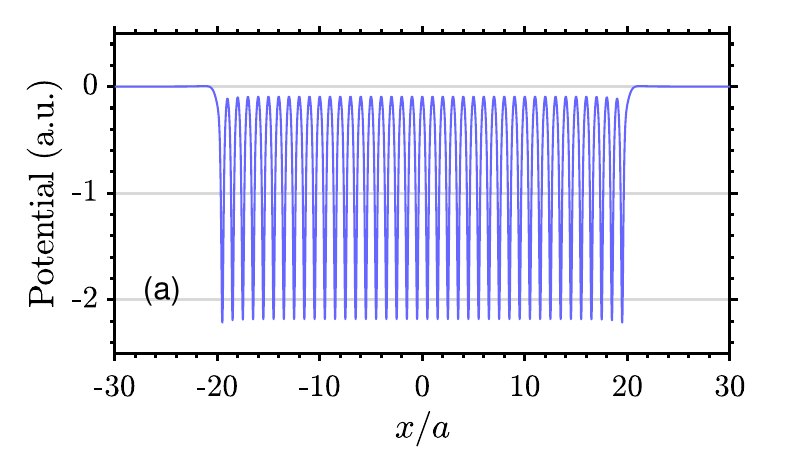}\\
\includegraphics[width=\linewidth]{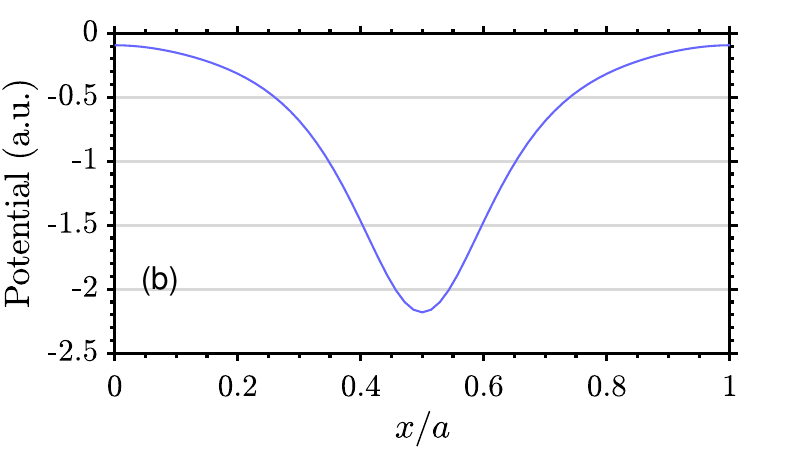}
\caption{(a) KS potential $V_{\mathrm{KS}}[\{n_{\mathrm{GS}}(x)\}](x)$ [Eq.~(\ref{eq:KSpot})] for the GS density $n_{\mathrm{GS}}(x)$ obtained by imaginary-time propagation of the TDKSE in the absence of external vector potentials, for a linear chain of $N_{\mathrm{at}} = 40$ atoms with nuclear charge $Z = 4$, lattice constant $a = 7\,\mathrm{a.u.}$, and softening parameter $\epsilon = 0.9\,\mathrm{a.u.}$ (b) Detail of the potential in panel~(a) in the bulk of the finite-size solid.}
\label{fig:potential}
\end{figure}

We use TDDFT to model a linear chain of $N_{\mathrm{at}} = 40$ atoms with nuclear charge $Z = 4$, assuming a lattice constant of $a = 7\,\mathrm{a.u.}$ and a softening parameter of $\epsilon = 0.9\,\mathrm{a.u.}$ The GS is found by imaginary-time propagation of the TDKSE, with a time step $\Delta t = 0.5\,\mathrm{a.u.}$ and on a space grid of 17,000 points with spacing $\Delta x = 0.1\,\mathrm{a.u.}$ Figure~\ref{fig:potential}(a) shows the resulting GS KS potential $V_{\mathrm{KS}}[\{n_{\mathrm{GS}}(x)\}](x)$ [Eq.~(\ref{eq:KSpot})] for the GS density $n_{\mathrm{GS}}(x)$. While the effect of the finite-size of the solid is visible at the edges of the linear chain, the potential becomes almost perfectly periodic in the bulk. In Fig.~\ref{fig:potential}(b), we highlight the potential within a single cell in the center of the solid.

\begin{figure}[t]
\centering
\includegraphics[width=\linewidth]{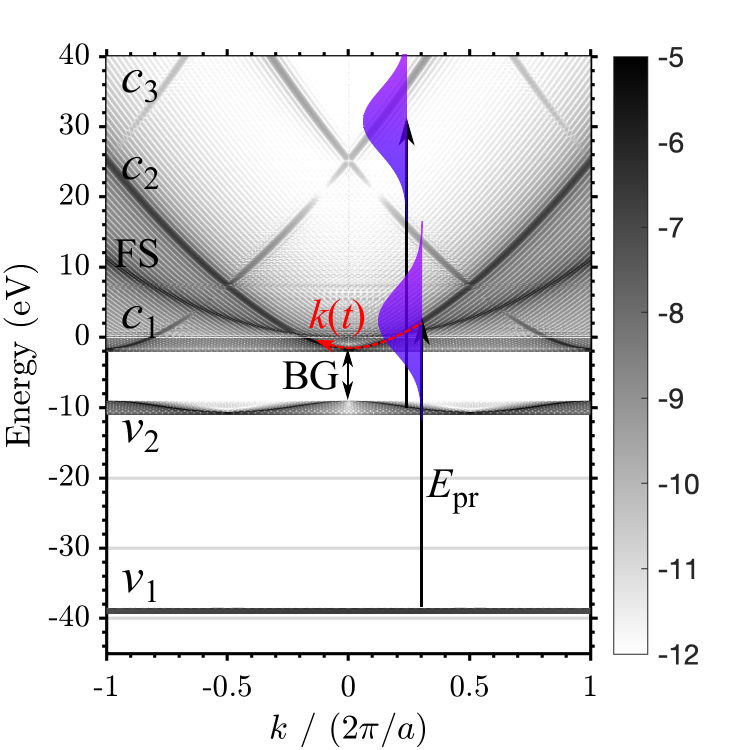}
\caption{Modulus square $|\tilde{\phi}_{\mathrm{GS},q}(k)|^2$ (displayed in logarithmic scale) of the Fourier transform $\tilde{\phi}_{\mathrm{GS},q}(k)$ of the eigenstates $\phi_{\mathrm{GS},q}(x)$ of the GS KS Hamiltonian, as a function of their associated eigenenergies $\mathcal{E}_{\mathrm{GS},q}$, for a linear chain of $N_{\mathrm{at}} = 40$ atoms with nuclear charge $Z = 4$, lattice constant $a = 7\,\mathrm{a.u.}$, and softening parameter $\epsilon = 0.9\,\mathrm{a.u.}$ The inner VB $v_1$, the more highly excited VB $v_2$, and the CBs $c_1$, $c_2$, and $c_3$ can be distinguished. BG denotes the bandgap between the highest VB $v_2$ and the lowest CB $c_1$. FS denotes the free-space dispersion curve, stemming from those parts of the eigenstates ${\phi}_{\mathrm{GS},q}(x)$ lying outside the boundaries of the solid. The black, continuous arrows indicate the action of the XUV probe pulse, exciting electrons from both VBs $v_1$ and $v_2$ into corresponding CBs. The broad spectral width of the pulse is also indicated, showing which CB states are excited. The red, dashed arrow depicts the intraband dynamics in the CB $c_1$ caused by interaction with the few-femtosecond pump pulse. The \textit{ab-initio} TDDFT simulations include all dynamics induced by the light pulses, beyond the $c_1$ intraband dynamics indicated in the figure.}
\label{fig:TDDFT_states}
\end{figure}

In order to visualize the properties of the GS, we diagonalize the single-particle Hamiltonian associated with the GS potential $V_{\mathrm{KS}}[\{n_{\mathrm{GS}}(x)\}](x)$, providing the GS eigenstates $\phi_{\mathrm{GS},q}(x)$ and the associated eigenenergies $\mathcal{E}_{\mathrm{GS},q}$. Note that the index $q$ can potentially go up to infinity: The first $ZN_{\mathrm{at}}/2$ eigenstates represent occupied VB orbitals, whereas the remaining ones are unoccupied CB orbitals. Figure~\ref{fig:TDDFT_states} displays the modulus square of the Fourier transform  $\tilde{\phi}_{\mathrm{GS},q}(k)$ of ${\phi}_{\mathrm{GS},q}(x)$ (in logarithmic scale), plotted for the corresponding eigenenergy $\mathcal{E}_{\mathrm{GS},q}$. The figure allows one to distinguish the two occupied VBs $v_1$ and $v_2$, as well as the first unoccupied CBs $c_1,\, c_2,\,\ldots$ The figure shows a wide $v_2$--$c_1$ direct bandgap of approximately $7\,\mathrm{eV}$. It also highlights a transition energy of approximately $37\,\mathrm{eV}$ between the lowest-lying VB $v_1$ and the bottom of the CB $c_1$, which renders this solid suitable for excitation by an attosecond XUV probe pulse, as shown by the black arrow. For $k = \pm \pi/a$, i.e., at the edges of the BZ, we can also observe an avoided crossing (AC) between the CBs $c_1$ and $c_2$: The transition energy between this $c_1$--$c_2$ AC and the VB $v_1$ is equal to approximately $47\,\mathrm{eV}$. Note that contributions from free-space dispersion, $\mathcal{E} = k^2/2$, can also be distinguished for eigenenergies $\mathcal{E}_{\mathrm{GS},q}>0$. These are a consequence of the finite-size of the solid considered here, as they stem from those parts of the eigenstates ${\phi}_{\mathrm{GS},q}(x)$ which lie outside the boundaries of the solid. The GS eigenstates ${\phi}_{\mathrm{GS},q}(x)$ also provide the associated dipole couplings
\begin{equation}
d_{q'q} = -\int \diff x\, {\phi}^*_{\mathrm{GS},q'}(x)\,x\,{\phi}_{\mathrm{GS},q}(x),
\label{eq:dipoledq'q}
\end{equation}
i.e., the transition matrix elements of the dipole operator $-\hat{x}$, where the integral runs over the entire simulation box.

\subsection{Bloch states of a solid with periodic boundary conditions imposed}
\label{Subsec:Bloch}
To compare results from TDDFT and the EH model, we investigate the infinite solid associated with the finite-size solid presented in the previous section. For this purpose, we consider a periodic system whose potential $U(x)$ [Eq.~(\ref{eq:singlepartHam})] is given by the periodic repetition of the potential shown in Fig.~\ref{fig:potential}(b). This is equivalent to imposing that $U(x)$ be given by a Fourier series
\begin{equation}
U(x) = \sum_{s} \tilde{U}_s\,\eu^{\uimm \frac{2\pi s}{a}x}
\end{equation}
where the Fourier coefficients $\tilde{U}_s$ are calculated from the GS KS potential $V_{\mathrm{KS}}[\{n_{\mathrm{GS}}(x)\}](x)$ within a single cell in the bulk of the solid:
\begin{equation}
\tilde{U}_s = \frac{1}{a}\int_0^a V_{\mathrm{KS}}[\{n_{\mathrm{GS}}(x)\}](x)\,\eu^{-\uimm \frac{2\pi s}{a}x}.
\end{equation}
The periodic component $u_{n,k}(x)$ of the Bloch states $\psi_{n,k}(x)$ of Eq.~(\ref{eq:Blochstates}) can be similarly expanded as a Fourier series
\begin{equation}
u_{n,k}(x) = \sum_{s} c_{k,s}^{(n)}\,\eu^{\uimm \frac{2\pi s}{a}x}
\label{eq:ukexpansion}
\end{equation}
with coefficients
\begin{equation}
c_{k,s}^{(n)} = \frac{1}{a}\int_0^a u_{n,k}(x)\,\eu^{-\uimm \frac{2\pi s}{a}x}.
\label{eq:ckexpansion}
\end{equation}
The associated Schr\"{o}dinger equation in Eq.~(\ref{eq:SchroedingerBloch}) can therefore be recast in terms of the following eigenvalue problem \cite{ashcroft_solid_1976},
\begin{equation}
\sum_{s'} \left[\frac{1}{2}\left( k - \frac{2\pi s'}{a}\right)^2\delta_{ss'} +\tilde{U}_{s'-s}\right]\,c_{k,s'}^{(n)} = \mathcal{E}_{n,k}\,c_{k,s}^{(n)},
\label{eq:discreteSchroed}
\end{equation}
whose solution provides the band structure $\mathcal{E}_{n,k}$, the coefficients $c_{k,s}^{(n)} $, and therefore the Bloch states $\psi_{n,k}(x)$. 

Numerically, this is achieved by sampling $V_{\mathrm{KS}}[\{n_{\mathrm{GS}}(x)\}](x)$ and $u_{n,k}(x)$ over $N_{\mathrm{samp}} = 70$ points $x_j = j\Delta x$ uniformly spanning a single unit cell, with $j = 0,\,1,\,\ldots,\,N_{\mathrm{samp}} - 1$ and grid spacing $\Delta x = a/N_{\mathrm{samp}}$. All the above Fourier coefficients $\tilde{U}_s = \tilde{U}_{s + N_{\mathrm{samp}}}$ and $c_{k,s}^{(n)} = c_{k,s + N_{\mathrm{samp}}}^{(n)} $ are therefore periodic in reciprocal space, and all sums in $s$ and $s'$ in the above identities are limited to $s = 0,\,1,\,\ldots,\,N_{\mathrm{samp}} - 1$. In solving Eq.~(\ref{eq:discreteSchroed}), to ensure that the kinetic-energy term $( k - {2\pi s'}/{a})^2$ is properly defined, we shift $s'\rightarrow (s' + rN_{\mathrm{samp}})$ by a suitable integer number $r$ so that the condition $-\pi N_{\mathrm{samp}} /a \leq [k - 2\pi  (s' + rN_{\mathrm{samp}})/a ]< \pi N_{\mathrm{samp}}/a$ is fulfilled. We also impose the normalization conditions 
\begin{equation}
\sum_s \left(c_{k,s}^{(n)}\right)^*c_{k,s}^{(n')} = \delta_{nn'}
\end{equation}
which, together with Eq.~(\ref{eq:ckexpansion}), allows us to ensure the normalization of $u_{n,k}(x)$ set by Eq.~(\ref{eq:normofu}).

\begin{figure}[t]
\centering
\includegraphics[width=\linewidth]{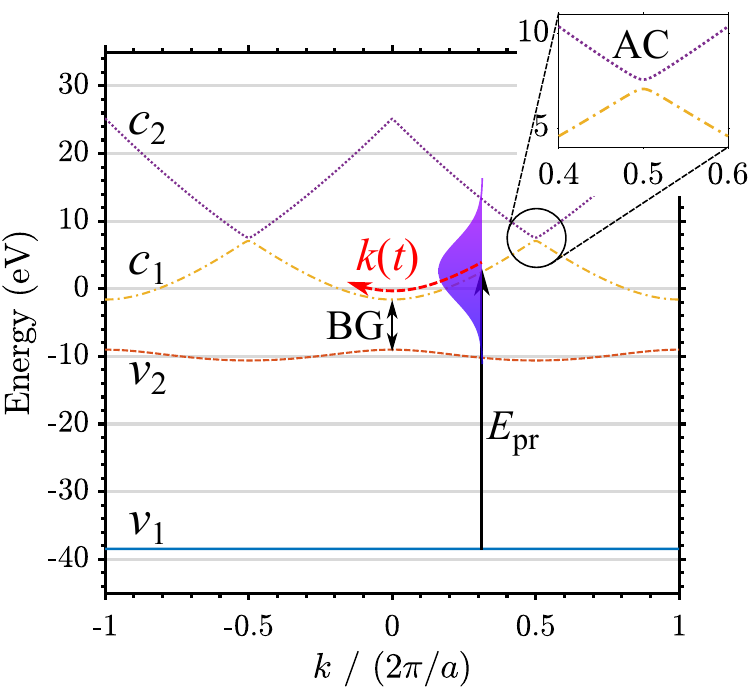}
\caption{First four energy bands $\mathcal{E}_{n,k}$ for a solid with Born--von K\'arm\'an periodic boundary conditions and with a potential within a single unit cell equal to the potential in Fig.~\ref{fig:potential}(b). The four bands depicted in the figure are those included in the numerical simulations based on the EH model and represent (blue, continuous) VB $v_1$, (orange, dashed) VB $v_2$, (yellow, dot-dashed) CB $c_1$, and (purple, dotted) CB $c_2$. BG denotes the bandgap between the highest VB $v_2$ and the lowest CB $c_1$ at $k = 0$. The inset highlights the AC between the two CBs $c_1$ and $c_2$ at $k = \pm \pi/a$. The black, continuous arrow indicates the action of the XUV probe pulse, exciting electrons from the VB $v_1$ into the CBs. The broad spectral width of the pulse is also indicated, showing which CB states are excited. Note that in this case, due to the absence of CBs $c_3$ and higher, the excitation of electrons from the VB $v_2$ is not included in the model. The red, dashed arrow depicts the intraband dynamics in the CB $c_1$ caused by interaction with the few-femtosecond pump pulse. More details on all the pump- and probe-pulse-induced dynamical processes included in our implementation of the EH model are discussed in the text.}
\label{fig:Bloch_states}
\end{figure}

Figure~\ref{fig:Bloch_states} displays the energy bands $\mathcal{E}_{n,k}$ for the two VBs and the first two CBs of the model, obtained by solving the eigenvalue problem of Eq.~(\ref{eq:discreteSchroed}). The energy bands of Fig.~\ref{fig:Bloch_states} are in perfect agreement with the band structure displayed in Fig.~\ref{fig:TDDFT_states} for the corresponding finite-size solid, reproducing the direct bandgap between the VB $v_2$ and the CB $c_1$ at $k = 0$, and the $c_1$--$c_2$ AC at the edges of the BZ for $k = \pm \pi/a$. Note that the free-space dispersion is absent in Fig.~\ref{fig:Bloch_states} due to the periodic boundary conditions imposed in this case and the ensuing absence of free-space eigenstates.

\subsection{Parallel-transport structure gauge and calculation of crystal-momentum-dependent interband couplings}

For each band $n$, the coefficients $c_{k,s}^{(n)}$ given by the eigenvalue problem of Eq.~(\ref{eq:discreteSchroed}) and the corresponding Bloch states $\psi_{n,k}(x)$ obtained via Eqs.~(\ref{eq:Blochstates}) and (\ref{eq:ukexpansion}) are defined up to a $k$-dependent free phase term $\eu^{\uimm\alpha(k)}$. This gauge freedom in the definition of the Bloch states is referred to as structure gauge \cite{Yue_tutorial_22}. The random structure gauge, in which the phases $\alpha(k)$ vary randomly as a function of $k$, is not a suitable choice for our EH model, which relies on the interband couplings $\xi_{n',n}(k_m(t))$ for a time-dependent crystal momentum $k_m(t) = k_m + A(t)$. We require a structure gauge in which the Bloch states and the associated couplings $\xi_{n',n}(k)$ are differentiable functions of $k$. Possible gauges addressing this issue have been studied and discussed before \cite{PhysRevA.101.053411, PhysRevA.94.013846, PhysRevB.102.155201}. Here, we follow the parallel-transport procedure outlined in Ref.~\cite{vanderbilt_berry_2018}, which we briefly summarize below.

In the parallel-transport procedure, all Berry connections are set to identically vanish:
\begin{equation}
\xi_{n,n}(k) = \frac{\uimm}{a}\int_0^a \diff x \,u^*_{n,k}(x)\frac{\diff u_{n,k}(x)}{\diff k} = 0.
\end{equation}
Numerically, this is achieved by sampling the $k$ axis of the BZ over $\tilde{N}$ points, $k_m = m\Delta k$, with $\Delta k = 2\pi/(a \tilde{N})$ and $m = -\tilde{N}/2,\,-\tilde{N}/2 + 1,\, \ldots,\,\tilde{N}/2-1$. We note that the number of sampling points $\tilde{N}$ for the parallel transport procedure and $N$ for the Born--von K\'arm\'an periodic boundary conditions (see Sec.~\ref{Subsec:singlepartHam}) need not be the same. For the remaining of this paper, we set $N = \tilde{N} = 900$. For this discrete set of $k_m$s, the above condition of vanishing Berry connections reduces to
\begin{equation}
\xi_{n,n}(k_m) \,\Delta k = -\mathrm{Im}\log \frac{1}{a}\int_0^a \diff x \,u^*_{n,k_m}(x)\,u_{n,k_{m+1}}(x) = 0,
\label{eq:condition}
\end{equation}
which is equivalent to requesting that
\begin{equation}
\frac{1}{a}\int_0^a \diff x \,u^*_{n,k_m}(x)\,u_{n,k_{m+1}}(x) \in \mathbb{R}_+\ \  \forall m,
\end{equation}
i.e., that the above integral be real and positive for every $m$. 

If we denote by $u_{n,k_{m}}^{\mathrm{rand}}(x)$ the functions obtained by Eq.~(\ref{eq:ukexpansion}) and (\ref{eq:discreteSchroed}) with random $k$-dependent phases, then the parallel-transport procedure which allows one to fulfill  Eq.~(\ref{eq:condition}) consists in the following steps:
\begin{equation}
\begin{aligned}
u_{n,k_{-\tilde{N}/2}}(x) &= u_{n,k_{-\tilde{N}/2}}^{\mathrm{rand}}(x),\\
u_{n,k_{m}}(x) &= u_{n,k_{m}}^{\mathrm{rand}}(x)\frac{\int_0^a \diff x \,u^*_{n,k_{m-1}}(x)\,u_{n,k_{m}}^{\mathrm{rand}}(x) }{\left| \int_0^a \diff x \,u^*_{n,k_{m-1}}(x)\,u_{n,k_{m}}^{\mathrm{rand}}(x) \right|},
\end{aligned}
\end{equation}
for $m = -\tilde{N}/2 + 1,\, \ldots,\,\tilde{N}/2-1$. The procedure imposes a specific $k$-dependence on the phases $\alpha(k_m)$ of each $u_{n,k_{m}}(x)$, and thus sets a specific structure gauge (up to the free, irrelevant initial phase of $u_{n,k_{-\tilde{N}/2}}(x)$).

In terms of the discrete $k_m$ grid and in the parallel-structure gauge, the Berry phase associated with the $n$th band [Eq.~(\ref{eq:Berryphase})] can be recast as \cite{vanderbilt_berry_2018}
\begin{equation}
\phi_n = - \mathrm{Im}\log \frac{1}{a}\int_0^a \diff x \,u^*_{n,k_{-\tilde{N}/2}}(x)\,u_{n,k_{\tilde{N}/2}}(x).
\end{equation}
In calculating these Berry phases, it is important to ensure the correct periodicity of the associated Bloch states over a complete BZ cycle, $\psi_{n,k_{\tilde{N}/2} }(x) = \psi_{n,k_{-\tilde{N}/2} }(x)$, which implies that
\begin{equation}
u_{n,k_{\tilde{N}/2} }(x) = \eu^{-\uimm\frac{2\pi}{a} x}\,u_{n,k_{-\tilde{N}/2} }(x).
\end{equation}
We implemented the parallel-transport procedure for the four bands displayed in Fig.~\ref{fig:Bloch_states} and verified that all four associated Berry phases vanish in our case. This ensures that, for our specific system, the parallel-transport gauge is also a periodic structure gauge \cite{PhysRevA.101.053411, Yue_tutorial_22}. 

In our chosen gauge, all Berry connections $\xi_{n,n}(k)$ vanish by definition. In order to compute the interband couplings $\xi_{n',n}(k)$ for $n\neq n'$, we exploit Eq.~(\ref{eq:xin'n}) and the relationship between position and momentum operators:
\begin{equation}
\xi_{n',n}(k) = -\uimm \frac{p_{n',n}(k)}{\mathcal{E}_{n',k} -\mathcal{E}_{n,k}},
\end{equation} 
where the transition matrix elements of the momentum operator $\hat{p} = -\uimm\diff/\diff x$ are given by
\begin{equation}
p_{n',n}(k) = \langle \psi_{n',k}|\hat{p}|\psi_{n,k}\rangle = -\frac{\uimm}{a}\int_0^a  \diff x\,u^*_{n,k}(x)\frac{\diff u_{n,k}(x)}{\diff x}
\end{equation}
for $n\neq n'$. Numerically, the momentum transition matrix elements are computed with the functions $u_{n,k}(x_j)$ obtained by the parallel-transport procedure, sampled over $N_{\mathrm{samp}} = 70$ points $x_j = j\Delta x$ uniformly spanning a single unit cell, with $j = 0,\,1,\,\ldots,\,N_{\mathrm{samp}} - 1$ and grid spacing $\Delta x = a/N_{\mathrm{samp}}$ (see also Sec.~\ref{Subsec:Bloch}). Over this discrete space grid $x_j$, the first-order space derivative operator is computed by the finite-difference two-point stencil formula
\begin{equation}
\left.\frac{\diff f(x)}{\diff x}\right|_{x_j} = \frac{f(x_{j+1}) - f(x_{j-1})}{2\Delta x} + O(\Delta x^2),
\end{equation}
with the unit-cell periodicity condition given by $f(x_{-1}) = f(x_{N_{\mathrm{samp}} - 1})$ and $f(x_{N_{\mathrm{samp}} }) = f(x_0)$. Over the grid $x_j$, the derivative operator thus reduces to the following $N_{\mathrm{samp}}\times N_{\mathrm{samp}}$ matrix \cite{PhysRevA.102.033105}:
\begin{equation}
\frac{\diff }{\diff x} \approx \frac{1}{2\Delta x}
\begin{pmatrix}
0   & 1          &           &           &-1  \\
-1 & 0          & 1        &           & 0   \\
     &\ddots  &\ddots &\ddots &      \\
     &             & -1       &0         &1    \\
1   &             &            &-1       &0    
\end{pmatrix}.
\end{equation}

\begin{figure*}[t]
\centering
\includegraphics[width=0.48\linewidth]{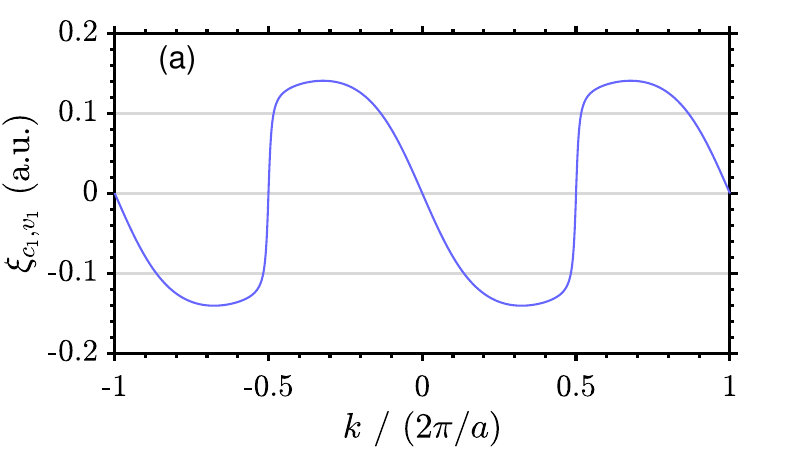}
\includegraphics[width=0.48\linewidth]{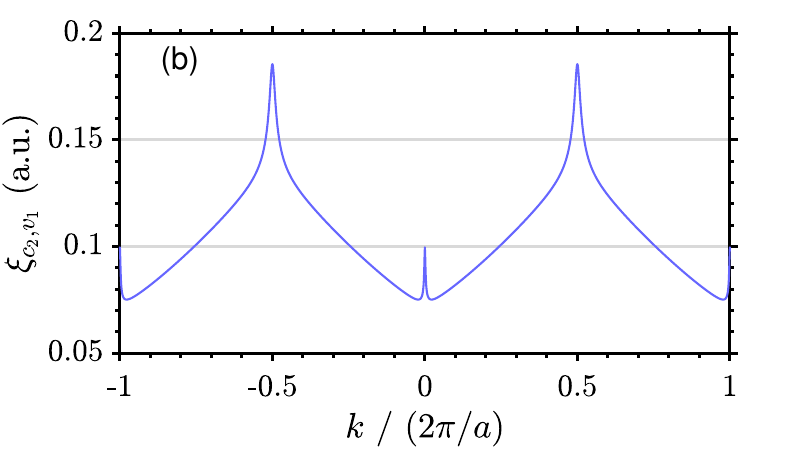}\\
\includegraphics[width=0.48\linewidth]{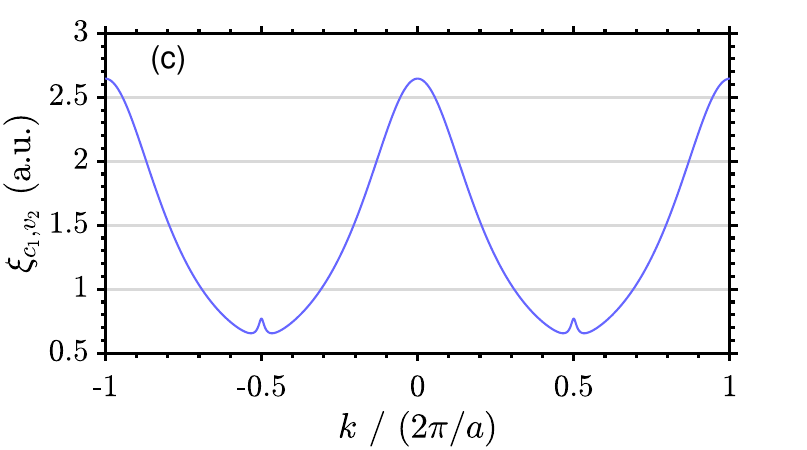}
\includegraphics[width=0.48\linewidth]{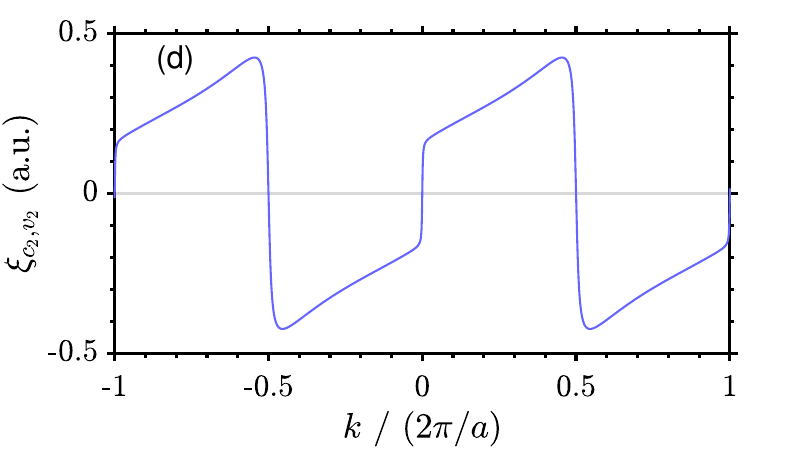}\\
\includegraphics[width=0.48\linewidth]{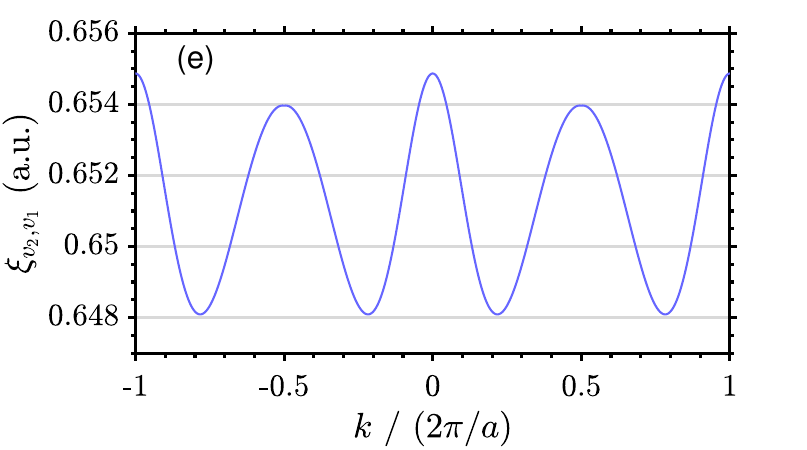}
\includegraphics[width=0.48\linewidth]{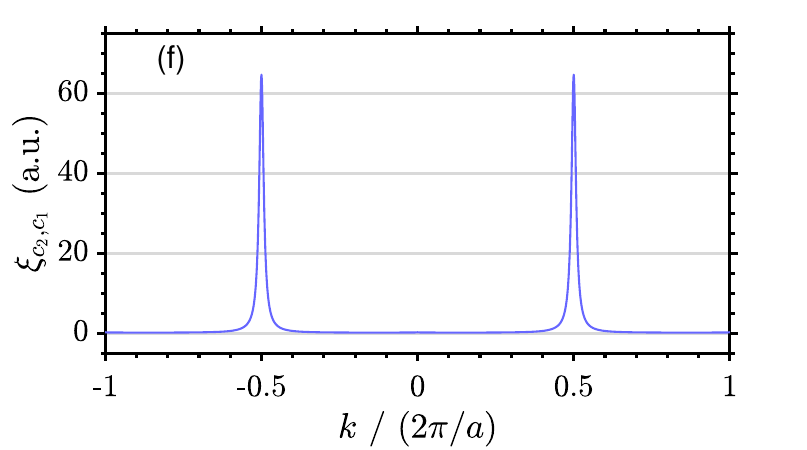}
\caption{Interband couplings calculated by the parallel-transport procedure for the model of a solid with periodic boundary conditions of Fig.~\ref{fig:Bloch_states} (see text for details). The panels exhibit (a) $\xi_{c_1,v_1}(k)$, (b) $\xi_{c_2,v_1}(k)$, (c) $\xi_{c_1,v_2}(k)$, (d) $\xi_{c_2,v_2}(k)$, (e) $\xi_{v_2,v_1}(k)$, and (f) $\xi_{c_2,c_1}(k)$.}
\label{fig:interband_couplings}
\end{figure*}

The resulting interband couplings $\xi_{n',n}(k)$ are displayed in Fig.~\ref{fig:interband_couplings}. One can see that all couplings are continuous functions of $k$ and periodic over a single BZ. We note in particular the very rapid change of sign of both $\xi_{c_1,v_1}(k)$ and $\xi_{c_2,v_2}(k)$ at the edges of the BZ, around $k = \pm \pi/a$. We also note that the $c_1$--$c_2$ coupling is mostly localized around values of $k$ at the edges of the BZ, corresponding to the AC between the two CBs $c_1$ and $c_2$. Although this coupling is relatively localized in $k$ space, for these $k$ values it reaches significantly large values, much larger than the strength of all the other interband couplings.

\section{Strong-field dynamics and attosecond transient absorption spectroscopy signals}
\label{Sec:ATAS}
\subsection{Pulse parameters}

We employ the EH model of Sec.~\ref{Sec:EHmodel} and the TDDFT model introduced in Sec.~\ref{Sec:TDDFT} to simulate the dynamics of the solid presented in Sec.~\ref{Sec:Solid}. For this purpose, for both Eqs.~(\ref{eq:EH-EOMs-tilde}) and (\ref{eq:TDKSE}), we assume a vector potential given by
\begin{equation}
A(t, \tau) = A_{\mathrm{pu}}(t) + A_{\mathrm{pr}}(t, \tau)
\label{eq:vectorpotential}
\end{equation}
with the pump pulse
\begin{equation}
A_{\mathrm{pu}}(t) = A_{0,\mathrm{pu}}\,f(t /T_{\mathrm{pu}})\,\cos(\omega_{\mathrm{pu}} t)
\end{equation}
centered at $t = 0$, and the XUV probe pulse
\begin{equation}
A_{\mathrm{pr}}(t, \tau) = A_{0,\mathrm{pr}}\,f((t - \tau)/T_{\mathrm{pr}})\,\cos(\omega_{\mathrm{pr}}(t-\tau))
\end{equation}
centered at the variable time delay $\tau$. Positive (negative) time delays describe experiments where the central time of the pump precedes (follows) the central time of the probe pulse. The same envelope function
\begin{equation}
f(x) = \left\{
\begin{aligned}
&\cos^2(\pi x), &\text{if $-1/2< x < 1/2$,}\\
&0, &\text{otherwise,} 
\end{aligned}
\right.
\end{equation}
is used to set the finite duration of both pump and probe pulses. We denote by $\omega_{\mathrm{pu}}$ and $\omega_{\mathrm{pr}}$ the central frequencies of the pump and probe pulses, respectively, whereas
\begin{equation}
T_j  = \frac{\pi}{2\arccos\sqrt[4]{1/2}}\,T_{\mathrm{FWHM},j}
\end{equation}
is the interval in which the $j$th pulse does not vanish, $j\in \{\mathrm{pu},\,\mathrm{pr}\}$. $T_j$ is related to the full width at half maximum (FWHM) $T_{\mathrm{FWHM}, j}$ of the intensity profile of the same $j$th pulse. The peak strengths $A_{0,\mathrm{pu}}$ and $A_{0,\mathrm{pr}}$ of the pump and probe vector potentials, respectively, are related to the respective pulse peak intensities via
\begin{equation}
I_j = \frac{|A_{0,j} \omega_j|^2}{8\pi\alpha},
\end{equation}
also in this case for $j\in \{\mathrm{pu},\,\mathrm{pr}\}$. 

\subsection{Computation of the attosecond transient absorption spectra by the electron-hole model and time-dependent density functional theory}

The ATAS signal is defined in terms of the difference between the spectral intensity of the XUV probe pulse following (out) and preceding (in) the interaction with the medium \cite{PhysRevA.91.043408}:
\begin{equation}
\begin{aligned}
\tilde{S}(\omega,\tau) &=  - \left( |\tilde{E}_{\mathrm{out}}(\omega, \tau)|^2  - |\tilde{E}_{\mathrm{in}}(\omega, \tau)|^2 \right)\\
&\approx  - 2\mathrm{Re}\left\{\tilde{E}^*_{\mathrm{pr}}(\omega, \tau)\tilde{E}_{\mathrm{gen}}(\omega, \tau)\right\}.
\end{aligned}
\end{equation}
Here, $\tilde{E}_{\mathrm{in}}(\omega, \tau) = \tilde{E}_{\mathrm{pr}}(\omega, \tau)$ and $\tilde{E}_{\mathrm{out}}(\omega, \tau) = \tilde{E}_{\mathrm{pr}}(\omega, \tau) + \tilde{E}_{\mathrm{gen}}(\omega, \tau)$ are the Fourier transforms of the incoming and outgoing pulses at a given time delay $\tau$, here defined as
\begin{equation}
\tilde{E}(\omega,\tau) = \int \diff t\,\eu^{\uimm\omega t}\,E(t,\tau).
\end{equation}
$\tilde{E}_{\mathrm{gen}}(\omega, \tau)$ represents the Fourier transform of the electromagnetic field generated upon interaction with the medium, whose interference with the incoming pulse gives rise to the spectral features of the absorption spectrum. This can be related to the electron dynamics in the medium,
\begin{equation}
\tilde{E}_{\mathrm{gen}}(\omega, \tau) = \uimm\,\frac{2\pi\omega}{c}\tilde{d}(\omega,\tau),
\end{equation}
where
\begin{equation}
\tilde{d}(\omega,\tau) = \int \diff t\,\eu^{\uimm\omega t}\,d(t,\tau)
\end{equation}
is the Fourier transform of the dipole response $d(t,\tau) $ of the system at a given time delay $\tau$,
\begin{equation}
d(t,\tau)  = - \sum_{i} \langle\varPsi(t,\tau)|\hat{x}_i|\varPsi(t,\tau)\rangle.
\label{eq:dttau}
\end{equation}
Here, $\hat{x}_i$ is the position operator associated with the $i$th electron and the minus sign accounts for the negative value of the electron charge. In Eq.~(\ref{eq:dttau}), $|\varPsi(t,\tau)\rangle$ is the total time-dependent many-body wavefunction of the system, given by Eq.~(\ref{eq:EHansatz}) or (\ref{eq:Psi-TDDFT}) for the EH model and TDDFT, respectively. With the above definitions, the ATAS signal then reduces to
\begin{equation}
\tilde{S}(\omega,\tau) = \frac{4\pi\omega}{c}\mathrm{Im}\left\{\tilde{E}^*_{\mathrm{pr}}(\omega, \tau)\tilde{d}(\omega,\tau) \right\}.
\end{equation}
Different expressions of the ATAS signal are discussed in Ref.~\cite{PhysRevA.91.043408}, see also Refs.~\cite{PhysRevA.83.013419, PhysRevA.83.033405}.

In the following, we compute the reduced signal
\begin{equation}
S(\omega,\tau) = \frac{\tilde{S}(\omega,\tau)}{L} = \frac{4\pi\omega}{cL}\mathrm{Im}\left\{\tilde{E}^*_{\mathrm{pr}}(\omega, \tau)\tilde{d}(\omega,\tau) \right\},
\label{eq:spectrumdef}
\end{equation}
normalized by the length $L$ of the solid, in order to have quantitatively comparable results from the EH model (where $L = 2\pi/\Delta k$ is related to the spacing in $k$ space employed for numerical simulations) and TDDFT (where $L = N_{\mathrm{at}}a$ is associated with the finite number of atoms in the solid). The sign of the absorption spectrum is defined such that the spectrum is positive when light is absorbed, and negative when light is emitted. Note that the ATAS signal could have been analogously defined in terms of the Fourier transforms of the current $j(t,\tau)$ and probe vector potential $A_{\mathrm{pr}}(t,\tau)$. We verified that the two definitions deliver the same results.

In the following calculations, we assume an attosecond XUV probe pulse of central frequency $\omega_{\mathrm{pr}} = 1.5\,\mathrm{a.u.} = 41\,\mathrm{eV}$, FWHM duration of $T_{\mathrm{FWHM},\mathrm{pr}} = 10\,\mathrm{a.u.} = 240\,\mathrm{as}$, and peak intensity of $I_{\mathrm{pr}} = 1.1\times 10^{12}\,\mathrm{W/cm^2}$. The central frequency is chosen such that the pulse is tuned to the transition between VB $v_1$ and the center of the CB $c_1$ for the system of interest, while its duration ensures that the pulse is sufficiently broad to also cover the $v_1$--$v_2$ and part of the $v_1$--$c_2$ transitions. The pulse intensity ensures that the probe pulse induces a linear excitation, as we verify below. The pump pulse is assumed to have a carrier frequency of $\omega_{\mathrm{pu}} = 0.026\,\mathrm{a.u.} = 0.70\,\mathrm{eV}$ and a FWHM duration of $T_{\mathrm{FWHM, pu}} = 1325\,\mathrm{a.u.}=32\,\mathrm{fs}$. Different pump intensities will be considered in the following, from the weak- to the strong-field regime.

In order to calculate the dipole response $d(t,\tau)$ from the EH model, we solve the EOMs of Eq.~(\ref{eq:EH-EOMs-tilde}) for a grid of $N = 900$ $k_m$s spanning the entire BZ, $\Delta k = 2\pi/(Na)$, and with the vector potentials and electric fields from Eqs.~(\ref{eq:vectorpotential}) and (\ref{eq:dAdt}), respectively. The resulting slowly varying variables $\tilde{b}_{0,k_m}(t)$ and $\tilde{b}_{c_i,v_j,k_m}(t)$ are then employed to calculate the dipole response of the system and the associated ATAS signal. Since we are interested in the fast oscillating components of the dipole response which can appear within the spectral observation window determined by the probe pulse $\tilde{E}_{\mathrm{pr}}(\omega)$, we compute the dipole response as
\begin{equation}
\begin{aligned}
d(t,\tau) =-\sum_{k_m} \sum_{i = 1}^2&\, \xi_{v_1,c_i}(k_m(t))\, \tilde{b}^*_{0,k_m}(t,\tau)\,\tilde{b}_{c_i,v_1,k_m}(t,\tau) \\
&\times \eu^{-\uimm\int_{t_0}^t\diff t' \left[\mathcal{E}_{c_i,k_m(t')} - \mathcal{E}_{v_{1},k_m(t')}\right]}\,+\, \mathrm{c.c.},
\end{aligned}
\end{equation}
where we include contributions involving the high-frequency evolution of $v_1$--$c_1$ and $v_1$--$c_2$ pairs. In solving Eq.~(\ref{eq:EH-EOMs-tilde}), due to the ultrashort duration of the probe pulse, one can safely neglect the contribution of the probe-pulse vector potential $A_{\mathrm{pr}}(t)$ to the evolution of $k_m(t)$ and only include the action of the pump, so that
\begin{equation}
k_m(t) = k_m + A_{\mathrm{pu}}(t).
\label{eq:k_mA_pu}
\end{equation}
For the solid considered here, with deep $v_1$ core states and more highly excited $v_2$ states, the main decay processes are those involving the excitation of a $v_1$ state. We account for this main decay mechanism by setting $\gamma_{c_1, v_1} = \gamma_{c_2, v_1} = 1/(5\,\mathrm{fs})$ in Eq.~(\ref{eq:EH-EOMs-tilde}), assuming this decay to be much faster then the decay following the excitation of a $v_2$ electron. The few-femtosecond decay time was set to suitably model the decay of an inner-core electron, see, e.g., Ref.~\cite{Picon2019}. The couplings $\xi_{n',n}(k)$ were calculated in the parallel-transport structure gauge, as presented in Sec.~\ref{Sec:Solid}, resulting in vanishing (intraband) Berry connections and in the interband couplings depicted in Fig.~\ref{fig:interband_couplings}. We note that the interband coupling $\xi_{c_1,v_1}(k)$ has a very steep variation around $k = -\pi/a$, passing from $\approx-0.15\,\mathrm{a.u.}$ to $\approx0.15\,\mathrm{a.u.}$ very abruptly. This implies that, in the EOMs of the EH model, the effective time-dependent coupling $\xi_{c_1,v_1}(k_m + A_{\mathrm{pu}}(t))$ can feature a very fast dependence on time, especially at high pump pulse intensities where $\diff k_m(t)/\diff t$ is very large. For the largest pump intensities considered in this work, we noticed that this leads to an artificially ultrabroadband coupling $\xi_{c_1,v_1}(k_m + A_{\mathrm{pu}}(t))$, which leads to the small, periodic excitations of $v_1$--$c_1$ EH pairs. This was imprinted as additional, yet artificial spectral features in the ATAS signal, which do not have any counterparts in the TDDFT result. In order to avoid these artifacts, which are clearly due to the combination of using Houston states and the steep $k$-dependence of $\xi_{c_1,v_1}(k)$, we employed for our computations dipole couplings given by
\begin{equation}
d(t,\tau) \rightarrow d(t,\tau) \,\theta(t -\tau-T_{\mathrm{pr}}/2 ),
\label{eq:remove_artifacts}
\end{equation}
in which we disregarded artificial contributions preceding the arrival of the XUV probe pulse by introducing the Heaviside step function $\theta(x)$. We stress that this procedure, albeit applied to all our computations, was required only for the highest pump intensities considered in this work and shown in Fig.~\ref{fig:spectra-024}. The EH results of Figs.~\ref{fig:spectra-0008} and \ref{fig:spectra-008} were completely unaffected by implementing Eq.~(\ref{eq:remove_artifacts}).

For the TDDFT approach , we perform real-time propagation of the TDKSE of Eq.~(\ref{eq:TDKSE}) with a time step $\Delta t = 0.1\,\mathrm{a.u.}$ and the same space grid of 17,000 points with spacing $\Delta x = 0.1\,\mathrm{a.u.}$ used for imaginary-time propagation. We employed a complex absorbing potential to avoid backscattering of the electrons reaching the boundaries of the simulation box. The KS orbitals $\phi_{i}(x, t,\tau)$ thus obtained, also dependent on the pump--probe delay $\tau$, were then employed to calculate the associated dipole response as the expectation value of the position operator
\begin{equation}
d(t,\tau) = - \sum_{i} \int \diff x\,\phi^*_{i}(x, t,\tau)\, x\, \phi_{i}(x, t,\tau),
\end{equation}
where the sum runs over all KS orbitals $i$ of spin-up electrons, and the integral covers the entire simulation box. While the dipole response $d(t,\tau)$ calculated by TDDFT includes the decay processes owing to the complex absorbing potential at the boundaries of the simulation box, it does not account for additional decoherence processes such as those due to the finite lifetime of the excited EH states. The fastest decay processes involved in ATAS are those following the excitation of a core $v_1$ electron: This is much faster than any other decay processes, such as those due, e.g., to the excitation of a $v_2$ electron, and determines the width of the ATAS spectral features. To account for this ultrafast decay, we multiply the dipole response obtained by TDDFT by an exponential decay following the XUV excitation,
\begin{equation}
d(t,\tau) \rightarrow d(t,\tau)\,\{[1 - \theta(t-\tau)] + \eu^{-\gamma (t-\tau)}\,\theta(t-\tau)\},
\end{equation}
with a decay rate of $\gamma = 1/(5\,\mathrm{fs})$.

\subsection{Static absorption spectra in the absence of pump excitation}

\begin{figure}[t]
\centering
\includegraphics[width=\linewidth]{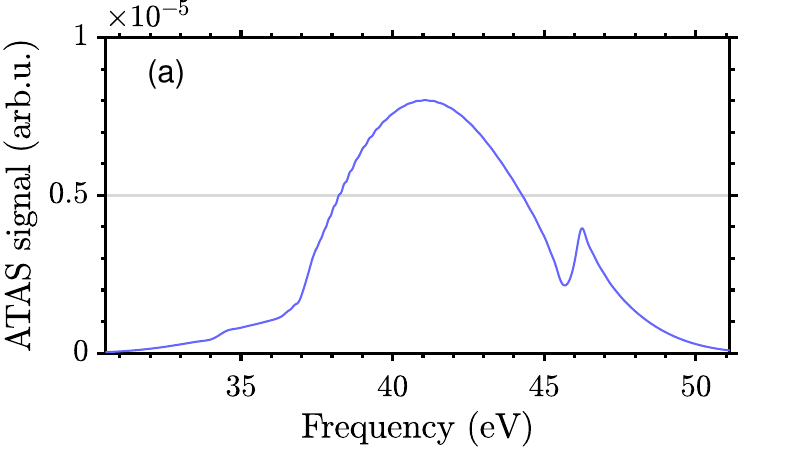}\\
\includegraphics[width=\linewidth]{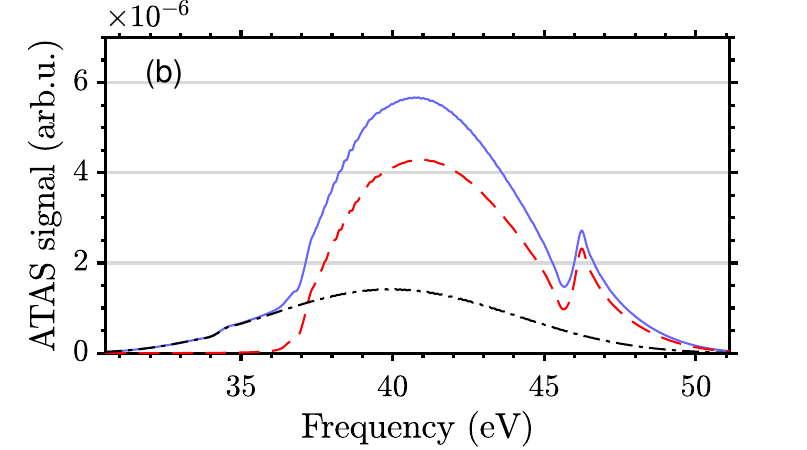}\\
\includegraphics[width=\linewidth]{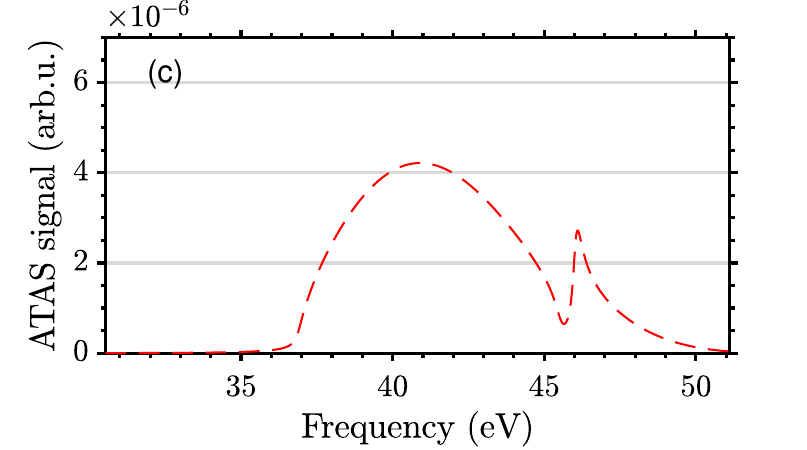}
\caption{Static absorption spectrum of a broadband XUV probe pulse exciting the solid. (a) Static absorption spectrum calculated by TDDFT, see also Fig.~\ref{fig:TDDFT_states}. (b) Static absorption spectrum calculated by linear-response theory with dipole couplings calculated based on the eigenstates of the GS TDDFT Hamiltonian: (blue, continuous) total signal, (red, dashed), contribution to the signal due to the excitation of electrons from the VB $v_1$, and (black, dot-dashed) contribution to the signal due to the excitation of electrons from the VB $v_2$. (c) Static absorption spectrum calculated by the EH model with the four bands depicted in Fig.~\ref{fig:Bloch_states}, thus only including the excitation of electrons from the VB $v_1$ into the CBs $c_1$ and $c_2$.}
\label{fig:staticspectra}
\end{figure}

In order to compare spectra obtained by the two methods and validate our EH model, we show here the static absorption spectrum computed in the absence of the pump pulse. Figure~\ref{fig:staticspectra}(a) displays the absorption spectrum obtained by TDDFT. This corresponds to the excitation mechanism depicted in Fig.~\ref{fig:TDDFT_states}, where the XUV pulse $E_{\mathrm{pr}}(t,\tau)$ excites electrons from both VBs $v_1$ and $v_2$ into correspondingly higher CBs. The absorption spectrum exhibited in Fig.~\ref{fig:staticspectra}(a) covers a frequency range from $32\,\mathrm{eV}$ to $50\,\mathrm{eV}$, reflecting the broad spectrum of the XUV probe pulse $\tilde{E}(\omega)$ also shown in Fig.~\ref{fig:TDDFT_states}. Figure~\ref{fig:staticspectra}(a) highlights a variation and increase in the strength of the absorption signal at $37\,\mathrm{eV}$, i.e., the transition energy at which the excitation of electrons from the VB $v_1$ into the CB $c_1$ starts to be possible. One can also observe an asymmetric absorption feature at $46\,\mathrm{eV}$, corresponding to the AC between the CBs $c_1$ and $c_2$.

In order to assign the features appearing in the spectrum, we compute the same spectrum based on linear-response theory \cite{haug_quantum_2009},
\begin{equation}
S(\omega) =-\frac{4\pi\omega}{cL}\sum_{e>g}\mathrm{Im}\left\{\frac{\tilde{E}^*_{\mathrm{pr}}(\omega)\,\tilde{E}_{\mathrm{pr}}(\omega_{eg})\,|d_{eg}|^2}{\omega - \omega_{eg}  +\uimm\gamma} \right\},
\label{eq:chi}
\end{equation}
where the indices $e$ and $g$ run over the eigenstates $|\phi_{\mathrm{GS},q}\rangle$ of the single-particle Hamiltonian associated with the GS potential $V_{\mathrm{KS}}[\{n_{\mathrm{GS}}(x)\}](x)$, and the dipole couplings $d_{eg}$ are calculated based on Eq.~(\ref{eq:dipoledq'q}). The red, dashed line in Fig.~\ref{fig:staticspectra}(b) displays the sum in Eq.~(\ref{eq:chi}) for $g$ running over all states in VB $v_1$, whereas the black, dot-dashed line in the same figure is obtained by summing over $g$ states in the VB $v_2$. For both lines, the sum in $e$ runs over all states in all CBs. The blue, continuous line shows the total signal from both VBs, whose shape is in very good agreement with the TDDFT results exhibited in Fig.~\ref{fig:staticspectra}(a). Figure~\ref{fig:staticspectra}(b) allows one to distinguish contributions from $v_1$ and $v_2$. One can clearly see that, for frequencies below $37\,\mathrm{eV}$, transitions from VB $v_1$ do not contribute to the signal, since this corresponds to the minimum transition energy between VB $v_1$ and CB $c_1$. The spectral features appearing at frequencies below $37\,\mathrm{eV}$ are therefore only due to the excitation of electrons from the VB $v_2$ to more highly excited CBs. As already discussed above, the asymmetric absorption feature at $46\,\mathrm{eV}$ corresponds to the AC between the CBs $c_1$ and $c_2$.  A similar feature is also present at $\approx 34\,\mathrm{eV}$, corresponding to the AC between the CBs $c_2$ and $c_3$ highlighted in Fig.~\ref{fig:TDDFT_states}. 

Figure~\ref{fig:staticspectra}(c) displays the absorption spectrum computed by the EH model including the four bands depicted in Fig.~\ref{fig:Bloch_states} and only accounting for the excitation of an electron from the VB $v_1$ into the CBs $c_1$ and $c_2$. The EM-model results of Fig.~\ref{fig:staticspectra}(c) display a very good agreement with the red, dashed line of Fig.~\ref{fig:staticspectra}(b). The comparison of static absorption spectra based on TDDFT and the EH model validates our approach, which we extend in the next subsection for the investigation of ATAS signals in the presence of an intense pump pulse. 

\subsection{Attosecond transient-absorption spectroscopy signal and its dependence on the pump-pulse intensity}

In this section, we compute the ATAS signal in the presence of an external pump pulse, from the low-intensity limit up to higher intensities. In particular, we compare our \textit{ab-initio} TDDFT results with results based on the EH model for the band structure of Fig.~\ref{fig:Bloch_states} and the interband couplings of Fig.\ref{fig:interband_couplings}.

We employ the same probe-pulse parameters already used for Fig.~\ref{fig:staticspectra}, i.e., a central frequency of $\omega_{\mathrm{pr}} = 1.5\,\mathrm{a.u.} = 41\,\mathrm{eV}$, a FWHM duration of $T_{\mathrm{FWHM},\mathrm{pr}} = 10\,\mathrm{a.u.} = 240\,\mathrm{as}$, and a pulse peak intensity of $I_{\mathrm{pr}} = 1.1\times 10^{12}\,\mathrm{W/cm^2}$. For the pump pulse, we assume a carrier frequency of $\omega_{\mathrm{pu}} = 0.026\,\mathrm{a.u.} = 0.70\,\mathrm{eV}$ and a FWHM duration of $T_{\mathrm{FWHM, pu}} = 1325\,\mathrm{a.u.}=32\,\mathrm{fs}$. The chosen pump central frequency is approximately a tenth of the $v_2$--$c_1$ bandgap, corresponding to an off-resonant pump excitation.

\begin{figure}[t]
\centering
\includegraphics[width=0.98\linewidth]{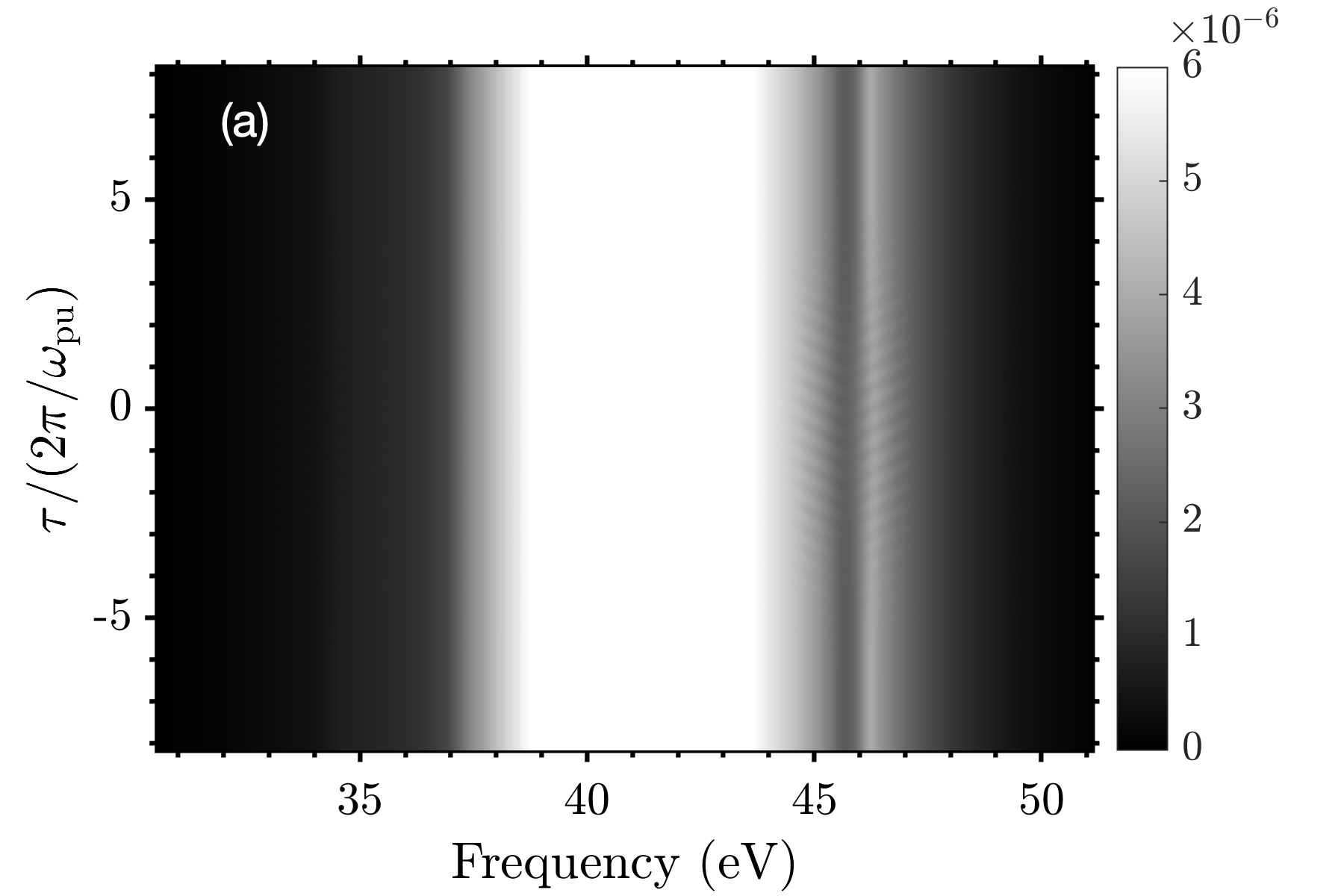}\\
\includegraphics[width=0.98\linewidth]{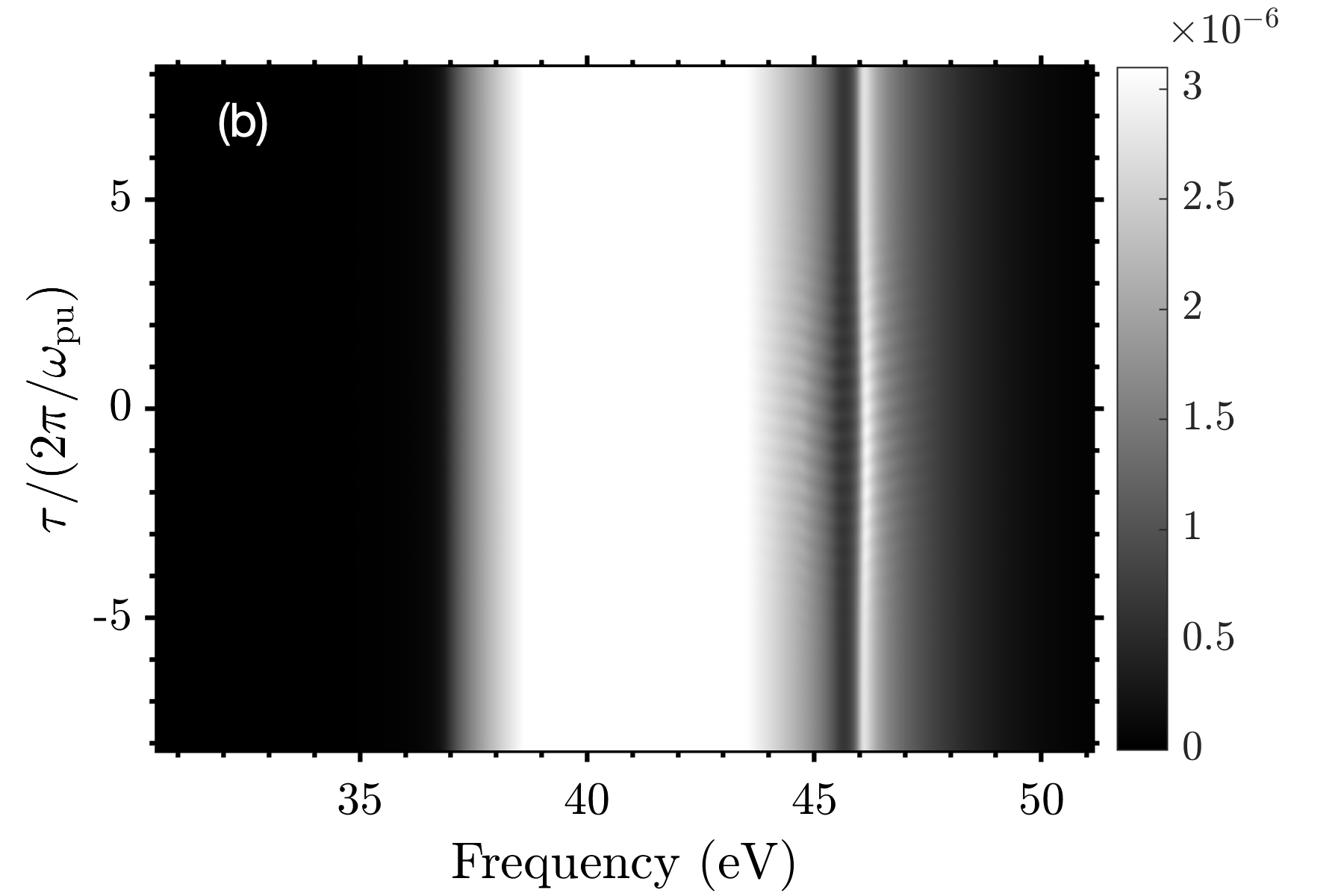}\\
\includegraphics[width=0.98\linewidth]{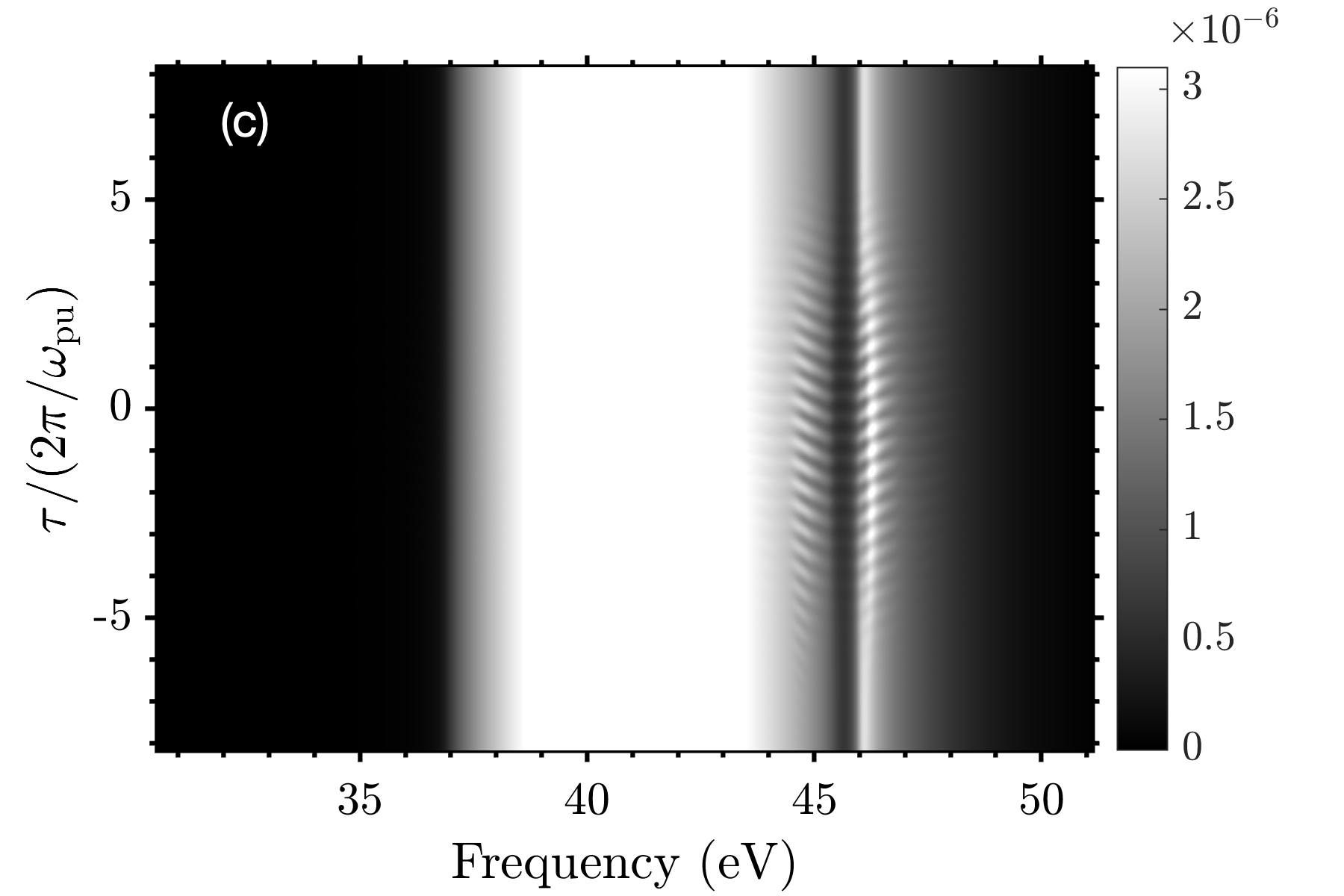}
\caption{Transient absorption spectrum for a peak strength of the pump vector potential $A_{0,\mathrm{pu}} = 0.008\,\mathrm{a.u.}$ The plots show a comparison of the transient absorption spectra computed (a) with TDDFT, (b) the EH model without including the $v_1$--$v_2$ and the $c_1$--$c_2$ couplings, and (c) the EH model including all the 6 interband couplings displayed in Fig.~\ref{fig:interband_couplings}. See the text for system and light parameters.}
\label{fig:spectra-0008}
\end{figure}

\begin{figure}[t]
\centering
\includegraphics[width=0.98\linewidth]{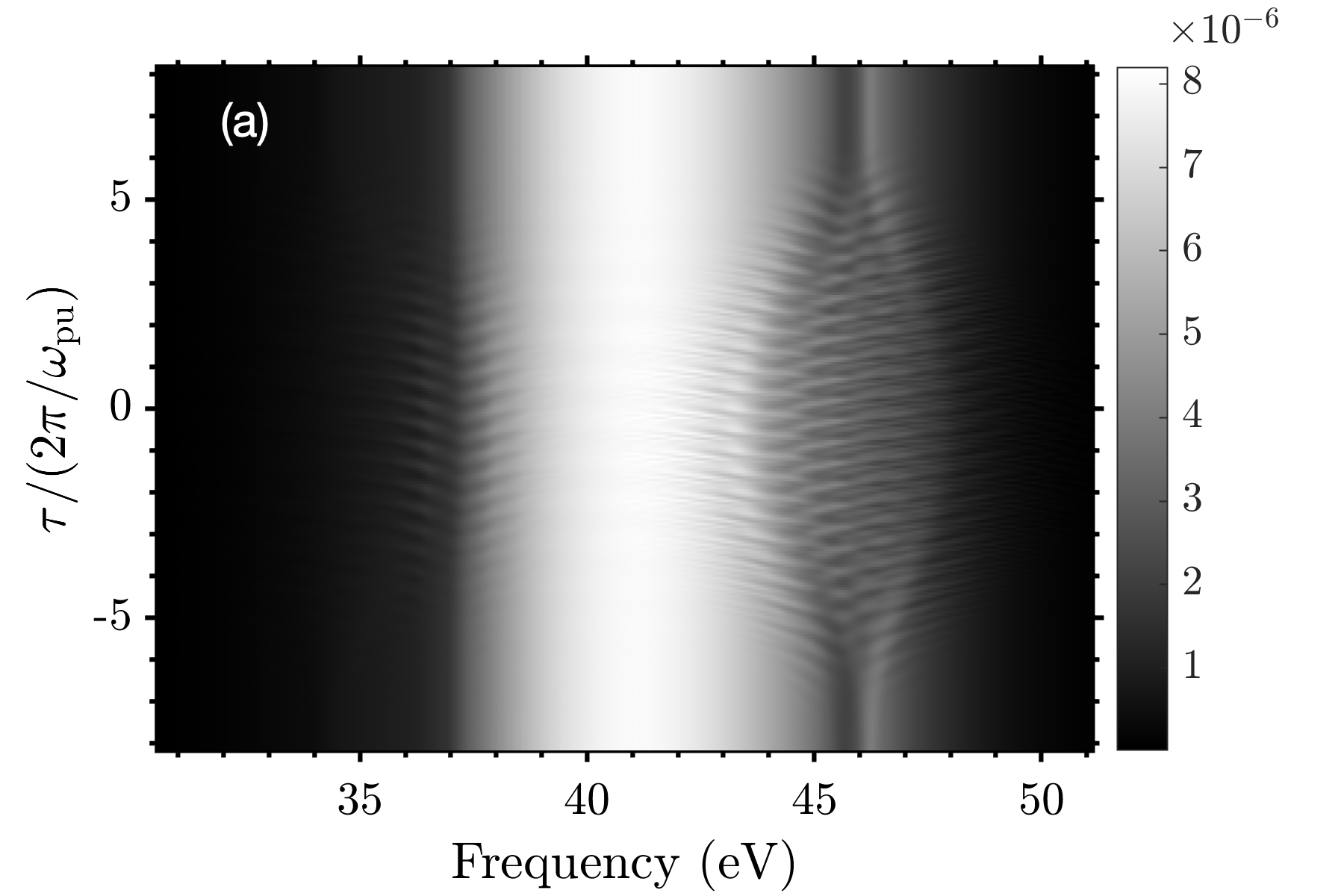}\\
\includegraphics[width=0.98\linewidth]{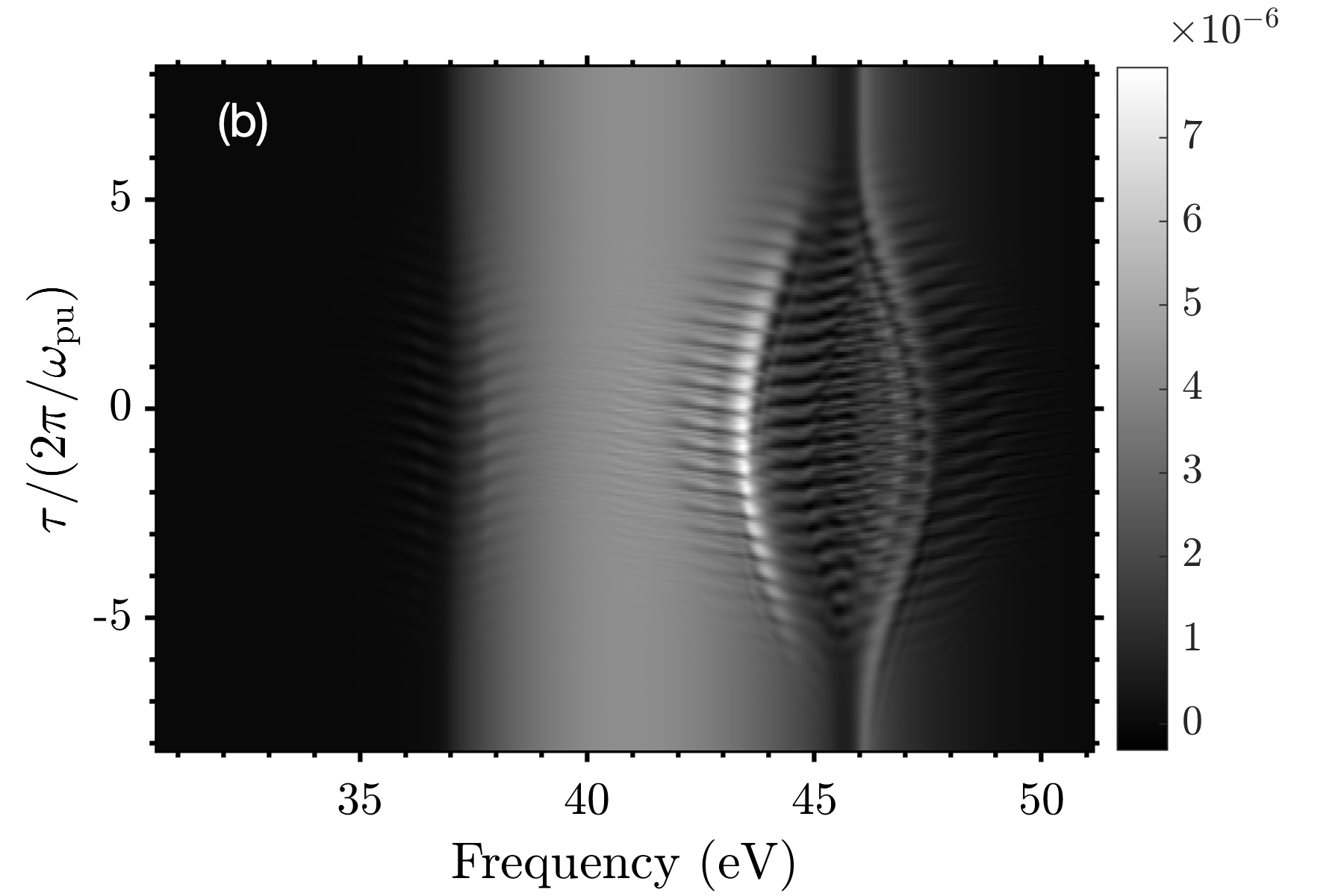}\\
\includegraphics[width=0.98\linewidth]{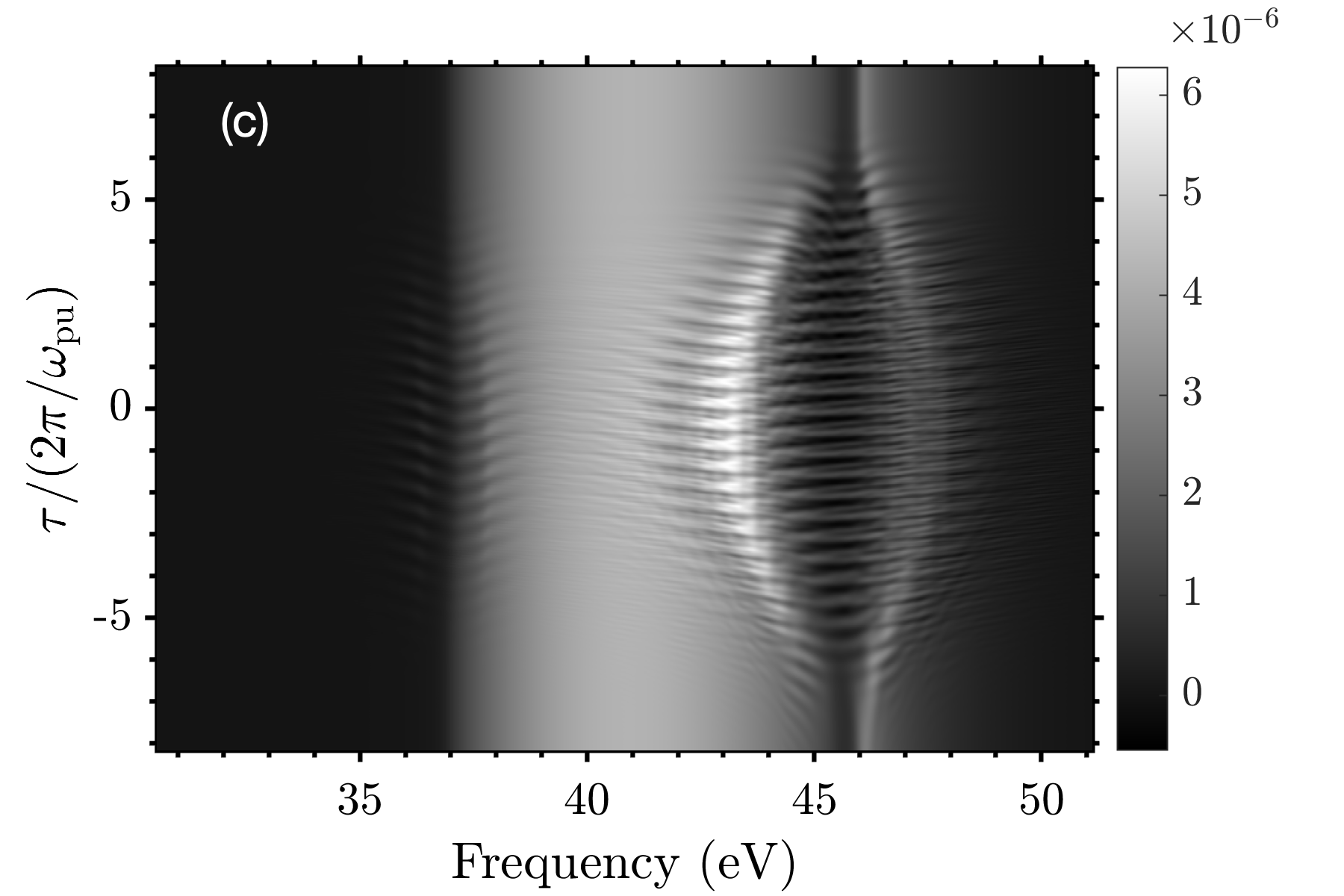}
\caption{Transient absorption spectrum for a peak strength of the pump vector potential $A_{0,\mathrm{pu}} = 0.08\,\mathrm{a.u.}$ The three panels (a)--(c) represent the same quantities as described in Fig.~\ref{fig:spectra-0008}.}
\label{fig:spectra-008}
\end{figure}

\begin{figure}[t]
\centering
\includegraphics[width=0.98\linewidth]{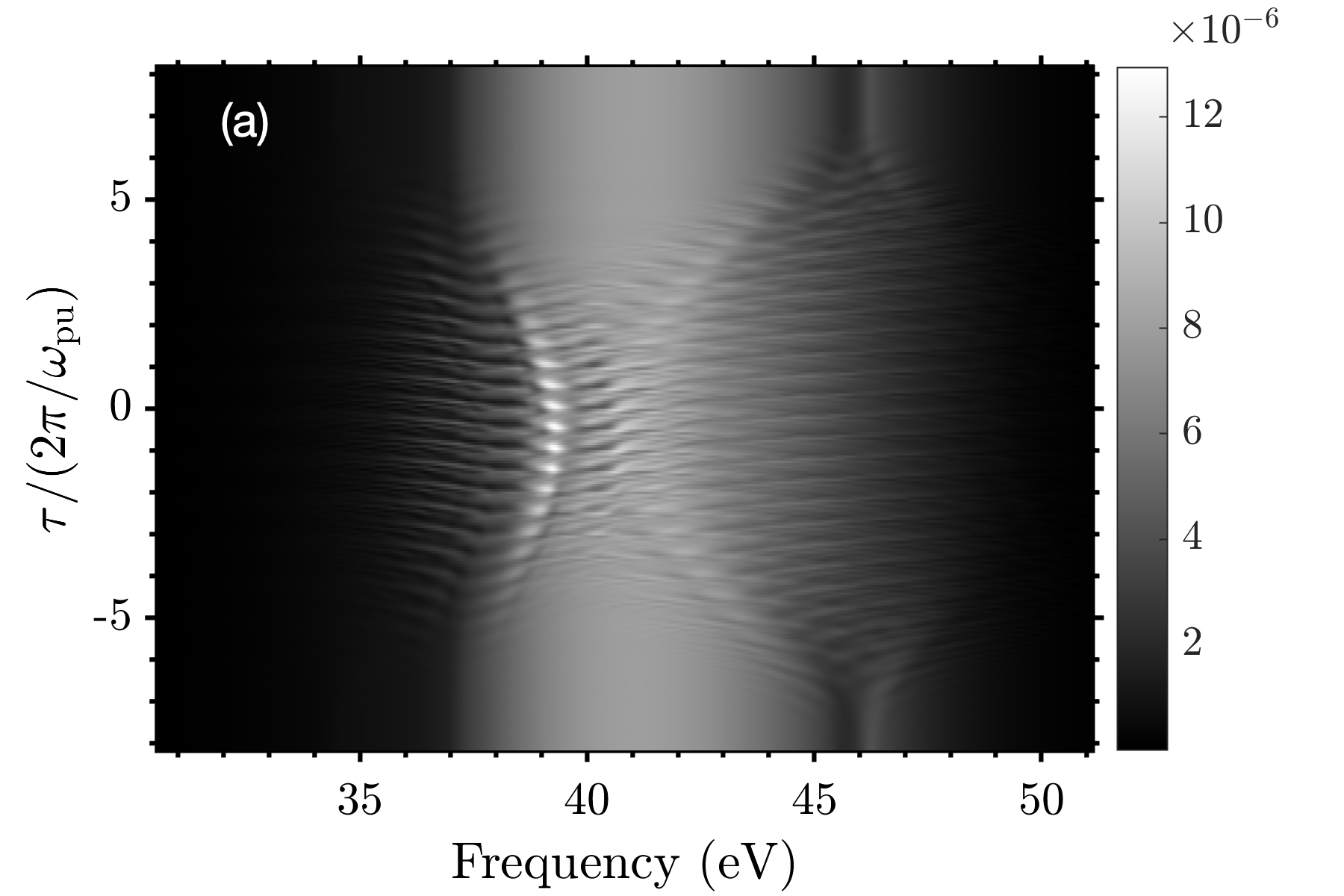}\\
\includegraphics[width=0.98\linewidth]{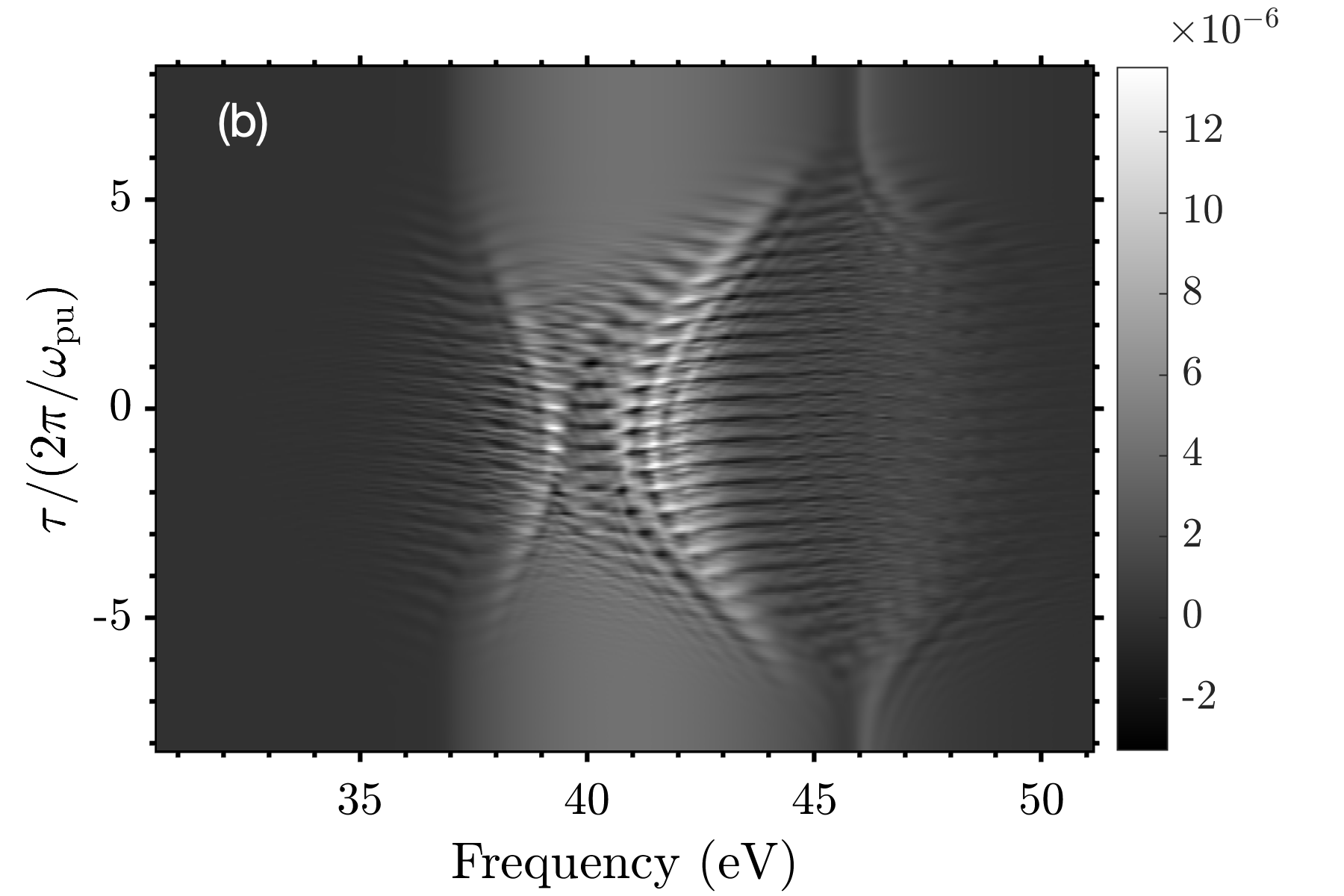}\\
\includegraphics[width=0.98\linewidth]{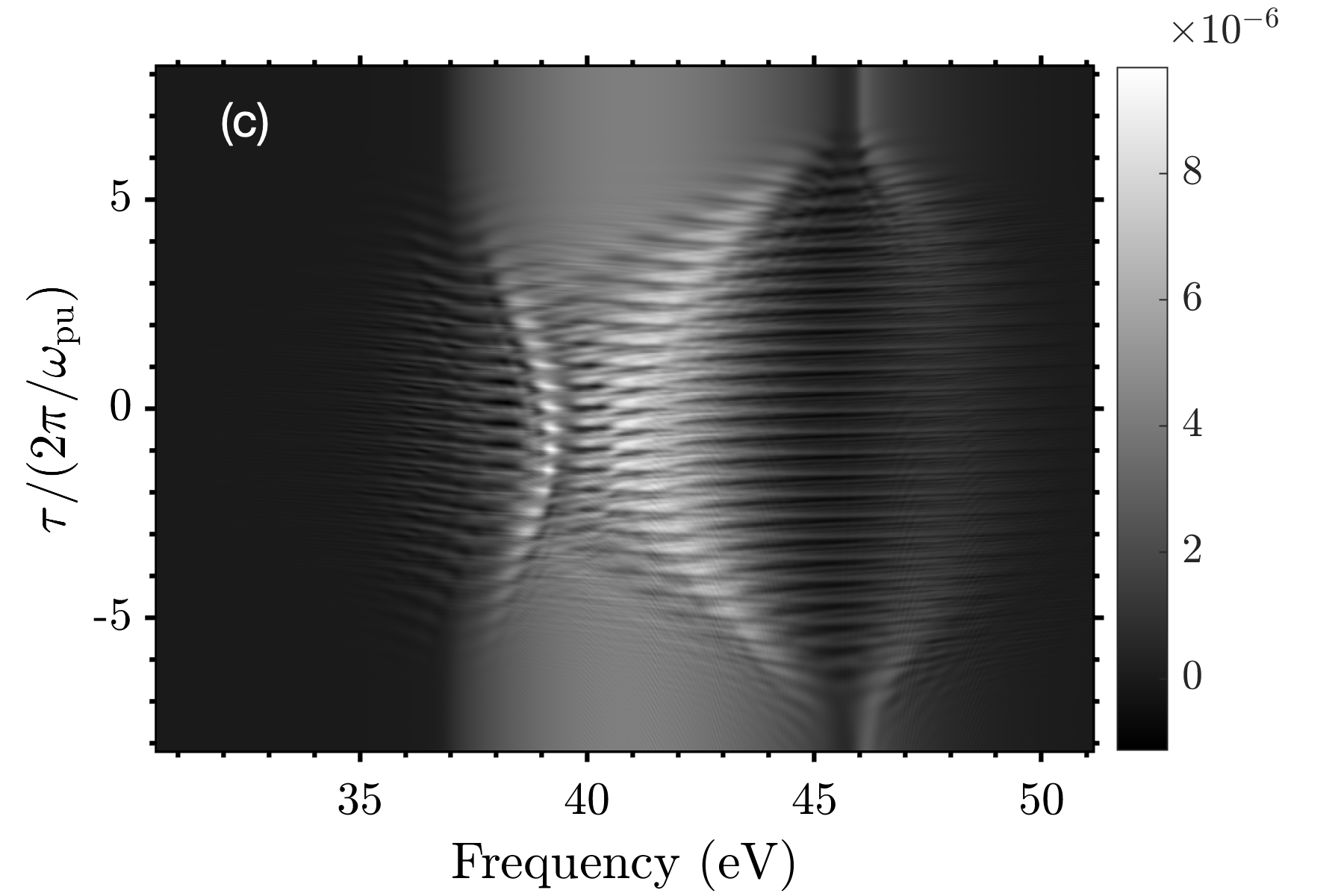}
\caption{Transient absorption spectrum for a peak strength of the pump vector potential $A_{0,\mathrm{pu}} = 0.24\,\mathrm{a.u.}$ The three panels (a)--(c) represent the same quantities as described in Fig.~\ref{fig:spectra-0008}.}
\label{fig:spectra-024}
\end{figure}

Figure~\ref{fig:spectra-0008} displays the ATAS signal for a pump pulse of $A_{0,\mathrm{pu}} = 0.008\,\mathrm{a.u.}$, corresponding to an intensity of $I_{\mathrm{pu}} = 1.51\times10^9\,\mathrm{W/cm^2}$. With the semiclassical evolution of the crystal momentum $k$ given in Eq.~(\ref{eq:k_mA_pu}), this corresponds to a maximum crystal-momentum variation of $2A_{0,\mathrm{pu}} = 0.018\times2\pi/a$ and thus an extremely small oscillation of the $k_m$s around their equilibrium values. Our TDDFT calculations, depicted in Fig.~\ref{fig:spectra-0008}(a), predict small diagonal fringes originating around the AC at $46\,\mathrm{eV}$ (transition energy between the VB $v_1$ and the $c_1$--$c_2$ AC), with a periodicity in $\tau$ of half pump laser period. This is in agreement with previous theory and experimental results, which have highlighted similar fishbone-like structures at the edges of the bands \cite{Lucchini2016, schlaepfer_attosecond_2018, Picon2019, cistaro_theoretical_2023, PhysRevA.100.043840, PhysRevA.106.063107, PhysRevB.107.184304}. These structures can be understood as the result of the periodic transition between Lorentzian and Fano-like lineshapes, resulting in V-shaped structures whose tilt angle is determined by the pump-pump properties \cite{PhysRevA.106.063107}. For our system, these fishbone-like structures are mostly present at the upper edge of the CB $c_1$ and lower edge of the CB $c_2$ for these lower intensities, while no significant features are apparent at the lower edge of the CB $c_1$. Simulations based on the EH model similarly predict fishbone-like structures in the spectral features of the ATAS signal. Figure~\ref{fig:spectra-0008}(b) shows the results of the EH model by only including interband couplings between a VB and a CB, but excluding the $v_1$--$v_2$ and the $c_1$--$c_2$ couplings. The signal features weak, yet visible fishbone-like structures at the $c_1$--$c_2$ AC. The periodicity and the shape of these spectral peaks agree with the TDDFT predictions. In Fig.~\ref{fig:spectra-0008}(c), we additionally included the contributions from the $v_1$--$v_2$ and the $c_1$--$c_2$ interband couplings. Although the $c_1$--$c_2$ interband couplings are very localized in $k$ space, their presence is expected to significantly influence the spectral features and fishbone-like structures around the $c_1$--$c_2$ AC, since this is the region in which these couplings are largest [see Fig.~\ref{fig:interband_couplings}(f)]. This is already visible when comparing Figs.~\ref{fig:spectra-0008}(b) and \ref{fig:spectra-0008}(c). When including these additional couplings, the fishbone-like structures appear to become more prominent. A more significant frequency bending of the main spectral peak as a function of time delay can also be observed, in agreement with the TDDFT predictions.

In order to investigate the dependence of the ATAS features on the pump pulse intensity, in Fig.~\ref{fig:spectra-008} we display ATAS signals for a pump pulse of a peak vector potential of $A_{0,\mathrm{pu}} = 0.08\,\mathrm{a.u.}$ and a peak intensity of $I_{\mathrm{pu}} = 1.51\times10^{11}\,\mathrm{W/cm^2}$. The ATAS signal calculated by TDDFT is displayed in Fig.~\ref{fig:spectra-008}(a). Three main features can be distinguished. First, the main spectral peaks at the $c_1$--$c_2$ AC are shifted by the pump pulse by a time-delay-dependent amplitude which is related to the strength of the vector potential at that delay. Second, in addition to the main time-delay-dependent peaks, one can also observe further spectral structure both to the left and right of the main spectral peaks. Third, new spectral features appear at the bottom of the CB $c_1$, both at frequencies larger and smaller than the minimum transition energy of $\approx 37\,\mathrm{eV}$ between $v_1$ and $c_1$. All these spectral features present a time-delay-dependent modulation with a periodicity of half laser period. 

We employ the EH model to understand and interpret the spectral features highlighted in the TDDFT simulations. ATAS signals calculated by the EH model for the same peak intensity are displayed in Figs.~\ref{fig:spectra-008}(b) and \ref{fig:spectra-008}(c). While the latter results are obtained by fully including all six couplings between the four energy bands in the model, the results of Fig.~\ref{fig:spectra-008}(b) were obtained by imposing that the $v_1$--$v_2$ and the $c_1$--$c_2$ couplings vanish. We first note that, for the highly off-resonant pump pulse employed here, the contributions due to the excitation of an EH pair from the VB $v_2$ into the CBs are significantly small. This is confirmed via EH simulations with vanishing $v_2$--$c_j$ couplings. The dynamics are therefore mostly determined by the interband evolution of $k_m(t)$, the $v_1$--$c_j$ couplings involving the excitation of an electron from the VB $v_1$, and the large $c_1$--$c_2$ coupling.

Figure~\ref{fig:spectra-008}(b) reproduces the same qualitative behavior highlighted in Fig.~\ref{fig:spectra-008}(a) via TDDFT. However, in Fig.~\ref{fig:spectra-008}(b) one can notice that the main spectral peaks originating from the $c_1$--$c_2$ AC are much more marked than in the TDDFT simulations. Especially the peaks of the CB $c_2$ for delays around $\tau = 0$ are clearly more marked in Fig.~\ref{fig:spectra-008}(b) then in Fig.~\ref{fig:spectra-008}(a). Similarly, the main peak of the CB $c_1$ in Fig.~\ref{fig:spectra-008}(b) displays a significantly more intense absorption than the surrounding peaks, in contrast with the TDDFT simulations of Fig.~\ref{fig:spectra-008}(a). By additionally including the $c_1$--$c_2$ coupling, all spectral peaks in Fig.~\ref{fig:spectra-008}(c) appear smoother than in Fig.~\ref{fig:spectra-008}(b), with a closer agreement with the results of Fig.~\ref{fig:spectra-008}(a). We note, however, that Fig.~\ref{fig:spectra-008}(c) features less visible spectral features at the center of the fishbone-like structure at the $c_1$--$c_2$ AC, in contrast to both Figs.~\ref{fig:spectra-008}(a) and \ref{fig:spectra-008}(b). We also note that the spectral features appearing at the bottom of the CB $c_1$ are essentially unaffected by the inclusion of the $c_1$--$c_2$ coupling and remain unaltered in both Figs.~\ref{fig:spectra-008}(b) and \ref{fig:spectra-008}(c).

In order to further interpret the time-delay dependence of the spectral features displayed in Fig.~\ref{fig:spectra-008}, we note that, due to the semiclassical evolution of the crystal momentum in Eq.~(\ref{eq:k_mA_pu}), the maximum amplitude of the crystal-momentum oscillations is equal to $2A_{0,\mathrm{pu}} = 0.18\times2\pi/a$. The amplitude of the vector potential allows an EH pair generated by the probe pulse to span a corresponding interval of crystal momenta, and thus to span a broader range of transition energies, as intuitively depicted by the red dashed line in both Figs.~\ref{fig:TDDFT_states} and \ref{fig:Bloch_states}. This explains why the peaks of the ATAS signal in Fig.~\ref{fig:spectra-008} at the $c_1$--$c_2$ AC span a broader energy range than in the lower-pump-intensity case shown in Fig.~\ref{fig:spectra-0008}. In order to understand why the position of these spectral peaks varies as a function of time delay, one needs to consider that an EH pair generated by the probe pulse at time $t = \tau$, i.e., at the central time of the probe pulse, will decay within a very short time window of a few femtoseconds, as set by the decay rate $\gamma = 1/(5\,\mathrm{fs})$. This is comparable with the pump laser period of $2\pi/\omega_{\mathrm{pu}} = 5.9\,\mathrm{fs}$. Within this very short time window, the amplitude of the oscillations in crystal momentum will be approximately set by the amplitude of the envelope of the vector potential around $t =\tau$, i.e., $A_{0,\mathrm{pu}}\,f(\tau /T_{\mathrm{pu}})$. With varying $\tau$, an EH pair will therefore be able to cover a $k$ interval determined by the envelope $f(\tau/T_{\mathrm{pu}})$ at that specific delay, and thus span an energy range correspondingly dependent on $\tau$. This dependence of the set of transition energies covered by an EH pair for a given time delay is reflected in the spectral features of the ATAS signal. 

The additional peaks visible in the signal, both at the $c_1$--$c_2$ AC and at the bottom of the CB $c_1$, are due to the Floquet dressing of the CBs by the pump pulse \cite{PhysRevA.100.013412, PhysRevB.107.184304}. Floquet dressing creates additional spectral features separated by the laser frequency $\omega_{\mathrm{pu}}$, whose strength increases with the strength of the laser field. These properties are apparent in Fig.~\ref{fig:spectra-008}, both at the center of the fishbone-like spectral feature at the $c_1$--$c_2$ AC, as well as at the bottom of the CB $c_1$. We note in particular the spectral features appearing at frequencies below the energy gap between the VB $v_1$ and the CB $c_1$, in a energy region in which no bare transitions are present and in which absorption is enabled by the dressing induced by the pump.

Finally, in Fig.~\ref{fig:spectra-024}, we show simulations for an intense pump pulse, with a peak vector potential strength of $A_{\mathrm{pu}} = 0.24\,\mathrm{a.u.}$ and a corresponding peak intensity of $I_{\mathrm{pu}} = 1.36\times 10^{12}\,\mathrm{W/cm^2}$. For the chosen vector potential strength, the maximum amplitude of the crystal-momentum oscillations is equal to $2A_{0,\mathrm{pu}} = 0.72\times2\pi/a$, i.e., more than half BZ, so that an EH pair generated by the probe pulse can potentially span all the allowed energy transitions between a given CB and the VB $v_1$. This is observed in the TDDFT simulations of Fig.~\ref{fig:spectra-024}(a), and is even more clearly visible in the EH simulations of Figs.~\ref{fig:spectra-024}(b) and \ref{fig:spectra-024}(c), similarly obtained without and with the inclusions of the $v_1$--$v_2$ and $c_1$--$c_2$ couplings, respectively. As we already discussed for Fig.~\ref{fig:spectra-008}, also in this case one can observe more marked peaks and spectral features in Fig.~\ref{fig:spectra-024}(b) than in the TDDFT simulations of Fig.~\ref{fig:spectra-024}(a). These features are smoothed and their intensity decreases when the $c_1$--$c_2$ coupling is included, as displayed in Fig.~\ref{fig:spectra-024}(c). Similarly to the case of Fig.~\ref{fig:spectra-008}, also here we can see that the spectral features at the bottom of the CB $c_1$ are essentially unaffected by the inclusion or exclusion of the $c_1$--$c_2$ interband coupling, with identical spectral features in both Figs.~\ref{fig:spectra-024}(b) and \ref{fig:spectra-024}(c).

\section{Conclusions}
\label{Sec:Conclusions}

In this paper, we studied theoretically the ATAS signal of a wide-bandgap semiconductor excited by an intense optical pump and probed by an attosecond XUV pulse. We employed a one-dimensional model providing a band structure with an inner core VB, a more excited VB, and several CBs, and assumed a pump off-resonant from the $v_2$--$c_1$ bandgap and a probe resonant with the transition between the deepest VB and the CBs. We employed both an EH theory model and TDDFT in order to compute the strong-field dynamics and predict the ATAS spectral features and their dependence on the pump--probe interpulse delay. In order to obtain quantitatively comparable results, we first set the properties of a finite-size solid in TDDFT, and then considered an associated solid with periodic boundary conditions in order to retrieve the transition energies and interband couplings in reciprocal space as a function of the crystal momentum $k$.

We showed fishbone-like features at the edges of the CBs emerging at low pump intensities, in agreement with previous theory and experimental studies. For larger intensities, we found that the broad energy range spanned by the EH pair in the semiclassical evolution of the crystal momentum $k(t)$ is imprinted in the range of frequencies spanned by the ATAS spectral features. We could show by TDDFT and interpret by the EH model that the spectral amplitude of these peaks varies with the interpulse time delay and encodes the local amplitude of the electric field strength at a given time delay. Additional features surrounding the main spectral peaks were ascribed to the Floquet dressing of the CBs by the periodic pump pulse. The comparison of ATAS signals computed by complementary simulation protocols allows us to interpret the \textit{ab-initio} TDDFT simulations in terms of crystal-momentum-dependent interband couplings, and to validate our EH model and corroborate its conclusions.

While our comparison between the EH model and TDDFT focused on a system of identical, equally spaced atoms, future work could investigate more complex systems, and the dependence of the strong-field dynamics and the ATAS spectral features on the size of the sample and type of ions. In order to compare results from different methods, it would also be interesting to include finite-size effects in the EH model, for instance in an effective way. Thereby, one could also simulate the effect of, e.g., topologically protected edge states. While our results focused on a linear chain of atoms, extensions to two- and three-dimensional systems can be envisaged.

\begin{acknowledgments}
Support from the European Union's Horizon 2020 research and innovation program through the Marie Sk{\l}odowska-Curie grant TReSFiDS (No.\ 886092) is gratefully acknowledged.
\end{acknowledgments}

%\bibliography{biblio}

\begin{thebibliography}{76}%
\makeatletter
\providecommand \@ifxundefined [1]{%
 \@ifx{#1\undefined}
}%
\providecommand \@ifnum [1]{%
 \ifnum #1\expandafter \@firstoftwo
 \else \expandafter \@secondoftwo
 \fi
}%
\providecommand \@ifx [1]{%
 \ifx #1\expandafter \@firstoftwo
 \else \expandafter \@secondoftwo
 \fi
}%
\providecommand \natexlab [1]{#1}%
\providecommand \enquote  [1]{``#1''}%
\providecommand \bibnamefont  [1]{#1}%
\providecommand \bibfnamefont [1]{#1}%
\providecommand \citenamefont [1]{#1}%
\providecommand \href@noop [0]{\@secondoftwo}%
\providecommand \href [0]{\begingroup \@sanitize@url \@href}%
\providecommand \@href[1]{\@@startlink{#1}\@@href}%
\providecommand \@@href[1]{\endgroup#1\@@endlink}%
\providecommand \@sanitize@url [0]{\catcode `\\12\catcode `\$12\catcode
  `\&12\catcode `\#12\catcode `\^12\catcode `\_12\catcode `\%12\relax}%
\providecommand \@@startlink[1]{}%
\providecommand \@@endlink[0]{}%
\providecommand \url  [0]{\begingroup\@sanitize@url \@url }%
\providecommand \@url [1]{\endgroup\@href {#1}{\urlprefix }}%
\providecommand \urlprefix  [0]{URL }%
\providecommand \Eprint [0]{\href }%
\providecommand \doibase [0]{http://dx.doi.org/}%
\providecommand \selectlanguage [0]{\@gobble}%
\providecommand \bibinfo  [0]{\@secondoftwo}%
\providecommand \bibfield  [0]{\@secondoftwo}%
\providecommand \translation [1]{[#1]}%
\providecommand \BibitemOpen [0]{}%
\providecommand \bibitemStop [0]{}%
\providecommand \bibitemNoStop [0]{.\EOS\space}%
\providecommand \EOS [0]{\spacefactor3000\relax}%
\providecommand \BibitemShut  [1]{\csname bibitem#1\endcsname}%
\let\auto@bib@innerbib\@empty
%</preamble>
\bibitem [{\citenamefont {Li}\ \emph {et~al.}(2020)\citenamefont {Li},
  \citenamefont {Lu}, \citenamefont {Chew}, \citenamefont {Han}, \citenamefont
  {Li}, \citenamefont {Wu}, \citenamefont {Wang}, \citenamefont {Ghimire},\
  and\ \citenamefont {Chang}}]{li_attosecond_2020}%
  \BibitemOpen
  \bibfield  {author} {\bibinfo {author} {\bibfnamefont {J.}~\bibnamefont
  {Li}}, \bibinfo {author} {\bibfnamefont {J.}~\bibnamefont {Lu}}, \bibinfo
  {author} {\bibfnamefont {A.}~\bibnamefont {Chew}}, \bibinfo {author}
  {\bibfnamefont {S.}~\bibnamefont {Han}}, \bibinfo {author} {\bibfnamefont
  {J.}~\bibnamefont {Li}}, \bibinfo {author} {\bibfnamefont {Y.}~\bibnamefont
  {Wu}}, \bibinfo {author} {\bibfnamefont {H.}~\bibnamefont {Wang}}, \bibinfo
  {author} {\bibfnamefont {S.}~\bibnamefont {Ghimire}}, \ and\ \bibinfo
  {author} {\bibfnamefont {Z.}~\bibnamefont {Chang}},\ }\bibfield  {title}
  {\enquote {\bibinfo {title} {Attosecond science based on high harmonic
  generation from gases and solids},}\ }\href {\doibase
  10.1038/s41467-020-16480-6} {\bibfield  {journal} {\bibinfo  {journal}
  {Nature Commun.}\ }\textbf {\bibinfo {volume} {11}},\ \bibinfo {pages} {2748}
  (\bibinfo {year} {2020})}\BibitemShut {NoStop}%
\bibitem [{\citenamefont {Geneaux}\ \emph {et~al.}(2019)\citenamefont
  {Geneaux}, \citenamefont {Marroux}, \citenamefont {Guggenmos}, \citenamefont
  {Neumark},\ and\ \citenamefont {Leone}}]{geneaux_transient_2019}%
  \BibitemOpen
  \bibfield  {author} {\bibinfo {author} {\bibfnamefont {R.}~\bibnamefont
  {Geneaux}}, \bibinfo {author} {\bibfnamefont {H.~J.~B.}\ \bibnamefont
  {Marroux}}, \bibinfo {author} {\bibfnamefont {A.}~\bibnamefont {Guggenmos}},
  \bibinfo {author} {\bibfnamefont {D.~M.}\ \bibnamefont {Neumark}}, \ and\
  \bibinfo {author} {\bibfnamefont {S.~R.}\ \bibnamefont {Leone}},\ }\bibfield
  {title} {\enquote {\bibinfo {title} {Transient absorption spectroscopy using
  high harmonic generation: a review of ultrafast {X}-ray dynamics in molecules
  and solids},}\ }\href {\doibase 10.1098/rsta.2017.0463} {\bibfield  {journal}
  {\bibinfo  {journal} {Phil. Trans. R. Soc. A}\ }\textbf {\bibinfo {volume}
  {377}},\ \bibinfo {pages} {20170463} (\bibinfo {year} {2019})}\BibitemShut
  {NoStop}%
\bibitem [{\citenamefont {Ghimire}\ \emph {et~al.}(2011)\citenamefont
  {Ghimire}, \citenamefont {DiChiara}, \citenamefont {Sistrunk}, \citenamefont
  {Agostini}, \citenamefont {DiMauro},\ and\ \citenamefont
  {Reis}}]{Ghimire2011}%
  \BibitemOpen
  \bibfield  {author} {\bibinfo {author} {\bibfnamefont {S.}~\bibnamefont
  {Ghimire}}, \bibinfo {author} {\bibfnamefont {A.~D.}\ \bibnamefont
  {DiChiara}}, \bibinfo {author} {\bibfnamefont {E.}~\bibnamefont {Sistrunk}},
  \bibinfo {author} {\bibfnamefont {P.}~\bibnamefont {Agostini}}, \bibinfo
  {author} {\bibfnamefont {L.~F.}\ \bibnamefont {DiMauro}}, \ and\ \bibinfo
  {author} {\bibfnamefont {D.~A.}\ \bibnamefont {Reis}},\ }\bibfield  {title}
  {\enquote {\bibinfo {title} {Observation of high-order harmonic generation in
  a bulk crystal},}\ }\href {\doibase 10.1038/nphys1847} {\bibfield  {journal}
  {\bibinfo  {journal} {Nature Phys.}\ }\textbf {\bibinfo {volume} {7}},\
  \bibinfo {pages} {138--141} (\bibinfo {year} {2011})}\BibitemShut {NoStop}%
\bibitem [{\citenamefont {Garg}\ \emph {et~al.}(2018)\citenamefont {Garg},
  \citenamefont {Kim},\ and\ \citenamefont
  {Goulielmakis}}]{garg_ultimate_2018}%
  \BibitemOpen
  \bibfield  {author} {\bibinfo {author} {\bibfnamefont {M.}~\bibnamefont
  {Garg}}, \bibinfo {author} {\bibfnamefont {H.~Y.}\ \bibnamefont {Kim}}, \
  and\ \bibinfo {author} {\bibfnamefont {E.}~\bibnamefont {Goulielmakis}},\
  }\bibfield  {title} {\enquote {\bibinfo {title} {Ultimate waveform
  reproducibility of extreme-ultraviolet pulses by high-harmonic generation in
  quartz},}\ }\href {\doibase 10.1038/s41566-018-0123-6} {\bibfield  {journal}
  {\bibinfo  {journal} {Nature Photon.}\ }\textbf {\bibinfo {volume} {12}},\
  \bibinfo {pages} {291--296} (\bibinfo {year} {2018})}\BibitemShut {NoStop}%
\bibitem [{\citenamefont {Luu}\ \emph {et~al.}(2015)\citenamefont {Luu},
  \citenamefont {Garg}, \citenamefont {Kruchinin}, \citenamefont {Moulet},
  \citenamefont {Hassan},\ and\ \citenamefont
  {Goulielmakis}}]{luu_extreme_2015}%
  \BibitemOpen
  \bibfield  {author} {\bibinfo {author} {\bibfnamefont {T.~T.}\ \bibnamefont
  {Luu}}, \bibinfo {author} {\bibfnamefont {M.}~\bibnamefont {Garg}}, \bibinfo
  {author} {\bibfnamefont {S.~Yu.}\ \bibnamefont {Kruchinin}}, \bibinfo
  {author} {\bibfnamefont {A.}~\bibnamefont {Moulet}}, \bibinfo {author}
  {\bibfnamefont {M.~Th.}\ \bibnamefont {Hassan}}, \ and\ \bibinfo {author}
  {\bibfnamefont {E.}~\bibnamefont {Goulielmakis}},\ }\bibfield  {title}
  {\enquote {\bibinfo {title} {Extreme ultraviolet high-harmonic spectroscopy
  of solids},}\ }\href {\doibase 10.1038/nature14456} {\bibfield  {journal}
  {\bibinfo  {journal} {Nature (London)}\ }\textbf {\bibinfo {volume} {521}},\
  \bibinfo {pages} {498--502} (\bibinfo {year} {2015})}\BibitemShut {NoStop}%
\bibitem [{\citenamefont {Vampa}\ \emph
  {et~al.}(2015{\natexlab{a}})\citenamefont {Vampa}, \citenamefont {Hammond},
  \citenamefont {Thir\'e}, \citenamefont {Schmidt}, \citenamefont {L\'egar\'e},
  \citenamefont {McDonald}, \citenamefont {Brabec}, \citenamefont {Klug},\ and\
  \citenamefont {Corkum}}]{vampa_all-optical_2015}%
  \BibitemOpen
  \bibfield  {author} {\bibinfo {author} {\bibfnamefont {G.}~\bibnamefont
  {Vampa}}, \bibinfo {author} {\bibfnamefont {T.~J.}\ \bibnamefont {Hammond}},
  \bibinfo {author} {\bibfnamefont {N.}~\bibnamefont {Thir\'e}}, \bibinfo
  {author} {\bibfnamefont {B.~E.}\ \bibnamefont {Schmidt}}, \bibinfo {author}
  {\bibfnamefont {F.}~\bibnamefont {L\'egar\'e}}, \bibinfo {author}
  {\bibfnamefont {C.~R.}\ \bibnamefont {McDonald}}, \bibinfo {author}
  {\bibfnamefont {T.}~\bibnamefont {Brabec}}, \bibinfo {author} {\bibfnamefont
  {D.~D.}\ \bibnamefont {Klug}}, \ and\ \bibinfo {author} {\bibfnamefont
  {P.~B.}\ \bibnamefont {Corkum}},\ }\bibfield  {title} {\enquote {\bibinfo
  {title} {All-{Optical} {Reconstruction} of {Crystal} {Band} {Structure}},}\
  }\href {\doibase 10.1103/PhysRevLett.115.193603} {\bibfield  {journal}
  {\bibinfo  {journal} {Phys. Rev. Lett.}\ }\textbf {\bibinfo {volume} {115}},\
  \bibinfo {pages} {193603} (\bibinfo {year} {2015}{\natexlab{a}})}\BibitemShut
  {NoStop}%
\bibitem [{\citenamefont {Schubert}\ \emph {et~al.}(2014)\citenamefont
  {Schubert}, \citenamefont {Hohenleutner}, \citenamefont {Langer},
  \citenamefont {Urbanek}, \citenamefont {Lange}, \citenamefont {Huttner},
  \citenamefont {Golde}, \citenamefont {Meier}, \citenamefont {Kira},
  \citenamefont {Koch},\ and\ \citenamefont {Huber}}]{schubert_sub-cycle_2014}%
  \BibitemOpen
  \bibfield  {author} {\bibinfo {author} {\bibfnamefont {O.}~\bibnamefont
  {Schubert}}, \bibinfo {author} {\bibfnamefont {M.}~\bibnamefont
  {Hohenleutner}}, \bibinfo {author} {\bibfnamefont {F.}~\bibnamefont
  {Langer}}, \bibinfo {author} {\bibfnamefont {B.}~\bibnamefont {Urbanek}},
  \bibinfo {author} {\bibfnamefont {C.}~\bibnamefont {Lange}}, \bibinfo
  {author} {\bibfnamefont {U.}~\bibnamefont {Huttner}}, \bibinfo {author}
  {\bibfnamefont {D.}~\bibnamefont {Golde}}, \bibinfo {author} {\bibfnamefont
  {T.}~\bibnamefont {Meier}}, \bibinfo {author} {\bibfnamefont
  {M.}~\bibnamefont {Kira}}, \bibinfo {author} {\bibfnamefont {S.~W.}\
  \bibnamefont {Koch}}, \ and\ \bibinfo {author} {\bibfnamefont
  {R.}~\bibnamefont {Huber}},\ }\bibfield  {title} {\enquote {\bibinfo {title}
  {Sub-cycle control of terahertz high-harmonic generation by dynamical {Bloch}
  oscillations},}\ }\href {\doibase 10.1038/nphoton.2013.349} {\bibfield
  {journal} {\bibinfo  {journal} {Nature Photon.}\ }\textbf {\bibinfo {volume}
  {8}},\ \bibinfo {pages} {119--123} (\bibinfo {year} {2014})}\BibitemShut
  {NoStop}%
\bibitem [{\citenamefont {Higuchi}\ \emph {et~al.}(2014)\citenamefont
  {Higuchi}, \citenamefont {Stockman},\ and\ \citenamefont
  {Hommelhoff}}]{higuchi2014strong}%
  \BibitemOpen
  \bibfield  {author} {\bibinfo {author} {\bibfnamefont {T.}~\bibnamefont
  {Higuchi}}, \bibinfo {author} {\bibfnamefont {M.~I.}\ \bibnamefont
  {Stockman}}, \ and\ \bibinfo {author} {\bibfnamefont {P.}~\bibnamefont
  {Hommelhoff}},\ }\bibfield  {title} {\enquote {\bibinfo {title} {Strong-field
  perspective on high-harmonic radiation from bulk solids},}\ }\href {\doibase
  10.1103/PhysRevLett.113.213901} {\bibfield  {journal} {\bibinfo  {journal}
  {Phys. Rev. Lett.}\ }\textbf {\bibinfo {volume} {113}},\ \bibinfo {pages}
  {213901} (\bibinfo {year} {2014})}\BibitemShut {NoStop}%
\bibitem [{\citenamefont {Wu}\ \emph {et~al.}(2015)\citenamefont {Wu},
  \citenamefont {Ghimire}, \citenamefont {Reis}, \citenamefont {Schafer},\ and\
  \citenamefont {Gaarde}}]{wu2015high}%
  \BibitemOpen
  \bibfield  {author} {\bibinfo {author} {\bibfnamefont {M.}~\bibnamefont
  {Wu}}, \bibinfo {author} {\bibfnamefont {S.}~\bibnamefont {Ghimire}},
  \bibinfo {author} {\bibfnamefont {D.~A}\ \bibnamefont {Reis}}, \bibinfo
  {author} {\bibfnamefont {K.~J.}\ \bibnamefont {Schafer}}, \ and\ \bibinfo
  {author} {\bibfnamefont {M.~B.}\ \bibnamefont {Gaarde}},\ }\bibfield  {title}
  {\enquote {\bibinfo {title} {High-harmonic generation from {Bloch} electrons
  in solids},}\ }\href {\doibase 10.1103/PhysRevA.91.043839} {\bibfield
  {journal} {\bibinfo  {journal} {Phys. Rev. A}\ }\textbf {\bibinfo {volume}
  {91}},\ \bibinfo {pages} {043839} (\bibinfo {year} {2015})}\BibitemShut
  {NoStop}%
\bibitem [{\citenamefont {McDonald}\ \emph {et~al.}(2015)\citenamefont
  {McDonald}, \citenamefont {Vampa}, \citenamefont {Corkum},\ and\
  \citenamefont {Brabec}}]{McDonald2015}%
  \BibitemOpen
  \bibfield  {author} {\bibinfo {author} {\bibfnamefont {C.~R.}\ \bibnamefont
  {McDonald}}, \bibinfo {author} {\bibfnamefont {G.}~\bibnamefont {Vampa}},
  \bibinfo {author} {\bibfnamefont {P.~B.}\ \bibnamefont {Corkum}}, \ and\
  \bibinfo {author} {\bibfnamefont {T.}~\bibnamefont {Brabec}},\ }\bibfield
  {title} {\enquote {\bibinfo {title} {Interband {Bloch} oscillation mechanism
  for high-harmonic generation in semiconductor crystals},}\ }\href {\doibase
  10.1103/PhysRevA.92.033845} {\bibfield  {journal} {\bibinfo  {journal} {Phys.
  Rev. A}\ }\textbf {\bibinfo {volume} {92}},\ \bibinfo {pages} {033845}
  (\bibinfo {year} {2015})}\BibitemShut {NoStop}%
\bibitem [{\citenamefont {Hohenleutner}\ \emph {et~al.}(2015)\citenamefont
  {Hohenleutner}, \citenamefont {Langer}, \citenamefont {Schubert},
  \citenamefont {Knorr}, \citenamefont {Huttner}, \citenamefont {Koch},
  \citenamefont {Kira},\ and\ \citenamefont {Huber}}]{hohenleutner2015real}%
  \BibitemOpen
  \bibfield  {author} {\bibinfo {author} {\bibfnamefont {M.}~\bibnamefont
  {Hohenleutner}}, \bibinfo {author} {\bibfnamefont {F.}~\bibnamefont
  {Langer}}, \bibinfo {author} {\bibfnamefont {O.}~\bibnamefont {Schubert}},
  \bibinfo {author} {\bibfnamefont {M.}~\bibnamefont {Knorr}}, \bibinfo
  {author} {\bibfnamefont {U.}~\bibnamefont {Huttner}}, \bibinfo {author}
  {\bibfnamefont {S.~W.}\ \bibnamefont {Koch}}, \bibinfo {author}
  {\bibfnamefont {M.}~\bibnamefont {Kira}}, \ and\ \bibinfo {author}
  {\bibfnamefont {R.}~\bibnamefont {Huber}},\ }\bibfield  {title} {\enquote
  {\bibinfo {title} {Real-time observation of interfering crystal electrons in
  high-harmonic generation},}\ }\href {\doibase 10.1038/nature14652} {\bibfield
   {journal} {\bibinfo  {journal} {Nature (London)}\ }\textbf {\bibinfo
  {volume} {523}},\ \bibinfo {pages} {572--575} (\bibinfo {year}
  {2015})}\BibitemShut {NoStop}%
\bibitem [{\citenamefont {Vampa}\ \emph
  {et~al.}(2015{\natexlab{b}})\citenamefont {Vampa}, \citenamefont {Hammond},
  \citenamefont {Thir{\'e}}, \citenamefont {Schmidt}, \citenamefont
  {L{\'e}gar{\'e}}, \citenamefont {McDonald}, \citenamefont {Brabec},
  \citenamefont {Klug},\ and\ \citenamefont {Corkum}}]{vampa2015all}%
  \BibitemOpen
  \bibfield  {author} {\bibinfo {author} {\bibfnamefont {G.}~\bibnamefont
  {Vampa}}, \bibinfo {author} {\bibfnamefont {T.~J.}\ \bibnamefont {Hammond}},
  \bibinfo {author} {\bibfnamefont {N.}~\bibnamefont {Thir{\'e}}}, \bibinfo
  {author} {\bibfnamefont {B.~E.}\ \bibnamefont {Schmidt}}, \bibinfo {author}
  {\bibfnamefont {F.}~\bibnamefont {L{\'e}gar{\'e}}}, \bibinfo {author}
  {\bibfnamefont {C.~R.}\ \bibnamefont {McDonald}}, \bibinfo {author}
  {\bibfnamefont {T.}~\bibnamefont {Brabec}}, \bibinfo {author} {\bibfnamefont
  {D.~D.}\ \bibnamefont {Klug}}, \ and\ \bibinfo {author} {\bibfnamefont
  {P.~B.}\ \bibnamefont {Corkum}},\ }\bibfield  {title} {\enquote {\bibinfo
  {title} {All-optical reconstruction of crystal band structure},}\ }\href
  {\doibase 10.1103/PhysRevLett.115.193603} {\bibfield  {journal} {\bibinfo
  {journal} {Phys. Rev. Lett.}\ }\textbf {\bibinfo {volume} {115}},\ \bibinfo
  {pages} {193603} (\bibinfo {year} {2015}{\natexlab{b}})}\BibitemShut
  {NoStop}%
\bibitem [{\citenamefont {Ndabashimiye}\ \emph {et~al.}(2016)\citenamefont
  {Ndabashimiye}, \citenamefont {Ghimire}, \citenamefont {Wu}, \citenamefont
  {Browne}, \citenamefont {Schafer}, \citenamefont {Gaarde},\ and\
  \citenamefont {Reis}}]{ndabashimiye2016solid}%
  \BibitemOpen
  \bibfield  {author} {\bibinfo {author} {\bibfnamefont {G.}~\bibnamefont
  {Ndabashimiye}}, \bibinfo {author} {\bibfnamefont {S.}~\bibnamefont
  {Ghimire}}, \bibinfo {author} {\bibfnamefont {M.}~\bibnamefont {Wu}},
  \bibinfo {author} {\bibfnamefont {D.~A.}\ \bibnamefont {Browne}}, \bibinfo
  {author} {\bibfnamefont {K.~J.}\ \bibnamefont {Schafer}}, \bibinfo {author}
  {\bibfnamefont {M.~B.}\ \bibnamefont {Gaarde}}, \ and\ \bibinfo {author}
  {\bibfnamefont {D.~A.}\ \bibnamefont {Reis}},\ }\bibfield  {title} {\enquote
  {\bibinfo {title} {Solid-state harmonics beyond the atomic limit},}\ }\href
  {\doibase 10.1038/nature17660} {\bibfield  {journal} {\bibinfo  {journal}
  {Nature (London)}\ }\textbf {\bibinfo {volume} {534}},\ \bibinfo {pages}
  {520--523} (\bibinfo {year} {2016})}\BibitemShut {NoStop}%
\bibitem [{\citenamefont {Garg}\ \emph {et~al.}(2016)\citenamefont {Garg},
  \citenamefont {Zhan}, \citenamefont {Luu}, \citenamefont {Lakhotia},
  \citenamefont {Klostermann}, \citenamefont {Guggenmos},\ and\ \citenamefont
  {Goulielmakis}}]{Garg2016}%
  \BibitemOpen
  \bibfield  {author} {\bibinfo {author} {\bibfnamefont {M.}~\bibnamefont
  {Garg}}, \bibinfo {author} {\bibfnamefont {M.}~\bibnamefont {Zhan}}, \bibinfo
  {author} {\bibfnamefont {T.~T.}\ \bibnamefont {Luu}}, \bibinfo {author}
  {\bibfnamefont {H.}~\bibnamefont {Lakhotia}}, \bibinfo {author}
  {\bibfnamefont {T.}~\bibnamefont {Klostermann}}, \bibinfo {author}
  {\bibfnamefont {A.}~\bibnamefont {Guggenmos}}, \ and\ \bibinfo {author}
  {\bibfnamefont {E.}~\bibnamefont {Goulielmakis}},\ }\bibfield  {title}
  {\enquote {\bibinfo {title} {Multi-petahertz electronic metrology},}\ }\href
  {\doibase 10.1038/nature19821} {\bibfield  {journal} {\bibinfo  {journal}
  {Nature (London)}\ }\textbf {\bibinfo {volume} {538}},\ \bibinfo {pages}
  {359--363} (\bibinfo {year} {2016})}\BibitemShut {NoStop}%
\bibitem [{\citenamefont {Schiffrin}\ \emph {et~al.}(2013)\citenamefont
  {Schiffrin}, \citenamefont {Paasch-Colberg}, \citenamefont {Karpowicz},
  \citenamefont {Apalkov}, \citenamefont {Gerster}, \citenamefont
  {M\"uhlbrandt}, \citenamefont {Korbman}, \citenamefont {Reichert},
  \citenamefont {Schultze}, \citenamefont {Holzner}, \citenamefont {Barth},
  \citenamefont {Kienberger}, \citenamefont {Ernstorfer}, \citenamefont
  {Yakovlev}, \citenamefont {Stockman},\ and\ \citenamefont
  {Krausz}}]{schiffrin_optical-field-induced_2013}%
  \BibitemOpen
  \bibfield  {author} {\bibinfo {author} {\bibfnamefont {A.}~\bibnamefont
  {Schiffrin}}, \bibinfo {author} {\bibfnamefont {T.}~\bibnamefont
  {Paasch-Colberg}}, \bibinfo {author} {\bibfnamefont {N.}~\bibnamefont
  {Karpowicz}}, \bibinfo {author} {\bibfnamefont {V.}~\bibnamefont {Apalkov}},
  \bibinfo {author} {\bibfnamefont {D.}~\bibnamefont {Gerster}}, \bibinfo
  {author} {\bibfnamefont {S.}~\bibnamefont {M\"uhlbrandt}}, \bibinfo {author}
  {\bibfnamefont {M.}~\bibnamefont {Korbman}}, \bibinfo {author} {\bibfnamefont
  {J.}~\bibnamefont {Reichert}}, \bibinfo {author} {\bibfnamefont
  {M.}~\bibnamefont {Schultze}}, \bibinfo {author} {\bibfnamefont
  {S.}~\bibnamefont {Holzner}}, \bibinfo {author} {\bibfnamefont {J.~V.}\
  \bibnamefont {Barth}}, \bibinfo {author} {\bibfnamefont {R.}~\bibnamefont
  {Kienberger}}, \bibinfo {author} {\bibfnamefont {R.}~\bibnamefont
  {Ernstorfer}}, \bibinfo {author} {\bibfnamefont {V.~S.}\ \bibnamefont
  {Yakovlev}}, \bibinfo {author} {\bibfnamefont {M.~I.}\ \bibnamefont
  {Stockman}}, \ and\ \bibinfo {author} {\bibfnamefont {F.}~\bibnamefont
  {Krausz}},\ }\bibfield  {title} {\enquote {\bibinfo {title}
  {Optical-field-induced current in dielectrics},}\ }\href {\doibase
  10.1038/nature11567} {\bibfield  {journal} {\bibinfo  {journal} {Nature
  (London)}\ }\textbf {\bibinfo {volume} {493}},\ \bibinfo {pages} {70--74}
  (\bibinfo {year} {2013})}\BibitemShut {NoStop}%
\bibitem [{\citenamefont {Schultze}\ \emph {et~al.}(2013)\citenamefont
  {Schultze}, \citenamefont {Bothschafter}, \citenamefont {Sommer},
  \citenamefont {Holzner}, \citenamefont {Schweinberger}, \citenamefont
  {Fiess}, \citenamefont {Hofstetter}, \citenamefont {Kienberger},
  \citenamefont {Apalkov}, \citenamefont {Yakovlev}, \citenamefont {Stockman},\
  and\ \citenamefont {Krausz}}]{Schultze2013}%
  \BibitemOpen
  \bibfield  {author} {\bibinfo {author} {\bibfnamefont {M.}~\bibnamefont
  {Schultze}}, \bibinfo {author} {\bibfnamefont {E.~M.}\ \bibnamefont
  {Bothschafter}}, \bibinfo {author} {\bibfnamefont {A.}~\bibnamefont
  {Sommer}}, \bibinfo {author} {\bibfnamefont {S.}~\bibnamefont {Holzner}},
  \bibinfo {author} {\bibfnamefont {W.}~\bibnamefont {Schweinberger}}, \bibinfo
  {author} {\bibfnamefont {M.}~\bibnamefont {Fiess}}, \bibinfo {author}
  {\bibfnamefont {M.}~\bibnamefont {Hofstetter}}, \bibinfo {author}
  {\bibfnamefont {R.}~\bibnamefont {Kienberger}}, \bibinfo {author}
  {\bibfnamefont {V.}~\bibnamefont {Apalkov}}, \bibinfo {author} {\bibfnamefont
  {V.~S.}\ \bibnamefont {Yakovlev}}, \bibinfo {author} {\bibfnamefont {M.~I.}\
  \bibnamefont {Stockman}}, \ and\ \bibinfo {author} {\bibfnamefont
  {F.}~\bibnamefont {Krausz}},\ }\bibfield  {title} {\enquote {\bibinfo {title}
  {Controlling dielectrics with the electric field of light},}\ }\href
  {\doibase 10.1038/nature11720} {\bibfield  {journal} {\bibinfo  {journal}
  {Nature (London)}\ }\textbf {\bibinfo {volume} {493}},\ \bibinfo {pages}
  {75--78} (\bibinfo {year} {2013})}\BibitemShut {NoStop}%
\bibitem [{\citenamefont {Mashiko}\ \emph {et~al.}(2016)\citenamefont
  {Mashiko}, \citenamefont {Oguri}, \citenamefont {Yamaguchi}, \citenamefont
  {Suda},\ and\ \citenamefont {Gotoh}}]{mashiko_petahertz_2016}%
  \BibitemOpen
  \bibfield  {author} {\bibinfo {author} {\bibfnamefont {H.}~\bibnamefont
  {Mashiko}}, \bibinfo {author} {\bibfnamefont {K.}~\bibnamefont {Oguri}},
  \bibinfo {author} {\bibfnamefont {T.}~\bibnamefont {Yamaguchi}}, \bibinfo
  {author} {\bibfnamefont {A.}~\bibnamefont {Suda}}, \ and\ \bibinfo {author}
  {\bibfnamefont {H.}~\bibnamefont {Gotoh}},\ }\bibfield  {title} {\enquote
  {\bibinfo {title} {Petahertz optical drive with wide-bandgap
  semiconductor},}\ }\href {\doibase 10.1038/nphys3711} {\bibfield  {journal}
  {\bibinfo  {journal} {Nature Phys.}\ }\textbf {\bibinfo {volume} {12}},\
  \bibinfo {pages} {741--745} (\bibinfo {year} {2016})}\BibitemShut {NoStop}%
\bibitem [{\citenamefont {Sommer}\ \emph {et~al.}(2016)\citenamefont {Sommer},
  \citenamefont {Bothschafter}, \citenamefont {Sato}, \citenamefont {Jakubeit},
  \citenamefont {Latka}, \citenamefont {Razskazovskaya}, \citenamefont
  {Fattahi}, \citenamefont {Jobst}, \citenamefont {Schweinberger},
  \citenamefont {Shirvanyan}, \citenamefont {Yakovlev}, \citenamefont
  {Kienberger}, \citenamefont {Yabana}, \citenamefont {Karpowicz},
  \citenamefont {Schultze},\ and\ \citenamefont
  {Krausz}}]{sommer_attosecond_2016}%
  \BibitemOpen
  \bibfield  {author} {\bibinfo {author} {\bibfnamefont {A.}~\bibnamefont
  {Sommer}}, \bibinfo {author} {\bibfnamefont {E.~M.}\ \bibnamefont
  {Bothschafter}}, \bibinfo {author} {\bibfnamefont {S.~A.}\ \bibnamefont
  {Sato}}, \bibinfo {author} {\bibfnamefont {C.}~\bibnamefont {Jakubeit}},
  \bibinfo {author} {\bibfnamefont {T.}~\bibnamefont {Latka}}, \bibinfo
  {author} {\bibfnamefont {O.}~\bibnamefont {Razskazovskaya}}, \bibinfo
  {author} {\bibfnamefont {H.}~\bibnamefont {Fattahi}}, \bibinfo {author}
  {\bibfnamefont {M.}~\bibnamefont {Jobst}}, \bibinfo {author} {\bibfnamefont
  {W.}~\bibnamefont {Schweinberger}}, \bibinfo {author} {\bibfnamefont
  {V.}~\bibnamefont {Shirvanyan}}, \bibinfo {author} {\bibfnamefont {V.~S.}\
  \bibnamefont {Yakovlev}}, \bibinfo {author} {\bibfnamefont {R.}~\bibnamefont
  {Kienberger}}, \bibinfo {author} {\bibfnamefont {K.}~\bibnamefont {Yabana}},
  \bibinfo {author} {\bibfnamefont {N.}~\bibnamefont {Karpowicz}}, \bibinfo
  {author} {\bibfnamefont {M.}~\bibnamefont {Schultze}}, \ and\ \bibinfo
  {author} {\bibfnamefont {F.}~\bibnamefont {Krausz}},\ }\bibfield  {title}
  {\enquote {\bibinfo {title} {Attosecond nonlinear polarization and
  light--matter energy transfer in solids},}\ }\href {\doibase
  10.1038/nature17650} {\bibfield  {journal} {\bibinfo  {journal} {Nature
  (London)}\ }\textbf {\bibinfo {volume} {534}},\ \bibinfo {pages} {86--90}
  (\bibinfo {year} {2016})}\BibitemShut {NoStop}%
\bibitem [{\citenamefont {Higuchi}\ \emph {et~al.}(2017)\citenamefont
  {Higuchi}, \citenamefont {Heide}, \citenamefont {Ullmann}, \citenamefont
  {Weber},\ and\ \citenamefont {Hommelhoff}}]{higuchi_light-field-driven_2017}%
  \BibitemOpen
  \bibfield  {author} {\bibinfo {author} {\bibfnamefont {T.}~\bibnamefont
  {Higuchi}}, \bibinfo {author} {\bibfnamefont {C.}~\bibnamefont {Heide}},
  \bibinfo {author} {\bibfnamefont {K.}~\bibnamefont {Ullmann}}, \bibinfo
  {author} {\bibfnamefont {H.~B.}\ \bibnamefont {Weber}}, \ and\ \bibinfo
  {author} {\bibfnamefont {P.}~\bibnamefont {Hommelhoff}},\ }\bibfield  {title}
  {\enquote {\bibinfo {title} {Light-field-driven currents in graphene},}\
  }\href {\doibase 10.1038/nature23900} {\bibfield  {journal} {\bibinfo
  {journal} {Nature (London)}\ }\textbf {\bibinfo {volume} {550}},\ \bibinfo
  {pages} {224--228} (\bibinfo {year} {2017})}\BibitemShut {NoStop}%
\bibitem [{\citenamefont {Bionta}\ \emph {et~al.}(2021)\citenamefont {Bionta},
  \citenamefont {Haddad}, \citenamefont {Leblanc}, \citenamefont {Gruson},
  \citenamefont {Lassonde}, \citenamefont {Ibrahim}, \citenamefont {Chaillou},
  \citenamefont {\'Emond}, \citenamefont {Otto}, \citenamefont
  {Jim\'enez-Gal\'an}, \citenamefont {Silva}, \citenamefont {Ivanov},
  \citenamefont {Siwick}, \citenamefont {Chaker},\ and\ \citenamefont
  {L\'egar\'e}}]{PhysRevResearch.3.023250}%
  \BibitemOpen
  \bibfield  {author} {\bibinfo {author} {\bibfnamefont {M.~R.}\ \bibnamefont
  {Bionta}}, \bibinfo {author} {\bibfnamefont {E.}~\bibnamefont {Haddad}},
  \bibinfo {author} {\bibfnamefont {A.}~\bibnamefont {Leblanc}}, \bibinfo
  {author} {\bibfnamefont {V.}~\bibnamefont {Gruson}}, \bibinfo {author}
  {\bibfnamefont {P.}~\bibnamefont {Lassonde}}, \bibinfo {author}
  {\bibfnamefont {H.}~\bibnamefont {Ibrahim}}, \bibinfo {author} {\bibfnamefont
  {J.}~\bibnamefont {Chaillou}}, \bibinfo {author} {\bibfnamefont
  {N.}~\bibnamefont {\'Emond}}, \bibinfo {author} {\bibfnamefont {M.n~R.}\
  \bibnamefont {Otto}}, \bibinfo {author} {\bibfnamefont {\'A.}\ \bibnamefont
  {Jim\'enez-Gal\'an}}, \bibinfo {author} {\bibfnamefont {R.~E.~F.}\
  \bibnamefont {Silva}}, \bibinfo {author} {\bibfnamefont {M.}~\bibnamefont
  {Ivanov}}, \bibinfo {author} {\bibfnamefont {B.~J.}\ \bibnamefont {Siwick}},
  \bibinfo {author} {\bibfnamefont {M.}~\bibnamefont {Chaker}}, \ and\ \bibinfo
  {author} {\bibfnamefont {F.}~\bibnamefont {L\'egar\'e}},\ }\bibfield  {title}
  {\enquote {\bibinfo {title} {Tracking ultrafast solid-state dynamics using
  high harmonic spectroscopy},}\ }\href {\doibase
  10.1103/PhysRevResearch.3.023250} {\bibfield  {journal} {\bibinfo  {journal}
  {Phys. Rev. Res.}\ }\textbf {\bibinfo {volume} {3}},\ \bibinfo {pages}
  {023250} (\bibinfo {year} {2021})}\BibitemShut {NoStop}%
\bibitem [{\citenamefont {Ossiander}\ \emph {et~al.}(2022)\citenamefont
  {Ossiander}, \citenamefont {Golyari}, \citenamefont {Scharl}, \citenamefont
  {Lehnert}, \citenamefont {Siegrist}, \citenamefont {B\"urger}, \citenamefont
  {Zimin}, \citenamefont {Gessner}, \citenamefont {Weidman}, \citenamefont
  {Floss}, \citenamefont {Smejkal}, \citenamefont {Donsa}, \citenamefont
  {Lemell}, \citenamefont {Libisch}, \citenamefont {Karpowicz}, \citenamefont
  {Burgd\"orfer}, \citenamefont {Krausz},\ and\ \citenamefont
  {Schultze}}]{ossiander_speed_2022}%
  \BibitemOpen
  \bibfield  {author} {\bibinfo {author} {\bibfnamefont {M.}~\bibnamefont
  {Ossiander}}, \bibinfo {author} {\bibfnamefont {K.}~\bibnamefont {Golyari}},
  \bibinfo {author} {\bibfnamefont {K.}~\bibnamefont {Scharl}}, \bibinfo
  {author} {\bibfnamefont {L.}~\bibnamefont {Lehnert}}, \bibinfo {author}
  {\bibfnamefont {F.}~\bibnamefont {Siegrist}}, \bibinfo {author}
  {\bibfnamefont {J.~P.}\ \bibnamefont {B\"urger}}, \bibinfo {author}
  {\bibfnamefont {D.}~\bibnamefont {Zimin}}, \bibinfo {author} {\bibfnamefont
  {J.~A.}\ \bibnamefont {Gessner}}, \bibinfo {author} {\bibfnamefont
  {M.}~\bibnamefont {Weidman}}, \bibinfo {author} {\bibfnamefont
  {I.}~\bibnamefont {Floss}}, \bibinfo {author} {\bibfnamefont
  {V.}~\bibnamefont {Smejkal}}, \bibinfo {author} {\bibfnamefont
  {S.}~\bibnamefont {Donsa}}, \bibinfo {author} {\bibfnamefont
  {C.}~\bibnamefont {Lemell}}, \bibinfo {author} {\bibfnamefont
  {F.}~\bibnamefont {Libisch}}, \bibinfo {author} {\bibfnamefont
  {N.}~\bibnamefont {Karpowicz}}, \bibinfo {author} {\bibfnamefont
  {J.}~\bibnamefont {Burgd\"orfer}}, \bibinfo {author} {\bibfnamefont
  {F.}~\bibnamefont {Krausz}}, \ and\ \bibinfo {author} {\bibfnamefont
  {M.}~\bibnamefont {Schultze}},\ }\bibfield  {title} {\enquote {\bibinfo
  {title} {The speed limit of optoelectronics},}\ }\href {\doibase
  10.1038/s41467-022-29252-1} {\bibfield  {journal} {\bibinfo  {journal}
  {Nature Commun.}\ }\textbf {\bibinfo {volume} {13}},\ \bibinfo {pages} {1620}
  (\bibinfo {year} {2022})}\BibitemShut {NoStop}%
\bibitem [{\citenamefont {Wang}\ \emph {et~al.}(2013)\citenamefont {Wang},
  \citenamefont {Steinberg}, \citenamefont {Jarillo-Herrero},\ and\
  \citenamefont {Gedik}}]{wang_observation_2013}%
  \BibitemOpen
  \bibfield  {author} {\bibinfo {author} {\bibfnamefont {Y.~H.}\ \bibnamefont
  {Wang}}, \bibinfo {author} {\bibfnamefont {H.}~\bibnamefont {Steinberg}},
  \bibinfo {author} {\bibfnamefont {P.}~\bibnamefont {Jarillo-Herrero}}, \ and\
  \bibinfo {author} {\bibfnamefont {N.}~\bibnamefont {Gedik}},\ }\bibfield
  {title} {\enquote {\bibinfo {title} {Observation of {Floquet}-{Bloch}
  {States} on the {Surface} of a {Topological} {Insulator}},}\ }\href {\doibase
  10.1126/science.1239834} {\bibfield  {journal} {\bibinfo  {journal}
  {Science}\ }\textbf {\bibinfo {volume} {342}},\ \bibinfo {pages} {453--457}
  (\bibinfo {year} {2013})}\BibitemShut {NoStop}%
\bibitem [{\citenamefont {Di~Palo}\ \emph {et~al.}(2024)\citenamefont
  {Di~Palo}, \citenamefont {Inzani}, \citenamefont {Dolso}, \citenamefont
  {Talarico}, \citenamefont {Bonetti},\ and\ \citenamefont
  {Lucchini}}]{DiPalo2024}%
  \BibitemOpen
  \bibfield  {author} {\bibinfo {author} {\bibfnamefont {N.}~\bibnamefont
  {Di~Palo}}, \bibinfo {author} {\bibfnamefont {G.}~\bibnamefont {Inzani}},
  \bibinfo {author} {\bibfnamefont {G.~L.}\ \bibnamefont {Dolso}}, \bibinfo
  {author} {\bibfnamefont {M.}~\bibnamefont {Talarico}}, \bibinfo {author}
  {\bibfnamefont {S.}~\bibnamefont {Bonetti}}, \ and\ \bibinfo {author}
  {\bibfnamefont {M.}~\bibnamefont {Lucchini}},\ }\bibfield  {title} {\enquote
  {\bibinfo {title} {Attosecond absorption and reflection spectroscopy of
  solids},}\ }\href {\doibase 10.1063/5.0176656} {\bibfield  {journal}
  {\bibinfo  {journal} {APL Photonics}\ }\textbf {\bibinfo {volume} {9}},\
  \bibinfo {pages} {020901} (\bibinfo {year} {2024})}\BibitemShut {NoStop}%
\bibitem [{\citenamefont {Gaynor}\ \emph {et~al.}(2021)\citenamefont {Gaynor},
  \citenamefont {Fidler}, \citenamefont {Lin}, \citenamefont {Chang},
  \citenamefont {Zuerch}, \citenamefont {Neumark},\ and\ \citenamefont
  {Leone}}]{PhysRevB.103.245140}%
  \BibitemOpen
  \bibfield  {author} {\bibinfo {author} {\bibfnamefont {J.~D.}\ \bibnamefont
  {Gaynor}}, \bibinfo {author} {\bibfnamefont {A.~P.}\ \bibnamefont {Fidler}},
  \bibinfo {author} {\bibfnamefont {Y.-C.}\ \bibnamefont {Lin}}, \bibinfo
  {author} {\bibfnamefont {H.-T.}\ \bibnamefont {Chang}}, \bibinfo {author}
  {\bibfnamefont {M.}~\bibnamefont {Zuerch}}, \bibinfo {author} {\bibfnamefont
  {D.~M.}\ \bibnamefont {Neumark}}, \ and\ \bibinfo {author} {\bibfnamefont
  {S.~R.}\ \bibnamefont {Leone}},\ }\bibfield  {title} {\enquote {\bibinfo
  {title} {Solid state core-exciton dynamics in {NaCl} observed by tabletop
  attosecond four-wave mixing spectroscopy},}\ }\href {\doibase
  10.1103/PhysRevB.103.245140} {\bibfield  {journal} {\bibinfo  {journal}
  {Phys. Rev. B}\ }\textbf {\bibinfo {volume} {103}},\ \bibinfo {pages}
  {245140} (\bibinfo {year} {2021})}\BibitemShut {NoStop}%
\bibitem [{\citenamefont {Schultze}\ \emph {et~al.}(2014)\citenamefont
  {Schultze}, \citenamefont {Ramasesha}, \citenamefont {Pemmaraju},
  \citenamefont {Sato}, \citenamefont {Whitmore}, \citenamefont {Gandman},
  \citenamefont {Prell}, \citenamefont {Borja}, \citenamefont {Prendergast},
  \citenamefont {Yabana}, \citenamefont {Neumark},\ and\ \citenamefont
  {Leone}}]{Schultze2014}%
  \BibitemOpen
  \bibfield  {author} {\bibinfo {author} {\bibfnamefont {M.}~\bibnamefont
  {Schultze}}, \bibinfo {author} {\bibfnamefont {K.}~\bibnamefont {Ramasesha}},
  \bibinfo {author} {\bibfnamefont {C.~D.}\ \bibnamefont {Pemmaraju}}, \bibinfo
  {author} {\bibfnamefont {S.~A.}\ \bibnamefont {Sato}}, \bibinfo {author}
  {\bibfnamefont {D.}~\bibnamefont {Whitmore}}, \bibinfo {author}
  {\bibfnamefont {A.}~\bibnamefont {Gandman}}, \bibinfo {author} {\bibfnamefont
  {J.~S.}\ \bibnamefont {Prell}}, \bibinfo {author} {\bibfnamefont {L.~J.}\
  \bibnamefont {Borja}}, \bibinfo {author} {\bibfnamefont {D.}~\bibnamefont
  {Prendergast}}, \bibinfo {author} {\bibfnamefont {K.}~\bibnamefont {Yabana}},
  \bibinfo {author} {\bibfnamefont {D.~M.}\ \bibnamefont {Neumark}}, \ and\
  \bibinfo {author} {\bibfnamefont {S.~R.}\ \bibnamefont {Leone}},\ }\bibfield
  {title} {\enquote {\bibinfo {title} {Attosecond band-gap dynamics in
  silicon},}\ }\href {\doibase 10.1126/science.1260311} {\bibfield  {journal}
  {\bibinfo  {journal} {Science}\ }\textbf {\bibinfo {volume} {346}},\ \bibinfo
  {pages} {1348--1352} (\bibinfo {year} {2014})}\BibitemShut {NoStop}%
\bibitem [{\citenamefont {Lucchini}\ \emph {et~al.}(2016)\citenamefont
  {Lucchini}, \citenamefont {Sato}, \citenamefont {Ludwig}, \citenamefont
  {Herrmann}, \citenamefont {Volkov}, \citenamefont {Kasmi}, \citenamefont
  {Shinohara}, \citenamefont {Yabana}, \citenamefont {Gallmann},\ and\
  \citenamefont {Keller}}]{Lucchini2016}%
  \BibitemOpen
  \bibfield  {author} {\bibinfo {author} {\bibfnamefont {M.}~\bibnamefont
  {Lucchini}}, \bibinfo {author} {\bibfnamefont {S.~A.}\ \bibnamefont {Sato}},
  \bibinfo {author} {\bibfnamefont {A.}~\bibnamefont {Ludwig}}, \bibinfo
  {author} {\bibfnamefont {J.}~\bibnamefont {Herrmann}}, \bibinfo {author}
  {\bibfnamefont {M.}~\bibnamefont {Volkov}}, \bibinfo {author} {\bibfnamefont
  {L.}~\bibnamefont {Kasmi}}, \bibinfo {author} {\bibfnamefont
  {Y.}~\bibnamefont {Shinohara}}, \bibinfo {author} {\bibfnamefont
  {K.}~\bibnamefont {Yabana}}, \bibinfo {author} {\bibfnamefont
  {L.}~\bibnamefont {Gallmann}}, \ and\ \bibinfo {author} {\bibfnamefont
  {U.}~\bibnamefont {Keller}},\ }\bibfield  {title} {\enquote {\bibinfo {title}
  {Attosecond dynamical {Franz}-{Keldysh} effect in polycrystalline diamond},}\
  }\href {\doibase 10.1126/science.aag1268} {\bibfield  {journal} {\bibinfo
  {journal} {Science}\ }\textbf {\bibinfo {volume} {353}},\ \bibinfo {pages}
  {916--919} (\bibinfo {year} {2016})}\BibitemShut {NoStop}%
\bibitem [{\citenamefont {Jager}\ \emph {et~al.}(2017)\citenamefont {Jager},
  \citenamefont {Ott}, \citenamefont {Kraus}, \citenamefont {Kaplan},
  \citenamefont {Pouse}, \citenamefont {Marvel}, \citenamefont {Haglund},
  \citenamefont {Neumark},\ and\ \citenamefont {Leone}}]{Jager2017}%
  \BibitemOpen
  \bibfield  {author} {\bibinfo {author} {\bibfnamefont {M.~F.}\ \bibnamefont
  {Jager}}, \bibinfo {author} {\bibfnamefont {C.}~\bibnamefont {Ott}}, \bibinfo
  {author} {\bibfnamefont {P.~M.}\ \bibnamefont {Kraus}}, \bibinfo {author}
  {\bibfnamefont {C.~J.}\ \bibnamefont {Kaplan}}, \bibinfo {author}
  {\bibfnamefont {W.}~\bibnamefont {Pouse}}, \bibinfo {author} {\bibfnamefont
  {R.~E.}\ \bibnamefont {Marvel}}, \bibinfo {author} {\bibfnamefont {R.~F.}\
  \bibnamefont {Haglund}}, \bibinfo {author} {\bibfnamefont {D.~M.}\
  \bibnamefont {Neumark}}, \ and\ \bibinfo {author} {\bibfnamefont {S.~R.}\
  \bibnamefont {Leone}},\ }\bibfield  {title} {\enquote {\bibinfo {title}
  {Tracking the insulator-to-metal phase transition in {VO$_2$} with
  few-femtosecond extreme {UV} transient absorption spectroscopy},}\ }\href
  {\doibase 10.1073/pnas.1707602114} {\bibfield  {journal} {\bibinfo  {journal}
  {Proc. Natl. Acad. Sci. U.S.A.}\ }\textbf {\bibinfo {volume} {114}},\
  \bibinfo {pages} {9558--9563} (\bibinfo {year} {2017})}\BibitemShut {NoStop}%
\bibitem [{\citenamefont {Z\"urch}\ \emph {et~al.}(2017)\citenamefont
  {Z\"urch}, \citenamefont {Chang}, \citenamefont {Borja}, \citenamefont
  {Kraus}, \citenamefont {Cushing}, \citenamefont {Gandman}, \citenamefont
  {Kaplan}, \citenamefont {Oh}, \citenamefont {Prell}, \citenamefont
  {Prendergast}, \citenamefont {Pemmaraju}, \citenamefont {Neumark},\ and\
  \citenamefont {Leone}}]{zurch_direct_2017}%
  \BibitemOpen
  \bibfield  {author} {\bibinfo {author} {\bibfnamefont {M.}~\bibnamefont
  {Z\"urch}}, \bibinfo {author} {\bibfnamefont {H.-T.}\ \bibnamefont {Chang}},
  \bibinfo {author} {\bibfnamefont {L.~J.}\ \bibnamefont {Borja}}, \bibinfo
  {author} {\bibfnamefont {P.~M.}\ \bibnamefont {Kraus}}, \bibinfo {author}
  {\bibfnamefont {S.~K.}\ \bibnamefont {Cushing}}, \bibinfo {author}
  {\bibfnamefont {A.}~\bibnamefont {Gandman}}, \bibinfo {author} {\bibfnamefont
  {C.~J.}\ \bibnamefont {Kaplan}}, \bibinfo {author} {\bibfnamefont {M.~H.}\
  \bibnamefont {Oh}}, \bibinfo {author} {\bibfnamefont {J.~S.}\ \bibnamefont
  {Prell}}, \bibinfo {author} {\bibfnamefont {D.}~\bibnamefont {Prendergast}},
  \bibinfo {author} {\bibfnamefont {C.~D.}\ \bibnamefont {Pemmaraju}}, \bibinfo
  {author} {\bibfnamefont {D.~M.}\ \bibnamefont {Neumark}}, \ and\ \bibinfo
  {author} {\bibfnamefont {S.~R.}\ \bibnamefont {Leone}},\ }\bibfield  {title}
  {\enquote {\bibinfo {title} {Direct and simultaneous observation of ultrafast
  electron and hole dynamics in germanium},}\ }\href {\doibase
  10.1038/ncomms15734} {\bibfield  {journal} {\bibinfo  {journal} {Nature
  Commun.}\ }\textbf {\bibinfo {volume} {8}},\ \bibinfo {pages} {15734}
  (\bibinfo {year} {2017})}\BibitemShut {NoStop}%
\bibitem [{\citenamefont {Moulet}\ \emph {et~al.}(2017)\citenamefont {Moulet},
  \citenamefont {Bertrand}, \citenamefont {Klostermann}, \citenamefont
  {Guggenmos}, \citenamefont {Karpowicz},\ and\ \citenamefont
  {Goulielmakis}}]{moulet_soft_2017}%
  \BibitemOpen
  \bibfield  {author} {\bibinfo {author} {\bibfnamefont {A.}~\bibnamefont
  {Moulet}}, \bibinfo {author} {\bibfnamefont {J.~B.}\ \bibnamefont
  {Bertrand}}, \bibinfo {author} {\bibfnamefont {T.}~\bibnamefont
  {Klostermann}}, \bibinfo {author} {\bibfnamefont {A.}~\bibnamefont
  {Guggenmos}}, \bibinfo {author} {\bibfnamefont {N.}~\bibnamefont
  {Karpowicz}}, \ and\ \bibinfo {author} {\bibfnamefont {E.}~\bibnamefont
  {Goulielmakis}},\ }\bibfield  {title} {\enquote {\bibinfo {title} {Soft x-ray
  excitonics},}\ }\href {\doibase 10.1126/science.aan4737} {\bibfield
  {journal} {\bibinfo  {journal} {Science}\ }\textbf {\bibinfo {volume}
  {357}},\ \bibinfo {pages} {1134--1138} (\bibinfo {year} {2017})}\BibitemShut
  {NoStop}%
\bibitem [{\citenamefont {Schlaepfer}\ \emph {et~al.}(2018)\citenamefont
  {Schlaepfer}, \citenamefont {Lucchini}, \citenamefont {Sato}, \citenamefont
  {Volkov}, \citenamefont {Kasmi}, \citenamefont {Hartmann}, \citenamefont
  {Rubio}, \citenamefont {Gallmann},\ and\ \citenamefont
  {Keller}}]{schlaepfer_attosecond_2018}%
  \BibitemOpen
  \bibfield  {author} {\bibinfo {author} {\bibfnamefont {F.}~\bibnamefont
  {Schlaepfer}}, \bibinfo {author} {\bibfnamefont {M.}~\bibnamefont
  {Lucchini}}, \bibinfo {author} {\bibfnamefont {S.~A.}\ \bibnamefont {Sato}},
  \bibinfo {author} {\bibfnamefont {M.}~\bibnamefont {Volkov}}, \bibinfo
  {author} {\bibfnamefont {L.}~\bibnamefont {Kasmi}}, \bibinfo {author}
  {\bibfnamefont {N.}~\bibnamefont {Hartmann}}, \bibinfo {author}
  {\bibfnamefont {A.}~\bibnamefont {Rubio}}, \bibinfo {author} {\bibfnamefont
  {L.}~\bibnamefont {Gallmann}}, \ and\ \bibinfo {author} {\bibfnamefont
  {U.}~\bibnamefont {Keller}},\ }\bibfield  {title} {\enquote {\bibinfo {title}
  {Attosecond optical-field-enhanced carrier injection into the {GaAs}
  conduction band},}\ }\href {\doibase 10.1038/s41567-018-0069-0} {\bibfield
  {journal} {\bibinfo  {journal} {Nature Phys.}\ }\textbf {\bibinfo {volume}
  {14}},\ \bibinfo {pages} {560--564} (\bibinfo {year} {2018})}\BibitemShut
  {NoStop}%
\bibitem [{\citenamefont {Volkov}\ \emph {et~al.}(2019)\citenamefont {Volkov},
  \citenamefont {Sato}, \citenamefont {Schlaepfer}, \citenamefont {Kasmi},
  \citenamefont {Hartmann}, \citenamefont {Lucchini}, \citenamefont {Gallmann},
  \citenamefont {Rubio},\ and\ \citenamefont
  {Keller}}]{volkov_attosecond_2019}%
  \BibitemOpen
  \bibfield  {author} {\bibinfo {author} {\bibfnamefont {M.}~\bibnamefont
  {Volkov}}, \bibinfo {author} {\bibfnamefont {S.~A.}\ \bibnamefont {Sato}},
  \bibinfo {author} {\bibfnamefont {F.}~\bibnamefont {Schlaepfer}}, \bibinfo
  {author} {\bibfnamefont {L.}~\bibnamefont {Kasmi}}, \bibinfo {author}
  {\bibfnamefont {N.}~\bibnamefont {Hartmann}}, \bibinfo {author}
  {\bibfnamefont {M.}~\bibnamefont {Lucchini}}, \bibinfo {author}
  {\bibfnamefont {L.}~\bibnamefont {Gallmann}}, \bibinfo {author}
  {\bibfnamefont {A.}~\bibnamefont {Rubio}}, \ and\ \bibinfo {author}
  {\bibfnamefont {U.}~\bibnamefont {Keller}},\ }\bibfield  {title} {\enquote
  {\bibinfo {title} {Attosecond screening dynamics mediated by electron
  localization in transition metals},}\ }\href {\doibase
  10.1038/s41567-019-0602-9} {\bibfield  {journal} {\bibinfo  {journal} {Nature
  Phys.}\ }\textbf {\bibinfo {volume} {15}},\ \bibinfo {pages} {1145--1149}
  (\bibinfo {year} {2019})}\BibitemShut {NoStop}%
\bibitem [{\citenamefont {Buades}\ \emph {et~al.}(2021)\citenamefont {Buades},
  \citenamefont {Pic\'on}, \citenamefont {Berger}, \citenamefont {Le\'on},
  \citenamefont {Di~Palo}, \citenamefont {Cousin}, \citenamefont {Cocchi},
  \citenamefont {Pellegrin}, \citenamefont {Martin}, \citenamefont {Ma\~nas
  Valero}, \citenamefont {Coronado}, \citenamefont {Danz}, \citenamefont
  {Draxl}, \citenamefont {Uemoto}, \citenamefont {Yabana}, \citenamefont
  {Schultze}, \citenamefont {Wall}, \citenamefont {Z\"urch},\ and\
  \citenamefont {Biegert}}]{buades_attosecond_2021}%
  \BibitemOpen
  \bibfield  {author} {\bibinfo {author} {\bibfnamefont {B.}~\bibnamefont
  {Buades}}, \bibinfo {author} {\bibfnamefont {A.}~\bibnamefont {Pic\'on}},
  \bibinfo {author} {\bibfnamefont {E.}~\bibnamefont {Berger}}, \bibinfo
  {author} {\bibfnamefont {I.}~\bibnamefont {Le\'on}}, \bibinfo {author}
  {\bibfnamefont {N.}~\bibnamefont {Di~Palo}}, \bibinfo {author} {\bibfnamefont
  {S.~L.}\ \bibnamefont {Cousin}}, \bibinfo {author} {\bibfnamefont
  {C.}~\bibnamefont {Cocchi}}, \bibinfo {author} {\bibfnamefont
  {E.}~\bibnamefont {Pellegrin}}, \bibinfo {author} {\bibfnamefont {J.~H.}\
  \bibnamefont {Martin}}, \bibinfo {author} {\bibfnamefont {S.}~\bibnamefont
  {Ma\~nas Valero}}, \bibinfo {author} {\bibfnamefont {E.}~\bibnamefont
  {Coronado}}, \bibinfo {author} {\bibfnamefont {T.}~\bibnamefont {Danz}},
  \bibinfo {author} {\bibfnamefont {C.}~\bibnamefont {Draxl}}, \bibinfo
  {author} {\bibfnamefont {M.}~\bibnamefont {Uemoto}}, \bibinfo {author}
  {\bibfnamefont {K.}~\bibnamefont {Yabana}}, \bibinfo {author} {\bibfnamefont
  {M.}~\bibnamefont {Schultze}}, \bibinfo {author} {\bibfnamefont
  {S.}~\bibnamefont {Wall}}, \bibinfo {author} {\bibfnamefont {M.}~\bibnamefont
  {Z\"urch}}, \ and\ \bibinfo {author} {\bibfnamefont {J.}~\bibnamefont
  {Biegert}},\ }\bibfield  {title} {\enquote {\bibinfo {title} {Attosecond
  state-resolved carrier motion in quantum materials probed by soft x-ray
  {XANES}},}\ }\href {\doibase 10.1063/5.0020649} {\bibfield  {journal}
  {\bibinfo  {journal} {Appl. Phys. Rev.}\ }\textbf {\bibinfo {volume} {8}},\
  \bibinfo {pages} {011408} (\bibinfo {year} {2021})}\BibitemShut {NoStop}%
\bibitem [{\citenamefont {Volkov}\ \emph {et~al.}(2023)\citenamefont {Volkov},
  \citenamefont {Sato}, \citenamefont {Niedermayr}, \citenamefont {Rubio},
  \citenamefont {Gallmann},\ and\ \citenamefont
  {Keller}}]{PhysRevB.107.184304}%
  \BibitemOpen
  \bibfield  {author} {\bibinfo {author} {\bibfnamefont {M.}~\bibnamefont
  {Volkov}}, \bibinfo {author} {\bibfnamefont {S.~A.}\ \bibnamefont {Sato}},
  \bibinfo {author} {\bibfnamefont {A.}~\bibnamefont {Niedermayr}}, \bibinfo
  {author} {\bibfnamefont {A.}~\bibnamefont {Rubio}}, \bibinfo {author}
  {\bibfnamefont {L.}~\bibnamefont {Gallmann}}, \ and\ \bibinfo {author}
  {\bibfnamefont {U.}~\bibnamefont {Keller}},\ }\bibfield  {title} {\enquote
  {\bibinfo {title} {Floquet-{Bloch} resonances in near-petahertz
  electroabsorption spectroscopy of {${\mathrm{SiO}}_{2}$}},}\ }\href {\doibase
  10.1103/PhysRevB.107.184304} {\bibfield  {journal} {\bibinfo  {journal}
  {Phys. Rev. B}\ }\textbf {\bibinfo {volume} {107}},\ \bibinfo {pages}
  {184304} (\bibinfo {year} {2023})}\BibitemShut {NoStop}%
\bibitem [{\citenamefont {Inzani}\ \emph {et~al.}(2023)\citenamefont {Inzani},
  \citenamefont {Adamska}, \citenamefont {Eskandari-asl}, \citenamefont
  {Di~Palo}, \citenamefont {Dolso}, \citenamefont {Moio}, \citenamefont
  {D'Onofrio}, \citenamefont {Lamperti}, \citenamefont {Molle}, \citenamefont
  {Borrego-Varillas}, \citenamefont {Nisoli}, \citenamefont {Pittalis},
  \citenamefont {Rozzi}, \citenamefont {Avella},\ and\ \citenamefont
  {Lucchini}}]{Inzani2023a}%
  \BibitemOpen
  \bibfield  {author} {\bibinfo {author} {\bibfnamefont {G.}~\bibnamefont
  {Inzani}}, \bibinfo {author} {\bibfnamefont {L.}~\bibnamefont {Adamska}},
  \bibinfo {author} {\bibfnamefont {A.}~\bibnamefont {Eskandari-asl}}, \bibinfo
  {author} {\bibfnamefont {N.}~\bibnamefont {Di~Palo}}, \bibinfo {author}
  {\bibfnamefont {G.~L.}\ \bibnamefont {Dolso}}, \bibinfo {author}
  {\bibfnamefont {B.}~\bibnamefont {Moio}}, \bibinfo {author} {\bibfnamefont
  {L.~J.}\ \bibnamefont {D'Onofrio}}, \bibinfo {author} {\bibfnamefont
  {A.}~\bibnamefont {Lamperti}}, \bibinfo {author} {\bibfnamefont
  {A.}~\bibnamefont {Molle}}, \bibinfo {author} {\bibfnamefont
  {R.}~\bibnamefont {Borrego-Varillas}}, \bibinfo {author} {\bibfnamefont
  {M.}~\bibnamefont {Nisoli}}, \bibinfo {author} {\bibfnamefont
  {S.}~\bibnamefont {Pittalis}}, \bibinfo {author} {\bibfnamefont {C.~A.}\
  \bibnamefont {Rozzi}}, \bibinfo {author} {\bibfnamefont {A.}~\bibnamefont
  {Avella}}, \ and\ \bibinfo {author} {\bibfnamefont {M.}~\bibnamefont
  {Lucchini}},\ }\bibfield  {title} {\enquote {\bibinfo {title} {Field-driven
  attosecond charge dynamics in germanium},}\ }\href {\doibase
  10.1038/s41566-023-01274-1} {\bibfield  {journal} {\bibinfo  {journal}
  {Nature Photon.}\ }\textbf {\bibinfo {volume} {17}},\ \bibinfo {pages}
  {1059--1065} (\bibinfo {year} {2023})}\BibitemShut {NoStop}%
\bibitem [{\citenamefont {Pellegrini}\ \emph {et~al.}(2016)\citenamefont
  {Pellegrini}, \citenamefont {Marinelli},\ and\ \citenamefont
  {Reiche}}]{pellegrini_physics_2016}%
  \BibitemOpen
  \bibfield  {author} {\bibinfo {author} {\bibfnamefont {C.}~\bibnamefont
  {Pellegrini}}, \bibinfo {author} {\bibfnamefont {A.}~\bibnamefont
  {Marinelli}}, \ and\ \bibinfo {author} {\bibfnamefont {S.}~\bibnamefont
  {Reiche}},\ }\bibfield  {title} {\enquote {\bibinfo {title} {The physics of
  x-ray free-electron lasers},}\ }\href {\doibase 10.1103/RevModPhys.88.015006}
  {\bibfield  {journal} {\bibinfo  {journal} {Rev. Mod. Phys.}\ }\textbf
  {\bibinfo {volume} {88}},\ \bibinfo {pages} {015006} (\bibinfo {year}
  {2016})}\BibitemShut {NoStop}%
\bibitem [{\citenamefont {Duris}\ \emph {et~al.}(2020)\citenamefont {Duris},
  \citenamefont {Li}, \citenamefont {Driver}, \citenamefont {Champenois},
  \citenamefont {MacArthur}, \citenamefont {Lutman}, \citenamefont {Zhang},
  \citenamefont {Rosenberger}, \citenamefont {Aldrich}, \citenamefont {Coffee},
  \citenamefont {Coslovich}, \citenamefont {Decker}, \citenamefont {Glownia},
  \citenamefont {Hartmann}, \citenamefont {Helml}, \citenamefont {Kamalov},
  \citenamefont {Knurr}, \citenamefont {Krzywinski}, \citenamefont {Lin},
  \citenamefont {Marangos}, \citenamefont {Nantel}, \citenamefont {Natan},
  \citenamefont {O'Neal}, \citenamefont {Shivaram}, \citenamefont {Walter},
  \citenamefont {Wang}, \citenamefont {Welch}, \citenamefont {Wolf},
  \citenamefont {Xu}, \citenamefont {Kling}, \citenamefont {Bucksbaum},
  \citenamefont {Zholents}, \citenamefont {Huang}, \citenamefont {Cryan},\ and\
  \citenamefont {Marinelli}}]{duris_tunable_2020}%
  \BibitemOpen
  \bibfield  {author} {\bibinfo {author} {\bibfnamefont {J.}~\bibnamefont
  {Duris}}, \bibinfo {author} {\bibfnamefont {S.}~\bibnamefont {Li}}, \bibinfo
  {author} {\bibfnamefont {T.}~\bibnamefont {Driver}}, \bibinfo {author}
  {\bibfnamefont {E.~G.}\ \bibnamefont {Champenois}}, \bibinfo {author}
  {\bibfnamefont {J.~P.}\ \bibnamefont {MacArthur}}, \bibinfo {author}
  {\bibfnamefont {A.~A.}\ \bibnamefont {Lutman}}, \bibinfo {author}
  {\bibfnamefont {Z.}~\bibnamefont {Zhang}}, \bibinfo {author} {\bibfnamefont
  {P.}~\bibnamefont {Rosenberger}}, \bibinfo {author} {\bibfnamefont {J.~W.}\
  \bibnamefont {Aldrich}}, \bibinfo {author} {\bibfnamefont {R.}~\bibnamefont
  {Coffee}}, \bibinfo {author} {\bibfnamefont {G.}~\bibnamefont {Coslovich}},
  \bibinfo {author} {\bibfnamefont {F.-J.}\ \bibnamefont {Decker}}, \bibinfo
  {author} {\bibfnamefont {J.~M.}\ \bibnamefont {Glownia}}, \bibinfo {author}
  {\bibfnamefont {G.}~\bibnamefont {Hartmann}}, \bibinfo {author}
  {\bibfnamefont {W.}~\bibnamefont {Helml}}, \bibinfo {author} {\bibfnamefont
  {A.}~\bibnamefont {Kamalov}}, \bibinfo {author} {\bibfnamefont
  {J.}~\bibnamefont {Knurr}}, \bibinfo {author} {\bibfnamefont
  {J.}~\bibnamefont {Krzywinski}}, \bibinfo {author} {\bibfnamefont {M.-F.}\
  \bibnamefont {Lin}}, \bibinfo {author} {\bibfnamefont {J.~P.}\ \bibnamefont
  {Marangos}}, \bibinfo {author} {\bibfnamefont {M.}~\bibnamefont {Nantel}},
  \bibinfo {author} {\bibfnamefont {A.}~\bibnamefont {Natan}}, \bibinfo
  {author} {\bibfnamefont {J.~T.}\ \bibnamefont {O'Neal}}, \bibinfo {author}
  {\bibfnamefont {N.}~\bibnamefont {Shivaram}}, \bibinfo {author}
  {\bibfnamefont {P.}~\bibnamefont {Walter}}, \bibinfo {author} {\bibfnamefont
  {A.~L.}\ \bibnamefont {Wang}}, \bibinfo {author} {\bibfnamefont {J.~J.}\
  \bibnamefont {Welch}}, \bibinfo {author} {\bibfnamefont {T.~J.~A.}\
  \bibnamefont {Wolf}}, \bibinfo {author} {\bibfnamefont {J.~Z.}\ \bibnamefont
  {Xu}}, \bibinfo {author} {\bibfnamefont {M.~F.}\ \bibnamefont {Kling}},
  \bibinfo {author} {\bibfnamefont {P.~H.}\ \bibnamefont {Bucksbaum}}, \bibinfo
  {author} {\bibfnamefont {A.}~\bibnamefont {Zholents}}, \bibinfo {author}
  {\bibfnamefont {Z.}~\bibnamefont {Huang}}, \bibinfo {author} {\bibfnamefont
  {J.~P.}\ \bibnamefont {Cryan}}, \ and\ \bibinfo {author} {\bibfnamefont
  {A.}~\bibnamefont {Marinelli}},\ }\bibfield  {title} {\enquote {\bibinfo
  {title} {Tunable isolated attosecond {X}-ray pulses with gigawatt peak power
  from a free-electron laser},}\ }\href {\doibase 10.1038/s41566-019-0549-5}
  {\bibfield  {journal} {\bibinfo  {journal} {Nature Photon.}\ }\textbf
  {\bibinfo {volume} {14}},\ \bibinfo {pages} {30--36} (\bibinfo {year}
  {2020})}\BibitemShut {NoStop}%
\bibitem [{\citenamefont {Malyzhenkov}\ \emph {et~al.}(2020)\citenamefont
  {Malyzhenkov}, \citenamefont {Arbelo}, \citenamefont {Craievich},
  \citenamefont {Dijkstal}, \citenamefont {Ferrari}, \citenamefont {Reiche},
  \citenamefont {Schietinger}, \citenamefont {Jurani\'c},\ and\ \citenamefont
  {Prat}}]{malyzhenkov_single-_2020}%
  \BibitemOpen
  \bibfield  {author} {\bibinfo {author} {\bibfnamefont {A.}~\bibnamefont
  {Malyzhenkov}}, \bibinfo {author} {\bibfnamefont {Y.~P.}\ \bibnamefont
  {Arbelo}}, \bibinfo {author} {\bibfnamefont {P.}~\bibnamefont {Craievich}},
  \bibinfo {author} {\bibfnamefont {P.}~\bibnamefont {Dijkstal}}, \bibinfo
  {author} {\bibfnamefont {E.}~\bibnamefont {Ferrari}}, \bibinfo {author}
  {\bibfnamefont {S.}~\bibnamefont {Reiche}}, \bibinfo {author} {\bibfnamefont
  {T.}~\bibnamefont {Schietinger}}, \bibinfo {author} {\bibfnamefont
  {P.}~\bibnamefont {Jurani\'c}}, \ and\ \bibinfo {author} {\bibfnamefont
  {E.}~\bibnamefont {Prat}},\ }\bibfield  {title} {\enquote {\bibinfo {title}
  {Single- and two-color attosecond hard x-ray free-electron laser pulses with
  nonlinear compression},}\ }\href {\doibase 10.1103/PhysRevResearch.2.042018}
  {\bibfield  {journal} {\bibinfo  {journal} {Phys. Rev. Research}\ }\textbf
  {\bibinfo {volume} {2}},\ \bibinfo {pages} {042018} (\bibinfo {year}
  {2020})}\BibitemShut {NoStop}%
\bibitem [{\citenamefont {Guo}\ \emph {et~al.}(2024)\citenamefont {Guo},
  \citenamefont {Driver}, \citenamefont {Beauvarlet}, \citenamefont {Cesar},
  \citenamefont {Duris}, \citenamefont {Franz}, \citenamefont {Alexander},
  \citenamefont {Bohler}, \citenamefont {Bostedt}, \citenamefont {Averbukh},
  \citenamefont {Cheng}, \citenamefont {DiMauro}, \citenamefont {Doumy},
  \citenamefont {Forbes}, \citenamefont {Gessner}, \citenamefont {Glownia},
  \citenamefont {Isele}, \citenamefont {Kamalov}, \citenamefont {Larsen},
  \citenamefont {Li}, \citenamefont {Li}, \citenamefont {Lin}, \citenamefont
  {McCracken}, \citenamefont {Obaid}, \citenamefont {O'Neal}, \citenamefont
  {Robles}, \citenamefont {Rolles}, \citenamefont {Ruberti}, \citenamefont
  {Rudenko}, \citenamefont {Slaughter}, \citenamefont {Sudar}, \citenamefont
  {Thierstein}, \citenamefont {Tuthill}, \citenamefont {Ueda}, \citenamefont
  {Wang}, \citenamefont {Wang}, \citenamefont {Wang}, \citenamefont {Weber},
  \citenamefont {Wolf}, \citenamefont {Young}, \citenamefont {Zhang},
  \citenamefont {Bucksbaum}, \citenamefont {Marangos}, \citenamefont {Kling},
  \citenamefont {Huang}, \citenamefont {Walter}, \citenamefont {Inhester},
  \citenamefont {Berrah}, \citenamefont {Cryan},\ and\ \citenamefont
  {Marinelli}}]{guo_experimental_2024}%
  \BibitemOpen
  \bibfield  {author} {\bibinfo {author} {\bibfnamefont {Z.}~\bibnamefont
  {Guo}}, \bibinfo {author} {\bibfnamefont {T.}~\bibnamefont {Driver}},
  \bibinfo {author} {\bibfnamefont {S.}~\bibnamefont {Beauvarlet}}, \bibinfo
  {author} {\bibfnamefont {D.}~\bibnamefont {Cesar}}, \bibinfo {author}
  {\bibfnamefont {J.}~\bibnamefont {Duris}}, \bibinfo {author} {\bibfnamefont
  {P.~L.}\ \bibnamefont {Franz}}, \bibinfo {author} {\bibfnamefont
  {O.}~\bibnamefont {Alexander}}, \bibinfo {author} {\bibfnamefont
  {D.}~\bibnamefont {Bohler}}, \bibinfo {author} {\bibfnamefont
  {C.}~\bibnamefont {Bostedt}}, \bibinfo {author} {\bibfnamefont
  {V.}~\bibnamefont {Averbukh}}, \bibinfo {author} {\bibfnamefont
  {X.}~\bibnamefont {Cheng}}, \bibinfo {author} {\bibfnamefont {L.~F.}\
  \bibnamefont {DiMauro}}, \bibinfo {author} {\bibfnamefont {G.}~\bibnamefont
  {Doumy}}, \bibinfo {author} {\bibfnamefont {R.}~\bibnamefont {Forbes}},
  \bibinfo {author} {\bibfnamefont {O.}~\bibnamefont {Gessner}}, \bibinfo
  {author} {\bibfnamefont {J.~M.}\ \bibnamefont {Glownia}}, \bibinfo {author}
  {\bibfnamefont {E.}~\bibnamefont {Isele}}, \bibinfo {author} {\bibfnamefont
  {A.}~\bibnamefont {Kamalov}}, \bibinfo {author} {\bibfnamefont {K.~A.}\
  \bibnamefont {Larsen}}, \bibinfo {author} {\bibfnamefont {S.}~\bibnamefont
  {Li}}, \bibinfo {author} {\bibfnamefont {X.}~\bibnamefont {Li}}, \bibinfo
  {author} {\bibfnamefont {M.-F.}\ \bibnamefont {Lin}}, \bibinfo {author}
  {\bibfnamefont {G.~A.}\ \bibnamefont {McCracken}}, \bibinfo {author}
  {\bibfnamefont {R.}~\bibnamefont {Obaid}}, \bibinfo {author} {\bibfnamefont
  {J.~T.}\ \bibnamefont {O'Neal}}, \bibinfo {author} {\bibfnamefont {R.~R.}\
  \bibnamefont {Robles}}, \bibinfo {author} {\bibfnamefont {D.}~\bibnamefont
  {Rolles}}, \bibinfo {author} {\bibfnamefont {M.}~\bibnamefont {Ruberti}},
  \bibinfo {author} {\bibfnamefont {A.}~\bibnamefont {Rudenko}}, \bibinfo
  {author} {\bibfnamefont {D.~S.}\ \bibnamefont {Slaughter}}, \bibinfo {author}
  {\bibfnamefont {N.~S.}\ \bibnamefont {Sudar}}, \bibinfo {author}
  {\bibfnamefont {E.}~\bibnamefont {Thierstein}}, \bibinfo {author}
  {\bibfnamefont {D.}~\bibnamefont {Tuthill}}, \bibinfo {author} {\bibfnamefont
  {K.}~\bibnamefont {Ueda}}, \bibinfo {author} {\bibfnamefont {E.}~\bibnamefont
  {Wang}}, \bibinfo {author} {\bibfnamefont {A.~L.}\ \bibnamefont {Wang}},
  \bibinfo {author} {\bibfnamefont {J.}~\bibnamefont {Wang}}, \bibinfo {author}
  {\bibfnamefont {T.}~\bibnamefont {Weber}}, \bibinfo {author} {\bibfnamefont
  {T.~J.~A.}\ \bibnamefont {Wolf}}, \bibinfo {author} {\bibfnamefont
  {L.}~\bibnamefont {Young}}, \bibinfo {author} {\bibfnamefont
  {Z.}~\bibnamefont {Zhang}}, \bibinfo {author} {\bibfnamefont {P.~H.}\
  \bibnamefont {Bucksbaum}}, \bibinfo {author} {\bibfnamefont {J.~P.}\
  \bibnamefont {Marangos}}, \bibinfo {author} {\bibfnamefont {M.~F.}\
  \bibnamefont {Kling}}, \bibinfo {author} {\bibfnamefont {Z.}~\bibnamefont
  {Huang}}, \bibinfo {author} {\bibfnamefont {P.}~\bibnamefont {Walter}},
  \bibinfo {author} {\bibfnamefont {L.}~\bibnamefont {Inhester}}, \bibinfo
  {author} {\bibfnamefont {N.}~\bibnamefont {Berrah}}, \bibinfo {author}
  {\bibfnamefont {J.~P.}\ \bibnamefont {Cryan}}, \ and\ \bibinfo {author}
  {\bibfnamefont {A.}~\bibnamefont {Marinelli}},\ }\bibfield  {title} {\enquote
  {\bibinfo {title} {Experimental demonstration of attosecond pump-probe
  spectroscopy with an x-ray free-electron laser},}\ }\href {\doibase
  10.1038/s41566-024-01419-w} {\bibfield  {journal} {\bibinfo  {journal}
  {Nature Photon.}\ }\textbf {\bibinfo {volume} {18}},\ \bibinfo {pages}
  {691--697} (\bibinfo {year} {2024})}\BibitemShut {NoStop}%
\bibitem [{\citenamefont {Runge}\ and\ \citenamefont
  {Gross}(1984)}]{PhysRevLett.52.997}%
  \BibitemOpen
  \bibfield  {author} {\bibinfo {author} {\bibfnamefont {Erich}\ \bibnamefont
  {Runge}}\ and\ \bibinfo {author} {\bibfnamefont {E.~K.~U.}\ \bibnamefont
  {Gross}},\ }\bibfield  {title} {\enquote {\bibinfo {title}
  {Density-functional theory for time-dependent systems},}\ }\href {\doibase
  10.1103/PhysRevLett.52.997} {\bibfield  {journal} {\bibinfo  {journal} {Phys.
  Rev. Lett.}\ }\textbf {\bibinfo {volume} {52}},\ \bibinfo {pages} {997--1000}
  (\bibinfo {year} {1984})}\BibitemShut {NoStop}%
\bibitem [{\citenamefont {Ullrich}(2011)}]{ullrich_time-dependent_2011}%
  \BibitemOpen
  \bibfield  {author} {\bibinfo {author} {\bibfnamefont {C.~A.}\ \bibnamefont
  {Ullrich}},\ }\href@noop {} {\emph {\bibinfo {title} {Time-{Dependent}
  {Density}-{Functional} {Theory}: {Concepts} and {Applications}}}}\ (\bibinfo
  {publisher} {Oxford University Press},\ \bibinfo {address} {Oxford},\
  \bibinfo {year} {2011})\BibitemShut {NoStop}%
\bibitem [{\citenamefont {Marques}\ \emph {et~al.}(2012)\citenamefont
  {Marques}, \citenamefont {Maitra}, \citenamefont {Nogueira}, \citenamefont
  {Gross},\ and\ \citenamefont {Rubio}}]{marques_fundamentals_2012}%
  \BibitemOpen
  \bibinfo {editor} {\bibfnamefont {M.}~\bibnamefont {Marques}}, \bibinfo
  {editor} {\bibfnamefont {N.~T.}\ \bibnamefont {Maitra}}, \bibinfo {editor}
  {\bibfnamefont {F.~M.~S.}\ \bibnamefont {Nogueira}}, \bibinfo {editor}
  {\bibfnamefont {E.~K.~U.}\ \bibnamefont {Gross}}, \ and\ \bibinfo {editor}
  {\bibfnamefont {A.}~\bibnamefont {Rubio}},\ eds.,\ \href@noop {} {\emph
  {\bibinfo {title} {Fundamentals of time-dependent density functional
  theory}}},\ \bibinfo {series} {Lecture notes in physics}\ No.\ \bibinfo
  {number} {Vol. 837}\ (\bibinfo  {publisher} {Springer},\ \bibinfo {address}
  {Berlin Heidelberg New York},\ \bibinfo {year} {2012})\BibitemShut {NoStop}%
\bibitem [{\citenamefont {Tancogne-Dejean}\ \emph {et~al.}(2017)\citenamefont
  {Tancogne-Dejean}, \citenamefont {M\"ucke}, \citenamefont {K\"artner},\ and\
  \citenamefont {Rubio}}]{PhysRevLett.118.087403}%
  \BibitemOpen
  \bibfield  {author} {\bibinfo {author} {\bibfnamefont {N.}~\bibnamefont
  {Tancogne-Dejean}}, \bibinfo {author} {\bibfnamefont {O.~D.}\ \bibnamefont
  {M\"ucke}}, \bibinfo {author} {\bibfnamefont {F.~X.}\ \bibnamefont
  {K\"artner}}, \ and\ \bibinfo {author} {\bibfnamefont {A.}~\bibnamefont
  {Rubio}},\ }\bibfield  {title} {\enquote {\bibinfo {title} {Impact of the
  electronic band structure in high-harmonic generation spectra of solids},}\
  }\href {\doibase 10.1103/PhysRevLett.118.087403} {\bibfield  {journal}
  {\bibinfo  {journal} {Phys. Rev. Lett.}\ }\textbf {\bibinfo {volume} {118}},\
  \bibinfo {pages} {087403} (\bibinfo {year} {2017})}\BibitemShut {NoStop}%
\bibitem [{\citenamefont {Floss}\ \emph {et~al.}(2018)\citenamefont {Floss},
  \citenamefont {Lemell}, \citenamefont {Wachter}, \citenamefont {Smejkal},
  \citenamefont {Sato}, \citenamefont {Tong}, \citenamefont {Yabana},\ and\
  \citenamefont {Burgd\"orfer}}]{PhysRevA.97.011401}%
  \BibitemOpen
  \bibfield  {author} {\bibinfo {author} {\bibfnamefont {I.}~\bibnamefont
  {Floss}}, \bibinfo {author} {\bibfnamefont {C.}~\bibnamefont {Lemell}},
  \bibinfo {author} {\bibfnamefont {G.}~\bibnamefont {Wachter}}, \bibinfo
  {author} {\bibfnamefont {V.}~\bibnamefont {Smejkal}}, \bibinfo {author}
  {\bibfnamefont {S.~A.}\ \bibnamefont {Sato}}, \bibinfo {author}
  {\bibfnamefont {X.-M.}\ \bibnamefont {Tong}}, \bibinfo {author}
  {\bibfnamefont {K.}~\bibnamefont {Yabana}}, \ and\ \bibinfo {author}
  {\bibfnamefont {J.}~\bibnamefont {Burgd\"orfer}},\ }\bibfield  {title}
  {\enquote {\bibinfo {title} {Ab initio multiscale simulation of high-order
  harmonic generation in solids},}\ }\href {\doibase
  10.1103/PhysRevA.97.011401} {\bibfield  {journal} {\bibinfo  {journal} {Phys.
  Rev. A}\ }\textbf {\bibinfo {volume} {97}},\ \bibinfo {pages} {011401}
  (\bibinfo {year} {2018})}\BibitemShut {NoStop}%
\bibitem [{\citenamefont {Hansen}\ \emph {et~al.}(2018)\citenamefont {Hansen},
  \citenamefont {Bauer},\ and\ \citenamefont
  {Madsen}}]{hansen_finite-system_2018}%
  \BibitemOpen
  \bibfield  {author} {\bibinfo {author} {\bibfnamefont {K.~K.}\ \bibnamefont
  {Hansen}}, \bibinfo {author} {\bibfnamefont {D.}~\bibnamefont {Bauer}}, \
  and\ \bibinfo {author} {\bibfnamefont {L.~B.}\ \bibnamefont {Madsen}},\
  }\bibfield  {title} {\enquote {\bibinfo {title} {Finite-system effects on
  high-order harmonic generation: {From} atoms to solids},}\ }\href {\doibase
  10.1103/PhysRevA.97.043424} {\bibfield  {journal} {\bibinfo  {journal} {Phys.
  Rev. A}\ }\textbf {\bibinfo {volume} {97}},\ \bibinfo {pages} {043424}
  (\bibinfo {year} {2018})}\BibitemShut {NoStop}%
\bibitem [{\citenamefont {Yu}\ \emph {et~al.}(2019{\natexlab{a}})\citenamefont
  {Yu}, \citenamefont {Hansen},\ and\ \citenamefont
  {Madsen}}]{yu_enhanced_2019}%
  \BibitemOpen
  \bibfield  {author} {\bibinfo {author} {\bibfnamefont {C.}~\bibnamefont
  {Yu}}, \bibinfo {author} {\bibfnamefont {K.~K.}\ \bibnamefont {Hansen}}, \
  and\ \bibinfo {author} {\bibfnamefont {L.~B.}\ \bibnamefont {Madsen}},\
  }\bibfield  {title} {\enquote {\bibinfo {title} {Enhanced high-order harmonic
  generation in donor-doped band-gap materials},}\ }\href {\doibase
  10.1103/PhysRevA.99.013435} {\bibfield  {journal} {\bibinfo  {journal} {Phys.
  Rev. A}\ }\textbf {\bibinfo {volume} {99}},\ \bibinfo {pages} {013435}
  (\bibinfo {year} {2019}{\natexlab{a}})}\BibitemShut {NoStop}%
\bibitem [{\citenamefont {Yu}\ \emph {et~al.}(2019{\natexlab{b}})\citenamefont
  {Yu}, \citenamefont {Hansen},\ and\ \citenamefont
  {Madsen}}]{yu_high-order_2019}%
  \BibitemOpen
  \bibfield  {author} {\bibinfo {author} {\bibfnamefont {C.}~\bibnamefont
  {Yu}}, \bibinfo {author} {\bibfnamefont {K.~K.}\ \bibnamefont {Hansen}}, \
  and\ \bibinfo {author} {\bibfnamefont {L.~B.}\ \bibnamefont {Madsen}},\
  }\bibfield  {title} {\enquote {\bibinfo {title} {High-order harmonic
  generation in imperfect crystals},}\ }\href {\doibase
  10.1103/PhysRevA.99.063408} {\bibfield  {journal} {\bibinfo  {journal} {Phys.
  Rev. A}\ }\textbf {\bibinfo {volume} {99}},\ \bibinfo {pages} {063408}
  (\bibinfo {year} {2019}{\natexlab{b}})}\BibitemShut {NoStop}%
\bibitem [{\citenamefont {Hansen}\ \emph {et~al.}(2017)\citenamefont {Hansen},
  \citenamefont {Deffge},\ and\ \citenamefont
  {Bauer}}]{hansen_high-order_2017}%
  \BibitemOpen
  \bibfield  {author} {\bibinfo {author} {\bibfnamefont {K.~K.}\ \bibnamefont
  {Hansen}}, \bibinfo {author} {\bibfnamefont {T.}~\bibnamefont {Deffge}}, \
  and\ \bibinfo {author} {\bibfnamefont {D.}~\bibnamefont {Bauer}},\ }\bibfield
   {title} {\enquote {\bibinfo {title} {High-order harmonic generation in solid
  slabs beyond the single-active-electron approximation},}\ }\href {\doibase
  10.1103/PhysRevA.96.053418} {\bibfield  {journal} {\bibinfo  {journal} {Phys.
  Rev. A}\ }\textbf {\bibinfo {volume} {96}},\ \bibinfo {pages} {053418}
  (\bibinfo {year} {2017})}\BibitemShut {NoStop}%
\bibitem [{\citenamefont {Bauer}\ and\ \citenamefont
  {Hansen}(2018)}]{bauer_high-harmonic_2018}%
  \BibitemOpen
  \bibfield  {author} {\bibinfo {author} {\bibfnamefont {D.}~\bibnamefont
  {Bauer}}\ and\ \bibinfo {author} {\bibfnamefont {K.~K.}\ \bibnamefont
  {Hansen}},\ }\bibfield  {title} {\enquote {\bibinfo {title} {High-{Harmonic}
  {Generation} in {Solids} with and without {Topological} {Edge} {States}},}\
  }\href {\doibase 10.1103/PhysRevLett.120.177401} {\bibfield  {journal}
  {\bibinfo  {journal} {Phys. Rev. Lett.}\ }\textbf {\bibinfo {volume} {120}},\
  \bibinfo {pages} {177401} (\bibinfo {year} {2018})}\BibitemShut {NoStop}%
\bibitem [{\citenamefont {Tancogne-Dejean}\ \emph {et~al.}(2018)\citenamefont
  {Tancogne-Dejean}, \citenamefont {Sentef},\ and\ \citenamefont
  {Rubio}}]{PhysRevLett.121.097402}%
  \BibitemOpen
  \bibfield  {author} {\bibinfo {author} {\bibfnamefont {N.}~\bibnamefont
  {Tancogne-Dejean}}, \bibinfo {author} {\bibfnamefont {M.~A.}\ \bibnamefont
  {Sentef}}, \ and\ \bibinfo {author} {\bibfnamefont {A.}~\bibnamefont
  {Rubio}},\ }\bibfield  {title} {\enquote {\bibinfo {title} {Ultrafast
  modification of hubbard $u$ in a strongly correlated material: Ab initio
  high-harmonic generation in nio},}\ }\href {\doibase
  10.1103/PhysRevLett.121.097402} {\bibfield  {journal} {\bibinfo  {journal}
  {Phys. Rev. Lett.}\ }\textbf {\bibinfo {volume} {121}},\ \bibinfo {pages}
  {097402} (\bibinfo {year} {2018})}\BibitemShut {NoStop}%
\bibitem [{\citenamefont {Tancogne-Dejean}\ \emph {et~al.}(2022)\citenamefont
  {Tancogne-Dejean}, \citenamefont {Eich},\ and\ \citenamefont
  {Rubio}}]{tancogne2022effect}%
  \BibitemOpen
  \bibfield  {author} {\bibinfo {author} {\bibfnamefont {N.}~\bibnamefont
  {Tancogne-Dejean}}, \bibinfo {author} {\bibfnamefont {F.~G.}\ \bibnamefont
  {Eich}}, \ and\ \bibinfo {author} {\bibfnamefont {A.}~\bibnamefont {Rubio}},\
  }\bibfield  {title} {\enquote {\bibinfo {title} {{Effect of spin-orbit
  coupling on the high harmonics from the topological Dirac semimetal
  Na3Bi}},}\ }\href {\doibase 10.1038/s41524-022-00831-6} {\bibfield  {journal}
  {\bibinfo  {journal} {npj Comput. Mater.}\ }\textbf {\bibinfo {volume} {8}},\
  \bibinfo {pages} {145} (\bibinfo {year} {2022})}\BibitemShut {NoStop}%
\bibitem [{\citenamefont {Neufeld}\ \emph {et~al.}(2023)\citenamefont
  {Neufeld}, \citenamefont {Tancogne-Dejean}, \citenamefont {H\"ubener},
  \citenamefont {De~Giovannini},\ and\ \citenamefont
  {Rubio}}]{PhysRevX.13.031011}%
  \BibitemOpen
  \bibfield  {author} {\bibinfo {author} {\bibfnamefont {O.}~\bibnamefont
  {Neufeld}}, \bibinfo {author} {\bibfnamefont {N.}~\bibnamefont
  {Tancogne-Dejean}}, \bibinfo {author} {\bibfnamefont {H.}~\bibnamefont
  {H\"ubener}}, \bibinfo {author} {\bibfnamefont {U.}~\bibnamefont
  {De~Giovannini}}, \ and\ \bibinfo {author} {\bibfnamefont {A.}~\bibnamefont
  {Rubio}},\ }\bibfield  {title} {\enquote {\bibinfo {title} {{Are There
  Universal Signatures of Topological Phases in High-Harmonic Generation?
  Probably Not.}}}\ }\href {\doibase 10.1103/PhysRevX.13.031011} {\bibfield
  {journal} {\bibinfo  {journal} {Phys. Rev. X}\ }\textbf {\bibinfo {volume}
  {13}},\ \bibinfo {pages} {031011} (\bibinfo {year} {2023})}\BibitemShut
  {NoStop}%
\bibitem [{\citenamefont {Meier}\ \emph {et~al.}(2007)\citenamefont {Meier},
  \citenamefont {Thomas},\ and\ \citenamefont {Koch}}]{meier_coherent_2007}%
  \BibitemOpen
  \bibfield  {author} {\bibinfo {author} {\bibfnamefont {T.}~\bibnamefont
  {Meier}}, \bibinfo {author} {\bibfnamefont {P.}~\bibnamefont {Thomas}}, \
  and\ \bibinfo {author} {\bibfnamefont {S.~W.}\ \bibnamefont {Koch}},\
  }\href@noop {} {\emph {\bibinfo {title} {Coherent semiconductor optics: from
  basic concepts to nanostructure applications}}}\ (\bibinfo  {publisher}
  {Springer},\ \bibinfo {address} {Berlin Heidelberg New York},\ \bibinfo
  {year} {2007})\BibitemShut {NoStop}%
\bibitem [{\citenamefont {Pic\'on}\ \emph {et~al.}(2019)\citenamefont
  {Pic\'on}, \citenamefont {Plaja},\ and\ \citenamefont {Biegert}}]{Picon2019}%
  \BibitemOpen
  \bibfield  {author} {\bibinfo {author} {\bibfnamefont {A.}~\bibnamefont
  {Pic\'on}}, \bibinfo {author} {\bibfnamefont {L.}~\bibnamefont {Plaja}}, \
  and\ \bibinfo {author} {\bibfnamefont {J.}~\bibnamefont {Biegert}},\
  }\bibfield  {title} {\enquote {\bibinfo {title} {Attosecond x-ray transient
  absorption in condensed-matter: a core-state-resolved {Bloch} model},}\
  }\href {\doibase 10.1088/1367-2630/ab1311} {\bibfield  {journal} {\bibinfo
  {journal} {New J. Phys.}\ }\textbf {\bibinfo {volume} {21}},\ \bibinfo
  {pages} {043029} (\bibinfo {year} {2019})}\BibitemShut {NoStop}%
\bibitem [{\citenamefont {Cistaro}\ \emph {et~al.}(2023)\citenamefont
  {Cistaro}, \citenamefont {Malakhov}, \citenamefont {Esteve-Paredes},
  \citenamefont {Ur\'ia-\'Alvarez}, \citenamefont {Silva}, \citenamefont
  {Mart\'in}, \citenamefont {Palacios},\ and\ \citenamefont
  {Pic\'on}}]{cistaro_theoretical_2023}%
  \BibitemOpen
  \bibfield  {author} {\bibinfo {author} {\bibfnamefont {G.}~\bibnamefont
  {Cistaro}}, \bibinfo {author} {\bibfnamefont {M.}~\bibnamefont {Malakhov}},
  \bibinfo {author} {\bibfnamefont {J.~J.}\ \bibnamefont {Esteve-Paredes}},
  \bibinfo {author} {\bibfnamefont {A.~J.}\ \bibnamefont {Ur\'ia-\'Alvarez}},
  \bibinfo {author} {\bibfnamefont {R.~E.~F.}\ \bibnamefont {Silva}}, \bibinfo
  {author} {\bibfnamefont {F.}~\bibnamefont {Mart\'in}}, \bibinfo {author}
  {\bibfnamefont {J.~J.}\ \bibnamefont {Palacios}}, \ and\ \bibinfo {author}
  {\bibfnamefont {A.}~\bibnamefont {Pic\'on}},\ }\bibfield  {title} {\enquote
  {\bibinfo {title} {Theoretical {Approach} for {Electron} {Dynamics} and
  {Ultrafast} {Spectroscopy} ({EDUS})},}\ }\href {\doibase
  10.1021/acs.jctc.2c00674} {\bibfield  {journal} {\bibinfo  {journal} {J.
  Chem. Theory Comput.}\ }\textbf {\bibinfo {volume} {19}},\ \bibinfo {pages}
  {333--348} (\bibinfo {year} {2023})}\BibitemShut {NoStop}%
\bibitem [{\citenamefont {Du}\ \emph {et~al.}(2019)\citenamefont {Du},
  \citenamefont {Liu}, \citenamefont {Zheng}, \citenamefont {Zeng},\ and\
  \citenamefont {Li}}]{PhysRevA.100.043840}%
  \BibitemOpen
  \bibfield  {author} {\bibinfo {author} {\bibfnamefont {M.}~\bibnamefont
  {Du}}, \bibinfo {author} {\bibfnamefont {C.}~\bibnamefont {Liu}}, \bibinfo
  {author} {\bibfnamefont {Y.}~\bibnamefont {Zheng}}, \bibinfo {author}
  {\bibfnamefont {Z.}~\bibnamefont {Zeng}}, \ and\ \bibinfo {author}
  {\bibfnamefont {R.}~\bibnamefont {Li}},\ }\bibfield  {title} {\enquote
  {\bibinfo {title} {Attosecond transient-absorption spectroscopy in
  one-dimensional periodic crystals},}\ }\href {\doibase
  10.1103/PhysRevA.100.043840} {\bibfield  {journal} {\bibinfo  {journal}
  {Phys. Rev. A}\ }\textbf {\bibinfo {volume} {100}},\ \bibinfo {pages}
  {043840} (\bibinfo {year} {2019})}\BibitemShut {NoStop}%
\bibitem [{\citenamefont {Dong}\ and\ \citenamefont
  {Liu}(2022)}]{PhysRevA.106.063107}%
  \BibitemOpen
  \bibfield  {author} {\bibinfo {author} {\bibfnamefont {F.}~\bibnamefont
  {Dong}}\ and\ \bibinfo {author} {\bibfnamefont {J.}~\bibnamefont {Liu}},\
  }\bibfield  {title} {\enquote {\bibinfo {title} {Fishbone resonance structure
  in the attosecond transient absorption spectrum of graphene},}\ }\href
  {\doibase 10.1103/PhysRevA.106.063107} {\bibfield  {journal} {\bibinfo
  {journal} {Phys. Rev. A}\ }\textbf {\bibinfo {volume} {106}},\ \bibinfo
  {pages} {063107} (\bibinfo {year} {2022})}\BibitemShut {NoStop}%
\bibitem [{\citenamefont {Jin}\ \emph {et~al.}(2019)\citenamefont {Jin},
  \citenamefont {Liang}, \citenamefont {Xiao}, \citenamefont {Wang},
  \citenamefont {Chen}, \citenamefont {Wu}, \citenamefont {Gong},\ and\
  \citenamefont {Peng}}]{PhysRevA.100.013412}%
  \BibitemOpen
  \bibfield  {author} {\bibinfo {author} {\bibfnamefont {J.-Z.}\ \bibnamefont
  {Jin}}, \bibinfo {author} {\bibfnamefont {H.}~\bibnamefont {Liang}}, \bibinfo
  {author} {\bibfnamefont {X.-R.}\ \bibnamefont {Xiao}}, \bibinfo {author}
  {\bibfnamefont {M.-X.}\ \bibnamefont {Wang}}, \bibinfo {author}
  {\bibfnamefont {S.-G.}\ \bibnamefont {Chen}}, \bibinfo {author}
  {\bibfnamefont {X.-Y.}\ \bibnamefont {Wu}}, \bibinfo {author} {\bibfnamefont
  {Q.}~\bibnamefont {Gong}}, \ and\ \bibinfo {author} {\bibfnamefont {L.-Y.}\
  \bibnamefont {Peng}},\ }\bibfield  {title} {\enquote {\bibinfo {title}
  {Contribution of {Floquet}-{Bloch} states to high-order harmonic generation
  in solids},}\ }\href {\doibase 10.1103/PhysRevA.100.013412} {\bibfield
  {journal} {\bibinfo  {journal} {Phys. Rev. A}\ }\textbf {\bibinfo {volume}
  {100}},\ \bibinfo {pages} {013412} (\bibinfo {year} {2019})}\BibitemShut
  {NoStop}%
\bibitem [{\citenamefont {Houston}(1940)}]{houston_acceleration_1940}%
  \BibitemOpen
  \bibfield  {author} {\bibinfo {author} {\bibfnamefont {W.~V.}\ \bibnamefont
  {Houston}},\ }\bibfield  {title} {\enquote {\bibinfo {title} {Acceleration of
  {Electrons} in a {Crystal} {Lattice}},}\ }\href {\doibase
  10.1103/PhysRev.57.184} {\bibfield  {journal} {\bibinfo  {journal} {Phys.
  Rev.}\ }\textbf {\bibinfo {volume} {57}},\ \bibinfo {pages} {184--186}
  (\bibinfo {year} {1940})}\BibitemShut {NoStop}%
\bibitem [{\citenamefont {Krieger}\ and\ \citenamefont
  {Iafrate}(1986)}]{krieger_time_1986}%
  \BibitemOpen
  \bibfield  {author} {\bibinfo {author} {\bibfnamefont {J.~B.}\ \bibnamefont
  {Krieger}}\ and\ \bibinfo {author} {\bibfnamefont {G.~J.}\ \bibnamefont
  {Iafrate}},\ }\bibfield  {title} {\enquote {\bibinfo {title} {Time evolution
  of {Bloch} electrons in a homogeneous electric field},}\ }\href {\doibase
  10.1103/PhysRevB.33.5494} {\bibfield  {journal} {\bibinfo  {journal} {Phys.
  Rev. B}\ }\textbf {\bibinfo {volume} {33}},\ \bibinfo {pages} {5494--5500}
  (\bibinfo {year} {1986})}\BibitemShut {NoStop}%
\bibitem [{\citenamefont {Li}\ \emph {et~al.}(2019)\citenamefont {Li},
  \citenamefont {Zhang}, \citenamefont {Fu}, \citenamefont {Feng},
  \citenamefont {Hu},\ and\ \citenamefont {Du}}]{PhysRevA.100.043404}%
  \BibitemOpen
  \bibfield  {author} {\bibinfo {author} {\bibfnamefont {J.}~\bibnamefont
  {Li}}, \bibinfo {author} {\bibfnamefont {X.}~\bibnamefont {Zhang}}, \bibinfo
  {author} {\bibfnamefont {S.}~\bibnamefont {Fu}}, \bibinfo {author}
  {\bibfnamefont {Y.}~\bibnamefont {Feng}}, \bibinfo {author} {\bibfnamefont
  {B.}~\bibnamefont {Hu}}, \ and\ \bibinfo {author} {\bibfnamefont
  {H.}~\bibnamefont {Du}},\ }\bibfield  {title} {\enquote {\bibinfo {title}
  {Phase invariance of the semiconductor {Bloch} equations},}\ }\href {\doibase
  10.1103/PhysRevA.100.043404} {\bibfield  {journal} {\bibinfo  {journal}
  {Phys. Rev. A}\ }\textbf {\bibinfo {volume} {100}},\ \bibinfo {pages}
  {043404} (\bibinfo {year} {2019})}\BibitemShut {NoStop}%
\bibitem [{\citenamefont {Jiang}\ \emph {et~al.}(2020)\citenamefont {Jiang},
  \citenamefont {Yu}, \citenamefont {Chen}, \citenamefont {Huang},
  \citenamefont {Lu},\ and\ \citenamefont {Lin}}]{PhysRevB.102.155201}%
  \BibitemOpen
  \bibfield  {author} {\bibinfo {author} {\bibfnamefont {S.}~\bibnamefont
  {Jiang}}, \bibinfo {author} {\bibfnamefont {C.}~\bibnamefont {Yu}}, \bibinfo
  {author} {\bibfnamefont {J.}~\bibnamefont {Chen}}, \bibinfo {author}
  {\bibfnamefont {Y.}~\bibnamefont {Huang}}, \bibinfo {author} {\bibfnamefont
  {R.}~\bibnamefont {Lu}}, \ and\ \bibinfo {author} {\bibfnamefont {C.~D.}\
  \bibnamefont {Lin}},\ }\bibfield  {title} {\enquote {\bibinfo {title} {Smooth
  periodic gauge satisfying crystal symmetry and periodicity to study
  high-harmonic generation in solids},}\ }\href {\doibase
  10.1103/PhysRevB.102.155201} {\bibfield  {journal} {\bibinfo  {journal}
  {Phys. Rev. B}\ }\textbf {\bibinfo {volume} {102}},\ \bibinfo {pages}
  {155201} (\bibinfo {year} {2020})}\BibitemShut {NoStop}%
\bibitem [{\citenamefont {Jiang}\ \emph {et~al.}(2018)\citenamefont {Jiang},
  \citenamefont {Chen}, \citenamefont {Wei}, \citenamefont {Yu}, \citenamefont
  {Lu},\ and\ \citenamefont {Lin}}]{PhysRevLett.120.253201}%
  \BibitemOpen
  \bibfield  {author} {\bibinfo {author} {\bibfnamefont {S.}~\bibnamefont
  {Jiang}}, \bibinfo {author} {\bibfnamefont {J.}~\bibnamefont {Chen}},
  \bibinfo {author} {\bibfnamefont {H.}~\bibnamefont {Wei}}, \bibinfo {author}
  {\bibfnamefont {C.}~\bibnamefont {Yu}}, \bibinfo {author} {\bibfnamefont
  {R.}~\bibnamefont {Lu}}, \ and\ \bibinfo {author} {\bibfnamefont {C.~D.}\
  \bibnamefont {Lin}},\ }\bibfield  {title} {\enquote {\bibinfo {title} {Role
  of the transition dipole amplitude and phase on the generation of odd and
  even high-order harmonics in crystals},}\ }\href {\doibase
  10.1103/PhysRevLett.120.253201} {\bibfield  {journal} {\bibinfo  {journal}
  {Phys. Rev. Lett.}\ }\textbf {\bibinfo {volume} {120}},\ \bibinfo {pages}
  {253201} (\bibinfo {year} {2018})}\BibitemShut {NoStop}%
\bibitem [{\citenamefont {Silva}\ \emph {et~al.}(2019)\citenamefont {Silva},
  \citenamefont {Mart\'in},\ and\ \citenamefont {Ivanov}}]{Silva2019}%
  \BibitemOpen
  \bibfield  {author} {\bibinfo {author} {\bibfnamefont {R.~E.~F.}\
  \bibnamefont {Silva}}, \bibinfo {author} {\bibfnamefont {F.}~\bibnamefont
  {Mart\'in}}, \ and\ \bibinfo {author} {\bibfnamefont {M.}~\bibnamefont
  {Ivanov}},\ }\bibfield  {title} {\enquote {\bibinfo {title} {High harmonic
  generation in crystals using maximally localized {Wannier} functions},}\
  }\href {\doibase 10.1103/PhysRevB.100.195201} {\bibfield  {journal} {\bibinfo
   {journal} {Phys. Rev. B}\ }\textbf {\bibinfo {volume} {100}},\ \bibinfo
  {pages} {195201} (\bibinfo {year} {2019})}\BibitemShut {NoStop}%
\bibitem [{\citenamefont {Yue}\ and\ \citenamefont
  {Gaarde}(2020)}]{PhysRevA.101.053411}%
  \BibitemOpen
  \bibfield  {author} {\bibinfo {author} {\bibfnamefont {L.}~\bibnamefont
  {Yue}}\ and\ \bibinfo {author} {\bibfnamefont {M.~B.}\ \bibnamefont
  {Gaarde}},\ }\bibfield  {title} {\enquote {\bibinfo {title} {Structure gauges
  and laser gauges for the semiconductor {Bloch} equations in high-order
  harmonic generation in solids},}\ }\href {\doibase
  10.1103/PhysRevA.101.053411} {\bibfield  {journal} {\bibinfo  {journal}
  {Phys. Rev. A}\ }\textbf {\bibinfo {volume} {101}},\ \bibinfo {pages}
  {053411} (\bibinfo {year} {2020})}\BibitemShut {NoStop}%
\bibitem [{\citenamefont {Yue}\ and\ \citenamefont
  {Gaarde}(2022)}]{Yue_tutorial_22}%
  \BibitemOpen
  \bibfield  {author} {\bibinfo {author} {\bibfnamefont {L.}~\bibnamefont
  {Yue}}\ and\ \bibinfo {author} {\bibfnamefont {M.~B.}\ \bibnamefont
  {Gaarde}},\ }\bibfield  {title} {\enquote {\bibinfo {title} {Introduction to
  theory of high-harmonic generation in solids: tutorial},}\ }\href {\doibase
  10.1364/JOSAB.448602} {\bibfield  {journal} {\bibinfo  {journal} {J. Opt.
  Soc. Am. B}\ }\textbf {\bibinfo {volume} {39}},\ \bibinfo {pages} {535--555}
  (\bibinfo {year} {2022})}\BibitemShut {NoStop}%
\bibitem [{\citenamefont {Blount}(1962)}]{blount_formalisms_1962}%
  \BibitemOpen
  \bibfield  {author} {\bibinfo {author} {\bibfnamefont {E.~I.}\ \bibnamefont
  {Blount}},\ }\bibfield  {title} {\enquote {\bibinfo {title} {Formalisms of
  {Band} {Theory}},}\ }in\ \href@noop {} {\emph {\bibinfo {booktitle} {Solid
  {State} {Physics}}}},\ Vol.~\bibinfo {volume} {13}\ (\bibinfo  {publisher}
  {Elsevier},\ \bibinfo {address} {Amsterdam},\ \bibinfo {year} {1962})\ pp.\
  \bibinfo {pages} {305--373}\BibitemShut {NoStop}%
\bibitem [{\citenamefont {Ashcroft}\ and\ \citenamefont
  {Mermin}(1976)}]{ashcroft_solid_1976}%
  \BibitemOpen
  \bibfield  {author} {\bibinfo {author} {\bibfnamefont {N.~W.}\ \bibnamefont
  {Ashcroft}}\ and\ \bibinfo {author} {\bibfnamefont {N.~D.}\ \bibnamefont
  {Mermin}},\ }\href@noop {} {\emph {\bibinfo {title} {Solid state physics}}}\
  (\bibinfo  {publisher} {Holt, Rinehart and Winston},\ \bibinfo {address} {New
  York},\ \bibinfo {year} {1976})\BibitemShut {NoStop}%
\bibitem [{\citenamefont {Vanderbilt}(2018)}]{vanderbilt_berry_2018}%
  \BibitemOpen
  \bibfield  {author} {\bibinfo {author} {\bibfnamefont {D.}~\bibnamefont
  {Vanderbilt}},\ }\href@noop {} {\emph {\bibinfo {title} {Berry {Phases} in
  {Electronic} {Structure} {Theory}: {Electric} {Polarization}, {Orbital}
  {Magnetization} and {Topological} {Insulators}}}},\ \bibinfo {edition} {1st}\
  ed.\ (\bibinfo  {publisher} {Cambridge University Press},\ \bibinfo {address}
  {Cambridge},\ \bibinfo {year} {2018})\BibitemShut {NoStop}%
\bibitem [{\citenamefont {Jensen}\ \emph {et~al.}(2021)\citenamefont {Jensen},
  \citenamefont {Iravani},\ and\ \citenamefont {Madsen}}]{PhysRevA.103.053121}%
  \BibitemOpen
  \bibfield  {author} {\bibinfo {author} {\bibfnamefont {S.~V.~B.}\
  \bibnamefont {Jensen}}, \bibinfo {author} {\bibfnamefont {H.}~\bibnamefont
  {Iravani}}, \ and\ \bibinfo {author} {\bibfnamefont {L.~B.}\ \bibnamefont
  {Madsen}},\ }\bibfield  {title} {\enquote {\bibinfo {title}
  {Edge-state-induced correlation effects in two-color pump-probe high-order
  harmonic generation},}\ }\href {\doibase 10.1103/PhysRevA.103.053121}
  {\bibfield  {journal} {\bibinfo  {journal} {Phys. Rev. A}\ }\textbf {\bibinfo
  {volume} {103}},\ \bibinfo {pages} {053121} (\bibinfo {year}
  {2021})}\BibitemShut {NoStop}%
\bibitem [{\citenamefont {Bauer}\ and\ \citenamefont
  {Koval}(2006)}]{bauer_qprop_2006}%
  \BibitemOpen
  \bibfield  {author} {\bibinfo {author} {\bibfnamefont {D.}~\bibnamefont
  {Bauer}}\ and\ \bibinfo {author} {\bibfnamefont {P.}~\bibnamefont {Koval}},\
  }\bibfield  {title} {\enquote {\bibinfo {title} {{QPROP}: {A}
  {Schr\"odinger}-solver for intense laser-atom interaction},}\ }\href
  {\doibase 10.1016/j.cpc.2005.11.001} {\bibfield  {journal} {\bibinfo
  {journal} {Comp. Phys. Commun.}\ }\textbf {\bibinfo {volume} {174}},\
  \bibinfo {pages} {396--421} (\bibinfo {year} {2006})}\BibitemShut {NoStop}%
\bibitem [{\citenamefont {Yu}\ \emph {et~al.}(2016)\citenamefont {Yu},
  \citenamefont {Zhang}, \citenamefont {Jiang}, \citenamefont {Cao},
  \citenamefont {Yuan}, \citenamefont {Wu}, \citenamefont {Bai},\ and\
  \citenamefont {Lu}}]{PhysRevA.94.013846}%
  \BibitemOpen
  \bibfield  {author} {\bibinfo {author} {\bibfnamefont {C.}~\bibnamefont
  {Yu}}, \bibinfo {author} {\bibfnamefont {X.}~\bibnamefont {Zhang}}, \bibinfo
  {author} {\bibfnamefont {S.}~\bibnamefont {Jiang}}, \bibinfo {author}
  {\bibfnamefont {X.}~\bibnamefont {Cao}}, \bibinfo {author} {\bibfnamefont
  {G.}~\bibnamefont {Yuan}}, \bibinfo {author} {\bibfnamefont {T.}~\bibnamefont
  {Wu}}, \bibinfo {author} {\bibfnamefont {L.}~\bibnamefont {Bai}}, \ and\
  \bibinfo {author} {\bibfnamefont {R.}~\bibnamefont {Lu}},\ }\bibfield
  {title} {\enquote {\bibinfo {title} {Dependence of high-order-harmonic
  generation on dipole moment in $\mathrm{Si}{\mathrm{o}}_{2}$ crystals},}\
  }\href {\doibase 10.1103/PhysRevA.94.013846} {\bibfield  {journal} {\bibinfo
  {journal} {Phys. Rev. A}\ }\textbf {\bibinfo {volume} {94}},\ \bibinfo
  {pages} {013846} (\bibinfo {year} {2016})}\BibitemShut {NoStop}%
\bibitem [{\citenamefont {Yu}\ \emph {et~al.}(2020)\citenamefont {Yu},
  \citenamefont {Iravani},\ and\ \citenamefont {Madsen}}]{PhysRevA.102.033105}%
  \BibitemOpen
  \bibfield  {author} {\bibinfo {author} {\bibfnamefont {C.}~\bibnamefont
  {Yu}}, \bibinfo {author} {\bibfnamefont {H.}~\bibnamefont {Iravani}}, \ and\
  \bibinfo {author} {\bibfnamefont {L.~B.}\ \bibnamefont {Madsen}},\ }\bibfield
   {title} {\enquote {\bibinfo {title} {Crystal-momentum-resolved contributions
  to multiple plateaus of high-order harmonic generation from band-gap
  materials},}\ }\href {\doibase 10.1103/PhysRevA.102.033105} {\bibfield
  {journal} {\bibinfo  {journal} {Phys. Rev. A}\ }\textbf {\bibinfo {volume}
  {102}},\ \bibinfo {pages} {033105} (\bibinfo {year} {2020})}\BibitemShut
  {NoStop}%
\bibitem [{\citenamefont {B\ae{}kh\o{}j}\ \emph {et~al.}(2015)\citenamefont
  {B\ae{}kh\o{}j}, \citenamefont {Yue},\ and\ \citenamefont
  {Madsen}}]{PhysRevA.91.043408}%
  \BibitemOpen
  \bibfield  {author} {\bibinfo {author} {\bibfnamefont {J.~E.}\ \bibnamefont
  {B\ae{}kh\o{}j}}, \bibinfo {author} {\bibfnamefont {L.}~\bibnamefont {Yue}},
  \ and\ \bibinfo {author} {\bibfnamefont {L.~B.}\ \bibnamefont {Madsen}},\
  }\bibfield  {title} {\enquote {\bibinfo {title} {Nuclear-motion effects in
  attosecond transient-absorption spectroscopy of molecules},}\ }\href
  {\doibase 10.1103/PhysRevA.91.043408} {\bibfield  {journal} {\bibinfo
  {journal} {Phys. Rev. A}\ }\textbf {\bibinfo {volume} {91}},\ \bibinfo
  {pages} {043408} (\bibinfo {year} {2015})}\BibitemShut {NoStop}%
\bibitem [{\citenamefont {Gaarde}\ \emph {et~al.}(2011)\citenamefont {Gaarde},
  \citenamefont {Buth}, \citenamefont {Tate},\ and\ \citenamefont
  {Schafer}}]{PhysRevA.83.013419}%
  \BibitemOpen
  \bibfield  {author} {\bibinfo {author} {\bibfnamefont {M.~B.}\ \bibnamefont
  {Gaarde}}, \bibinfo {author} {\bibfnamefont {C.}~\bibnamefont {Buth}},
  \bibinfo {author} {\bibfnamefont {J.~L.}\ \bibnamefont {Tate}}, \ and\
  \bibinfo {author} {\bibfnamefont {K.~J.}\ \bibnamefont {Schafer}},\
  }\bibfield  {title} {\enquote {\bibinfo {title} {Transient absorption and
  reshaping of ultrafast xuv light by laser-dressed helium},}\ }\href {\doibase
  10.1103/PhysRevA.83.013419} {\bibfield  {journal} {\bibinfo  {journal} {Phys.
  Rev. A}\ }\textbf {\bibinfo {volume} {83}},\ \bibinfo {pages} {013419}
  (\bibinfo {year} {2011})}\BibitemShut {NoStop}%
\bibitem [{\citenamefont {Santra}\ \emph {et~al.}(2011)\citenamefont {Santra},
  \citenamefont {Yakovlev}, \citenamefont {Pfeifer},\ and\ \citenamefont
  {Loh}}]{PhysRevA.83.033405}%
  \BibitemOpen
  \bibfield  {author} {\bibinfo {author} {\bibfnamefont {R.}~\bibnamefont
  {Santra}}, \bibinfo {author} {\bibfnamefont {V.~S.}\ \bibnamefont
  {Yakovlev}}, \bibinfo {author} {\bibfnamefont {T.}~\bibnamefont {Pfeifer}}, \
  and\ \bibinfo {author} {\bibfnamefont {Z.-H.}\ \bibnamefont {Loh}},\
  }\bibfield  {title} {\enquote {\bibinfo {title} {Theory of attosecond
  transient absorption spectroscopy of strong-field-generated ions},}\ }\href
  {\doibase 10.1103/PhysRevA.83.033405} {\bibfield  {journal} {\bibinfo
  {journal} {Phys. Rev. A}\ }\textbf {\bibinfo {volume} {83}},\ \bibinfo
  {pages} {033405} (\bibinfo {year} {2011})}\BibitemShut {NoStop}%
\bibitem [{\citenamefont {Haug}\ and\ \citenamefont
  {Koch}(2009)}]{haug_quantum_2009}%
  \BibitemOpen
  \bibfield  {author} {\bibinfo {author} {\bibfnamefont {H.}~\bibnamefont
  {Haug}}\ and\ \bibinfo {author} {\bibfnamefont {S.~W.}\ \bibnamefont
  {Koch}},\ }\href@noop {} {\emph {\bibinfo {title} {Quantum {Theory} of the
  {Optical} and {Electronic} {Properties} of {Semiconductors}}}},\ \bibinfo
  {edition} {5th}\ ed.\ (\bibinfo  {publisher} {World Scientific},\ \bibinfo
  {address} {Singapore},\ \bibinfo {year} {2009})\BibitemShut {NoStop}%
\end{thebibliography}

%merlin.mbs apsrev4-1.bst 2010-07-25 4.21a (PWD, AO, DPC) hacked
%Control: key (0)
%Control: author (0) dotless jnrlst
%Control: editor formatted (1) identically to author
%Control: production of article title (0) allowed
%Control: page (1) range
%Control: year (0) verbatim
%Control: production of eprint (0) enabled
%

\end{document}